%% file: main.tex
\documentclass[a4paper]{article}
\usepackage{graphicx} 
\usepackage{amsmath} 
\usepackage{amsfonts} 
\usepackage{amssymb} 
\usepackage{amsthm} 
\usepackage{mathrsfs}
\usepackage{booktabs}
\usepackage{graphicx}
\usepackage{titlesec}
\usepackage[labelfont=bf]{caption}
\usepackage{authblk}
\usepackage{natbib}
\usepackage[a4paper, total={6in, 8in}]{geometry}
\usepackage{caption}
\usepackage[disable]{todonotes}

\usepackage[toc]{appendix}
\usepackage[hyperfootnotes=false,colorlinks=true,citecolor=magenta,linkcolor=black]{hyperref}
\usepackage[section]{placeins}
\usepackage[utf8]{inputenc}

\DeclareMathOperator{\Diag}{Diag}
\DeclareMathOperator{\Cs}{CS}

\DeclareMathOperator*{\argmin}{argmin}
\newcommand{\Q}{\mathbb{Q}}
\newcommand{\E}{\mathbb{E}}
\newcommand{\ii}{\mathrm{i}}
\newcommand{\dd}{\mathrm{d}}
\DeclareMathOperator{\Imag}{Im}

\renewcommand{\Re}{\operatorname{Re}}
\renewcommand{\Im}{\operatorname{Im}}

\newtheorem{assum}{Assumption}
\newtheorem{prop}{Proposition}

\theoremstyle{remark}
\newtheorem{remark}{Remark}

\title{Arithmetic Models for Commodity Futures, Vanilla, and Calendar Spread Options\footnote{
We thank conference participants at the EURO 2024 conference in July 2024, the CEMA conference in June 2025, and the QMF conference in December 2025. Support from the Danish Finance Institute (DFI) is gratefully acknowledged.}}
\author[1]{Malthe Rauh Johansen}
\author[1]{Nina Lange}
\author[1]{David Skovmand}
\affil[1]{Department of Mathematical Sciences, University of Copenhagen, Universitetsparken 5, 2100 København Ø, Denmark}
\date{\today}

\renewcommand{\topfraction}{0.85}\renewcommand{\bottomfraction}{0.6}\renewcommand{\textfraction}{0.1}\renewcommand{\floatpagefraction}{0.8}
\begin{document}

\maketitle
\begin{abstract}
We develop an arithmetic multi-factor stochastic volatility model for commodity futures and options. The arithmetic specification yields a low-dimensional Markovian representation of the futures curve and semi-closed-form pricing formulas for vanilla and basket options, including calendar spread options. These properties make the model suitable for large-scale empirical estimation.

We estimate the model using an unscented Kalman filter on a panel of WTI crude oil futures, vanilla options, and calendar spread options spanning 2013–2024. We find that three stochastic volatility factors are needed to capture both the level and slope of the volatility term structure, and that hedges based on these factors remove almost all of the daily risk of a portfolio of vanilla options. Calendar spread options provide information about the dependence structure of futures prices that is not captured by futures and vanilla options alone. A model estimated without calendar spread options systematically overprices them by overstating the decorrelation between adjacent futures contracts. Including these options in the estimation brings the model's decorrelation close to the observed decorrelation, which removes the overpricing and improves their hedges, at only a small loss of accuracy in the pricing and hedging of vanilla options.  Finally, we show that the arithmetic model fits futures and vanilla options better than its geometric counterpart and hedges the vanilla options slightly better.
\end{abstract}

\input{introduction}

\input{model}

\input{data}
\input{estimation}
\input{results}
\input{conclusion}

\newpage
\subsection*{Disclosure of interest}
The authors declare no conflict of interest.
\subsection*{Funding}
No funding was received.
\subsection*{Generative AI declaration}
The authors have used Claude Opus 5.5 and Claude Fable 5.1 for coding assistance, language and grammar improvements, and idea exploration.
\bibliographystyle{apalike}
\bibliography{bib}

\begin{appendix}
\renewcommand{\topfraction}{0.9}\renewcommand{\bottomfraction}{0.7}\renewcommand{\textfraction}{0.07}\renewcommand{\floatpagefraction}{0.5}
\setcounter{topnumber}{3}\setcounter{bottomnumber}{2}\setcounter{totalnumber}{5}
\input{appendix_estimation}
\input{appendix_simulation}
\let\SavedFloatBarrier\FloatBarrier\let\FloatBarrier\relax
\input{appendix_proofs}
\let\FloatBarrier\SavedFloatBarrier

\FloatBarrier
\input{appendix_results}

\end{appendix}

\end{document}

%% file: introduction.tex
\section{Introduction}
During the past 20 years, several HJM-type \citep{Heath1992} models have been proposed to model the term structure of commodity futures prices. The HJM framework's ability to incorporate a wide range of volatility structures, such as stochastic volatility, seasonality and jumps, while maintaining consistency with no-arbitrage conditions makes the approach suitable for modeling commodity futures markets, such as oil, where prices exhibit both short-term volatility and long-term trends; see \cite{trolle2009} and \cite{chiarella2013}
for extensive empirical studies and \cite{clewlow1999multi}, \cite{Gibson1990} and \cite{yan2002valuation} for more foundational work, and \cite{Andersen2010} for finite-dimensional Markovian HJM models.

For more recent models, see \cite{schone2017fourfactor}, \cite{ballestra2026multifactor}, \cite{hughen2010maximal}, \cite{Schneider2018} and \cite{Schneider2015}; the book \cite{benth2023stochastic} presents a general discussion of the literature. 
Common to most of the approaches in the literature is that they assume a level dependence in the volatility that implies futures price dynamics of the generalized \textit{geometric} Brownian motion type. 

In this paper we will instead focus on so-called \textit{arithmetic} models, i.e., models without explicit level dependence in the volatility, the Bachelier model being the simplest example. Arithmetic models and their advantages are well known, and have been studied in many constructions before, but mostly in the context of modeling electricity spot prices and futures. See for example \cite{LuciaSchwartz2002}, \cite{BenthPiccirilliVargiolu2019}, \cite{LatiniPiccirilliVargiolu2019}, \cite{PiccirilliSchmeckVargiolu2021}, \cite{Piccirilli2018} and \cite{BenthDiNunnoKhedherSchmeck2015}. Instead, our paper focuses on building an arithmetic model for the WTI futures market. We also take a simpler approach than most of the electricity models listed above which are heavily focused on modeling seasonality and jumps in prices. Jumps in WTI futures prices appear to be relatively rare and, when present, largely confined to the very short end of the curve, with a limited contribution to overall variance, see \cite{ZikesBarunik2016}. Similarly, seasonality in crude oil is generally found to be weak \citep[see, e.g.,][]{BorovkovaGeman2006}. This motivates the use of a simpler driving process, and we construct a model driven by a continuous affine process with stochastic volatility in the sense of \cite{duffie2003affine}, on the state space $\mathbb{R}_+^k\times \mathbb{R}^n$. Here, $k$ is the number of stochastic volatility factors and $n$ is the number of factors driving the futures curve. Our model can thus be seen as an arithmetic version of the geometric models studied in \cite{trolle2009} and \cite{chiarella2013}. The arithmetic approach has a series of advantages over traditional geometric models. The primary practical advantages over the classical geometric approach with relevance to our study are listed below:

\begin{table}[!h]
\centering\small
\setlength{\aboverulesep}{0pt}\setlength{\belowrulesep}{0pt}%
\renewcommand{\arraystretch}{1.15}%
\begin{tabular}{@{}l|cc@{}}
\toprule
&Arithmetic models & Geometric models \\
\midrule
Number of state variables  & $k+2n$ & $k+5n$\\
Vanilla options & Semi-closed-form solution & Semi-closed-form solution\\
Basket options & Semi-closed-form solution & Approximation solution\\
Calendar spread options & Semi-closed-form solution & Approximation solution\\
Cross-commodity spread options & Semi-closed-form solution & Approximation solution\\
Average price options& Semi-closed-form solution & Approximation solution\\
Flow futures options & Semi-closed-form solution & Approximation solution \\
\bottomrule
\end{tabular}
\caption[Analytical features of arithmetic and geometric models]{\textbf{Analytical features of arithmetic and geometric models}\\ The first row refers to the number of state variables in the Markov representation of the futures curve with a hump-shaped volatility function in the sense of \cite{chiarella2013}. A semi-closed-form solution is one in which the price can be calculated as a one-dimensional numerical integral given the Laplace transform of the driving process.}
\label{tab:aritvsgeom}
\end{table}
The first thing to note in \autoref{tab:aritvsgeom} is that the Markov representation of the futures curve in the arithmetic framework has $3n$ fewer state variables than a similar geometric framework. This means that for realistic values of $k$ and $n$, the arithmetic models are much more amenable to large-scale empirical estimation than the geometric framework with the same driving process.  While the geometric approach admits semi-closed-form pricing for vanilla options, pricing options involving multiple futures contracts generally requires approximations or more computationally demanding numerical methods. On the other hand, arithmetic models produce semi-closed-form solutions for a wide variety of options.  
 
Beyond the practical advantages of arithmetic models, there are also some desirable empirical advantages.  Firstly, it is well documented \citep{Bouchouev2023} that the Bachelier volatility smile implied from options written on WTI futures is roughly symmetric. This implies a symmetric distribution of the futures price \textit{level}. This feature is produced by arithmetic models in a straightforward manner, but is significantly more complicated to do in a geometric model where symmetry in the distribution of the \textit{log-price} is the natural starting point. Furthermore, arithmetic models naturally produce negative prices of futures. This was previously seen as a deficiency, but since the negative price event observed in front-month WTI futures on 20 April 2020, this is now a realistic, but also necessary, feature to price options with negative strikes.\footnote{To accommodate negative prices, the CME switched its pricing and valuation model for options on WTI futures to the Bachelier model on 22 April 2020; see \cite{CME2020Bachelier}.}

The existence of semi-closed-form solutions for a wider class of options allows for a tractable inclusion of calendar spread options in addition to vanilla options and futures in the data panel. A calendar spread option allows the purchaser the option to roll the position of an existing futures position at a predetermined price, and thus is often referred to as a \textit{virtual storage} option.  In particular, calendar spread options on WTI futures have seen a substantial increase in liquidity in the last 15 years, due to the increasingly storage-driven nature of the WTI futures market.  We refer to \cite{Bouchouev2023}, chapter 13, for an overview of the increasing importance and underlying economics of calendar spread options.

While there exists a considerable literature on pricing and hedging spread options, in particular calendar spreads \citep[see for example][for an overview]{Schneider2018}, the empirical literature on calendar spread options is more modestly sized. \cite{Schneider2018} focus on the fact that calendar spread options provide additional information about the joint dynamics of futures contracts at different maturities and, in particular, their dependence structure. They demonstrate the importance of this information by jointly calibrating a multi-factor stochastic volatility model for crude oil futures to vanilla and calendar spread options, using the latter to investigate the term structure of dependence and the so-called Samuelson correlation effect. Their empirical analysis is based on calibration to three individual trading days. More recently, \cite{FRAU2025108809} develop a general geometric framework for pricing calendar spread options in energy markets under stochastic volatility and jumps. Their focus, however, is on calibration, which they do for a single day of data. Other articles with an empirical focus include \cite{SeokBrorsenNiyibizi2018} and \cite{SchmitzWangKimn2013}, both considering option pricing of calendar spread options on agricultural grain commodities. 

Our empirical analysis takes a different approach than what has been done in the existing literature. Since the arithmetic modeling framework provides a semi-closed-form solution for calendar spread options, we incorporate these prices directly into the estimation alongside futures and vanilla options. 

We estimate the proposed model jointly under the risk-neutral and real-world measures using the unscented Kalman filter in conjunction with the prediction error decomposition on an extensive data panel on WTI futures spanning January 2013 to February 2024, with a total of 36{,}530 futures prices, 235{,}277 vanilla option prices and 70{,}830 calendar spread option prices observed over 2{,}810 business days. The options are selected to ensure a representative sample of the daily observed volatility surface. To our knowledge, this is the first large-scale empirical investigation to include time-series of calendar spread option prices in the dynamic estimation of an arbitrage-free commodity futures term structure model. Furthermore, it is the first large-scale empirical study of the performance of arithmetic term structure models for crude oil markets. 
 
 Our results show that three stochastic volatility factors are needed to reproduce both the level and slope of the volatility term structure. We also show that calendar spread options carry information on the dependence structure of futures prices that is not provided by futures or vanilla options alone. A model estimated on futures and vanilla options alone substantially overprices calendar spread options because it overstates the decorrelation between adjacent futures contracts. Including calendar spread options in the estimation substantially improves their pricing fit and changes how volatility is distributed across the factors, allowing the model to capture the observed curve decorrelation. We also find that the model's constant leverage correlations capture only a small part of the observed variation in skew, while forcing the volatility factors to account for the skew worsens the fit to the volatility term structure.

We also investigate the model's hedging performance. Hedges built on the three factors remove almost all of the daily risk of a portfolio of vanilla options, but provide only incremental improvements when hedging calendar spread options. Finally, we compare the pricing and hedging performance of the arithmetic model with that of its geometric counterpart. The arithmetic model provides both a better fit to futures and vanilla option prices and modestly improved hedging performance. 

The paper is structured as follows: In Section~\ref{sec:framework}, the model framework is presented and a general option pricing formula is derived. In Section~\ref{sec:data}, we present the data and the estimation method. Section~\ref{sec:results} presents the estimates and the pricing and hedging results. Finally, Section~\ref{sec:conclusion} concludes.

%% file: model.tex
\newcommand{\tf}{\psi}
\newcommand{\etf}{\chi}
\newcommand{\fvol}{\varphi}
\newcommand{\condwarning}{\color{red} [Needs technical conditions?] \color{black}}
\newcommand{\claim}{\color{red} [EXPAND UPON THIS?]\color{black}}
\newcommand{\refs}{\color{red} [CITE SOME PAPERS HERE]\color{black}}
\section{Arithmetic separable HJM framework}%
\label{sec:framework}
Consider a probability space
$ \left( \Omega, \mathcal{F}, \mathbb{Q} \right) $
equipped with a filtration $\left\{ \mathcal{F}_t\right\}_{t\ge 0}$ of 
$\mathcal{F}$.
We denote by $\mathbb{Q}$ the risk-neutral measure defined by the 
continuously compounded
numeraire $B(t) \triangleq e^{\int_0^t r(u) \mathrm{d}u}$,
where $r(t)$ is the risk-free short rate.
Denote the time-$t$ futures price of a commodity 
with delivery at time $T\geq t$ as $F(t,T)$.
The time-$t$ spot price of the commodity, denoted by $S(t)$, will then
be equal to the price of the instant-delivery futures contract, i.e., $S(t) = F(t,t)$.
Furthermore, the futures price with delivery
at time $T$ is given as the risk-neutral expectation of the time-$T$ spot price \citep[see e.g.,][]{Duffie2001},
such that
\begin{equation*}
	F(t,T) = 
	\mathbb{E}^\mathbb{Q}_t \left\{ F(T,T) \right\}
	=
	\mathbb{E}^\mathbb{Q}_t \left\{ S(T) \right\}.
\end{equation*}
Hence, it follows that commodity futures prices are martingales under the risk-neutral
measure. 
We take as given the initial futures curve $\left\{F(0,T)\right\}_{T>0}$
and make our first key assumption.
\begin{assum}[Arithmetic dynamics]
We assume that the commodity futures price process is an Itô process
with dynamics given by
\begin{equation}
	\label{eq:fut_price_dynamics}
	\mathrm{d}F(t,T) = \sigma(t,T) \cdot \mathrm{d}W^\mathbb{Q}(t),
\end{equation}
where $W^\mathbb{Q}$ is an $n$-dimensional Brownian motion and 
the futures price volatility process $\sigma(t,T)$ is an $n$-dimensional
progressive process for all $T \geq t$. 
\end{assum}
$F(t,T)$ is a true $\mathbb{Q}$-martingale for all $T$ under standard integrability assumptions on $\sigma(t,T)$ \citep[see for example][]{filipovic2009term}. Models of the above type are generally infinite-dimensional, which makes
them difficult to apply. However, it is possible to make simplifying assumptions on
the form of $\sigma(t,T)$ that allow for 
a finite-dimensional Markovian representation of the system, from which we can reconstitute the futures curve.
Drawing inspiration from the literature on finite-dimensional Markovian 
realizations of HJM models, in particular \cite{Bjork2006} for term structure models in general and \cite{Andersen2010} for commodity markets, we make the
following assumption on the form of the futures price volatility process.
\begin{assum}[Separable volatility]
	The stochastic volatility process in \eqref{eq:fut_price_dynamics} is
	of the form
	\begin{equation}
		\label{eq:seperable_vol}
		\sigma(t,T) = H(t) \fvol(T-t),
	\end{equation}
	for a (well-behaved) matrix-valued 
	$n\times n$-dimensional Itô process $H(t)$
	and a deterministic vector-valued function $\fvol$.
\end{assum}
Traditionally, the volatility process $\sigma(t,T)$ has been assumed to be
proportional to the price level, such that
$\sigma(t,T) = F(t,T) \widetilde \sigma(t,T)$
with $\widetilde \sigma(t,T)$ satisfying Assumption 2 \citep[see e.g.,][]{trolle2009,chiarella2013}.
This price-volatility proportionality gives the futures price 
dynamics a geometric structure, and we will throughout the paper refer to models having this property as \textit{geometric} models. We will, however, depart from this assumption and instead consider a
specification of $H(t)$ that includes stochastic volatility but does not
include the price-proportionality, and we will refer to this model class as \textit{arithmetic}. This class of models has to our knowledge mostly been applied in the context of modeling electricity  markets \citep[see the books][for an overview]{benth2008stochastic,benth2023stochastic}, but this is to our knowledge the first attempt at using this model structure for modeling the term structure in the market for crude oil. 
We will throughout this paper highlight its many advantages over the traditional geometric approach.

The specific arithmetic volatility structure $H(t)$ is assumed given by the following multivariate Heston-type form:
\begin{assum}[Heston-type stochastic volatility]
\label{assumption3}
	We assume that $H(t)$ in \eqref{eq:seperable_vol} is of the form
	\begin{equation*}
		H(t) = D_v(v(t)) \triangleq 
		\Diag \left( \sqrt{v_1(t)}, \dots, \sqrt{v_n(t)} \right)
	\end{equation*}
	where the $n$-dimensional stochastic volatility process $v(t)$ is
	given by 
	\begin{equation}
		\label{eq:cir_dynamics}
		\mathrm{d}v(t) 
		= \Gamma \left( v(t) - \theta^\mathbb{Q} \right)\mathrm{d}t
		+ \Diag(\sigma_v) D_v(v(t)) \mathrm{d}Z^\mathbb{Q}(t),
	\end{equation}
    where $\theta^\mathbb{Q}$ and $\sigma_v$ are $n$-dimensional vectors, $\Gamma$
    is an $n \times n$ matrix and
	$Z^\mathbb{Q}(t)$ is an $n$-dimensional 
	standard Brownian motion.
	We write $\vartheta \triangleq -\Gamma\theta^\mathbb{Q}$ for the constant part of
	the drift. Existence of the process requires that
	$\vartheta_i \geq 0$ for all $i$ and that the
	non-diagonal elements of $\Gamma$ are non-negative. Furthermore, to ensure
    that a stationary
    distribution for the volatility process
    exists, we will assume that the eigenvalues of $\Gamma$ are strictly
    negative \citep[see][for both existence and stationarity]{Dai2000}.
    We assume that correlation is only allowed between 
	the Brownian motions $W^\mathbb{Q}(t)$ and 
	$Z^\mathbb{Q}(t)$ element-wise, such that
\begin{equation*}
	\mathrm{d}
	\langle 
	W_{i}^\mathbb{Q}(t)
	,
	Z_{j}^\mathbb{Q}(t)
	\rangle
	= 
	\begin{cases}
		\rho_i \mathrm{d}t &\text{ for } i = j,\\
		0\, \mathrm{d}t & \text{ for } i\neq j.
	\end{cases}
\end{equation*}
\end{assum}
By modeling $\fvol$ as a function of time to maturity
$T-t$ rather than calendar time $t$ and maturity $T$, and by assuming time-homogeneous dynamics for $v(t)$,
the resulting term structure of volatility
becomes stationary over time. While this assumption may be restrictive in general commodity markets with pronounced seasonality patterns \citep[see][]{Schneider2015}, such seasonality can in principle be incorporated in the volatility specification in an arithmetic framework along the lines of \cite{Andersen2010}. However, for crude oil, seasonality is generally not considered a dominant feature of the futures term structure; see \cite{BorovkovaGeman2006}, \cite{carmonaludkovski2004}, \cite{gabillon1991} and \cite{MIRANTES2013}.

Finally, we need to further specify the form of 
$\fvol$. We follow 
\cite{Bjork2006} and \cite{chiarella2013},
and assume that the futures price volatility
loadings are given by the following.
\begin{assum}[Exponential polynomial volatility loadings]
	The vector-valued deterministic function $\fvol$ in \eqref{eq:seperable_vol} 
    is given element-wise by
	\begin{equation}
		\label{eq:fvol_form}
		\fvol_i(\tau) = 
		\sum^{m_i}_{j=1} 
		p_{ij}(\tau)
		e^{-\kappa_{ij} \tau} ,
	\end{equation}
	where $p_{ij}$ are real polynomials and $\kappa_{ij}>0$.
\end{assum}
Generally, this form allows for enough flexibility to capture the observed behavior of the term structure of volatility while allowing for a finite-dimensional realization of the futures curve.
As the time to maturity increases, the volatility of the futures contract converges to zero, i.e.,
$\lim_{\tau \rightarrow \infty}\fvol_i(\tau) = 0$. 
This specification also complies with the Samuelson effect, which states
that the volatility of futures prices 
increases as time to maturity decreases.
Typically, setting $m_i=1$ and using polynomials
of degree zero (constant) or one (linear) is enough for empirical applications.\footnote{
	Indeed, \cite{trolle2009} and \cite{Schneider2018} use constant (degree zero) polynomials,
	while \cite{chiarella2013} and \cite{Cheng2018} use the first order polynomial.
}

Inserting the above volatility specifications into the dynamics of the 
futures price processes, we obtain
\begin{equation}
\mathrm{d} F(t,T) = 
\sigma(t,T) \cdot \mathrm{d}
W^\mathbb{Q}(t)
=
\fvol^\top(T-t) D_v(v(t)) \mathrm{d}
W^\mathbb{Q}(t),
\label{eq:Fdynamics}
\end{equation}
such that the time-$t$ instantaneous volatility of a futures contract with maturity $T>t$ is given by
\begin{equation}
\label{eq:fut_inst_vol}
	\sigma_F(t,T) = 
	\sqrt{
	\sigma^\top(t,T) 
	\sigma(t,T)
	}
	= 
	\sqrt{
		\sum^{n}_{i=1} 
		v_i(t)
		\fvol_i^2(T-t)
	}.
\end{equation}
The instantaneous correlation between two futures with maturities $T_1$ and $T_2$, respectively, is given by
\begin{equation*}
	\varrho_F(t,T_1, T_2) = 
	\frac{
		\sum^{n}_{i=1} 
		v_i(t)
		\fvol_i(T_1 -t)
		\fvol_i(T_2 -t)
	}{
	\sigma_F(t,T_1)
	\sigma_F(t,T_2)
}.
\end{equation*}
Notice that for $n=1$, the correlation is deterministic and equal to one, but
for $n\geq 2$ the model exhibits stochastic correlation as a function of the stochastic volatility
process $v(t)$. The ability to generate 
stochastic correlation is especially important when fitting to correlation-dependent derivatives, such as the calendar
spread options we will consider in the
empirical application.
Another reason for using multifactor
stochastic volatility is to model multiple time scales with varying degrees of autocorrelation.
For equities, these properties are discussed
in e.g., \cite{Duffie2000},
while for crude oil specifically, \cite{trolle2009} develop
a class of (geometric)
stochastic volatility models and investigate model specifications with one
or multiple stochastic volatility factors, where the multidimensional
models achieve a significantly better
fit to crude oil option prices.
Their estimated model consists of 
both a persistent (slowly mean-reverting)
and a transitory (quickly mean-reverting) process,
corroborating the need for multiple time scales to fit the dynamics of the observed volatility surface for crude oil markets. In line with these considerations, the general specification
of the volatility process $v(t)$ in 
\eqref{eq:cir_dynamics} allows for a rich
structure of dependence between the individual stochastic volatility state variables, especially through the matrix $\Gamma$.

\subsection{Markov representation}%
Equation \eqref{eq:Fdynamics} reveals that the driving process of the model is $X(t)=\left(v(t),\int_0^tD_v(v(s))\, \mathrm{d}
W^\mathbb{Q}(s)\right)$ and furthermore $X$ is easily seen to be an affine process on $\mathbb{R}^n_+\times \mathbb{R}^n$. However, due to the presence of the volatility scaling term $\fvol(T-t)$ in \eqref{eq:Fdynamics}, reconstituting the entire futures curve $\{F(t,T)\}_{T\geq t}$ at time $t$ cannot be done from a realization of $X(t)$ alone. The state space needs to be extended to get a so-called Markov representation for the futures curve. In this section, we will derive this Markov representation of the futures curve for arithmetic models. Similar calculations can be found for geometric models in \cite{trolle2009} and \cite{chiarella2013}.

For our proposed models, we will use the form \eqref{eq:fvol_form}
with $m_i=1$ for all $i$.
In this case, the expression for $\fvol$ can be simplified to
\begin{equation*}
	\fvol(\tau) = e^{-K \tau} p(\tau),
\end{equation*}
where
$K \triangleq \Diag \left( \kappa_1, \dots,\kappa_n \right)$ and
$p(\tau) \triangleq \left( p_1(\tau),\dots, p_n(\tau) \right)^\top$.
Using polynomials of the first degree, such that
$p_i(\tau) = \beta_{0,i} + \beta_{1,i}\tau$
for $i=1,\dots,n$, the
finite-dimensional realization of the futures
curve is described below:
\begin{prop}
	Assuming that the volatility process of the futures price  is given as in \eqref{eq:cir_dynamics},
	the time-$t$ futures price with maturity $T\geq t$ can be reconstituted
	from $2n$ state variables 
	\begin{equation}
		F(t,T) = F(0,T) 
		+ \fvol(T-t) \cdot x(t)
		+ \fvol_z(T-t) \cdot z(t),
		\label{eq:pol0_markov}
	\end{equation}
	where $\fvol_z(\tau) \triangleq e^{-K\tau}\beta_1$,
	and the $n$-dimensional processes $x(t)$ and $z(t)$ are given
	by
	\begin{subequations}
	\begin{align}
		\label{eq:pol1_markov1}
		x(t) &\triangleq
		\int_0^t 
		e^{-K (t-u)}
		D_v(v(u))
		\mathrm{d}W^\mathbb{Q}(u),\\
		z(t) &\triangleq
		\int_0^t
		(t-u) 
		e^{-K(t-u)}
		D_v(v(u))
		\mathrm{d}W^\mathbb{Q}(u),
		\label{eq:pol1_markov2}
	\end{align}
	\end{subequations}
	with dynamics
	\begin{align*}
		\mathrm{d}x(t) &= -K x(t) \mathrm{d}t 
		+ D_v(v(t)) \mathrm{d}W^\mathbb{Q}(t),\\
		\mathrm{d}z(t) &= 
		\left( x(t) - K z(t) \right) \mathrm{d}t.
	\end{align*}
\end{prop}
\begin{proof}
See Appendix \ref{sub:proof1}.
\end{proof}
For the model proposed above, futures prices can be reconstituted from the $2n$-dimensional system
$(x(t), z(t))$, where $z(t)$ is of finite variation. The full system $(x(t), z(t), v(t))$ driving the dynamics of the whole futures curve is an affine process with a total of $3n$ state variables. An analogous geometric model has a total of $6n$ state variables \citep[see for example Proposition 1 in][]{chiarella2013}, a substantial difference that makes estimation significantly more computationally challenging. Furthermore, notice that for 
$\beta_{1,i} = 0$, the $z(t)$ variables become 
redundant, making the system only $2n$-dimensional instead. This corresponds to the simplest
polynomial specification, where $p_i(\tau) = \beta_{0,i}$, but it has the disadvantage of not being able to produce a humped volatility structure.

The futures curve depends on the state only through the factors $(x,z)$, whose shocks are generated by $W^\mathbb{Q}$ alone, while the volatility factors $v$ enter only the volatility of $x$. Futures can therefore hedge only risk spanned by $W^\mathbb{Q}$. Since each $v_i$ is driven by $Z_i^\mathbb{Q}$, which is correlated only with $W_i^\mathbb{Q}$, shocks to $v_i$ are spanned by $W^\mathbb{Q}$, and can thus be hedged with futures, if and only if $|\rho_i| = 1$. Whenever $|\rho_i| < 1$ for some $i$, volatility risk is not fully spanned by futures, so options on futures are not redundant and the model exhibits unspanned stochastic volatility in the manner of \cite{trolle2009}.

\subsection{Option pricing}
\label{Sec:Option pricing}
Due to the arithmetic framework, a large class
of option types can be priced using the same semi-closed-form pricing method as for vanilla
options on futures contracts.
This class importantly includes exotic options encountered in commodity
markets, such as the so-called \emph{calendar spread options}.
We define the time-$t$ price of a $\Delta$-lag calendar spread as
\begin{equation}
    \Cs(t,T_1,\Delta) = 
    F(t,T_1) - F(t,T_1+\Delta).
\end{equation}
A calendar spread call with maturity $T\leq T_1$ is then defined by the time-$T$ payoff
$\left[ \Cs(T,T_1,\Delta) - K \right]^+$ 
(or likewise for a put, $\left[ K -
\Cs(T,T_1,\Delta)\right]^+$). 
To develop the general pricing formula, we first 
consider a basket of $L$ different futures given by
\begin{equation*}
	Y(t) \triangleq \sum^{L}_{j=1} w_j F(t,T_j),
\end{equation*}
where $w_j$ are the individual weights.
Specific instances of $Y(t)$ include $F(t,T_1)$, the underlying for
a vanilla option on a $T_1$ futures contract, 
as well
as $F(t,T_1) - F(t,T_2)$ for $T_1<T_2$, the underlying for a calendar spread option
with lag $\Delta = T_2 - T_1$.
Now, for $T \leq \min_j T_j$ we can define 
$\fvol_Y(T-t) \triangleq \sum^{L}_{j=1}w_j\fvol(\Delta T_j + T -t)$ with 
$\Delta T_j \triangleq T_j - T$, such that
\begin{equation*}
	\mathrm{d}Y(t) =
	\left( \sum^{L}_{j=1} w_j \fvol(T_j -t)  \right)^\top
	D_v(v(t))
	\mathrm{d}W^\mathbb{Q}(t)
	= 
	\fvol_Y^\top(T-t)
	D_v(v(t)) 
	\mathrm{d}W^\mathbb{Q}(t).
\end{equation*}
Note that the dynamics of $Y(t)$ have the same structure as the individual futures price dynamics in \eqref{eq:Fdynamics}, a highly desirable feature specific to the arithmetic framework.

If we make the simplifying assumption that interest rates are independent of futures prices, we can write the formula for a European put option on a generic basket $Y(T)$ as 
\begin{equation*}
		\mathcal{P}(t,T,K)
		=
		\mathbb{E}^\mathbb{Q}_t
		\left\{ 
			e^{-\int_t^T r(s) \mathrm{d}s}
			\left( K - Y(T) \right)^+
		\right\}
		=
		P(t,T)
		\mathbb{E}^\mathbb{Q}_t
		\left\{ 
			\left( K - Y(T) \right)^+
		\right\},
\end{equation*}
where $P(t,T)$ is the time-$t$ price of a zero-coupon bond maturing at time $T$.
Indeed, \cite{Cheng2018} investigate the empirical option pricing performance of 
commodity futures models with and without stochastic interest rates dependent on the futures price. They find that the explicit modeling of the joint distribution of interest rates and futures prices is mainly relevant for pricing longer-dated options.
Since our main interest is in the short and medium term, we do not consider the assumption of independence to be problematic. Note that the setting with stochastic dependent interest rates is equally tractable. Further details are given in Appendix~\ref{sub:proof2}.   
 
Drawing from the work by \cite{Duffie2000} and \cite{CollinDufresne2003}, we describe a general pricing formula for any process $Y(t)$ with dynamics of the form above. \begin{prop}
\label{prop:optionprice}
	Consider the time-$t$ price of a European put option expiring at time
	$T$ on $Y(t)$ with dynamics of the form
	\begin{subequations}
	\begin{align}
		\mathrm{d}Y(t) &= 
	\varphi_Y^\top(T-t)
	D_v(v(t)) 
	\mathrm{d}W^\mathbb{Q}(t),\label{eq:prop2Y}\\
	\mathrm{d}v(t)
			       &= \Gamma \left( 
		v(t) - \theta^\mathbb{Q}
	\right)\mathrm{d}t
	+ \Diag(\sigma_v) D_v(v(t)) \mathrm{d}Z^\mathbb{Q}(t),\label{eq:prop2v}
	\end{align}
	\end{subequations}
	where $\varphi_Y$ is an $n$-dimensional deterministic function.
	The time-$t$ price of a put option with maturity $T>t$ and strike $K$ on $Y(t)$ is then given by
	\begin{align}
		\nonumber
		\mathcal{P}(t,T,K)
		&=
		P(t,T)
		\mathbb{E}^\mathbb{Q}_t
		\left\{ 
			\left( K - Y(T) \right)^+
		\right\}\\
		&=
		\label{eq:put_option_price}
		P(t,T)
		\left[ 
			\frac{K - Y(t)}{2} 
		-
		\int_0^\infty
		\frac{\Im \left\{
				\left[
					K - \etf(\omega)
				\right]
				e^{-\ii\omega K + \tf(\omega)}
		\right\}}{\pi\omega} 
		\mathrm{d}\omega
		\right],
	\end{align}
	where $\ii = \sqrt{-1}$, $e^{\tf(\omega)}=\mathbb{E}^\mathbb{Q}_t\{e^{\ii\omega Y(T)}\}$ is the conditional
	characteristic function of $Y(T)$, and $\etf(\omega)=-\ii\,\partial\tf(\omega)/\partial\omega$.
	These functions are given by 
	\begin{align*}
		\tf 
		\left[ 
		\omega;
		\tau, v(t), Y(t)
		\right]
		&= 
		A_\tf (\tau;\omega) + B_\tf(\tau;\omega) \cdot v(t)
		+ \mathrm{i} \omega Y(t), \\
		\chi
		\left[ 
			\omega;
			\tau, v(t), Y(t) 
		\right]
		&=
		A_\etf(\tau;\omega) + B_\etf(\tau;\omega) \cdot v(t)
		+ Y(t),
	\end{align*}
	where $\tau = T-t$. $(A_\tf, B_\tf)$ and $(A_\etf, B_\etf)$
	are solutions to the $2(n+1)$-dimensional system of complex-valued ODEs
	given by
	\begin{subequations}
	\begin{align}
		\label{eq:tfA_ode}
		\frac{\mathrm{d}A_\tf(\tau;\omega)}{\mathrm{d}\tau} 
		&= -(\Gamma\theta^\mathbb{Q}) \cdot B_\tf(\tau; \omega),
		& A_\tf(0;\omega) = 0; \\
		\label{eq:tfB_ode}
		\frac{\mathrm{d}B_{\tf}(\tau;\omega)}{\mathrm{d}\tau} 
	&=
	\Gamma^\top B_{\tf}(\tau;\omega)
	+ R_\tf \left[ \tau, B_\tf(\tau;\omega); \omega \right], 
		& B_\tf(0;\omega) = 0; \\
		\label{eq:etfA_ode}
	\frac{\mathrm{d}A_\etf(\tau;\omega)}{\mathrm{d}\tau} 
		&= -(\Gamma\theta^\mathbb{Q}) \cdot B_\etf(\tau;\omega),
		& A_\etf(0;\omega) = 0; \\
		\label{eq:etfB_ode}
		\frac{\mathrm{d}B_{\etf}(\tau;\omega)}{\mathrm{d}\tau} 
	&=
	\Gamma^\top B_{\etf}(\tau;\omega)
+ R_\etf\left[\tau, B_\tf(\tau;\omega), B_\etf(\tau;\omega); \omega \right],
		& B_\etf(0;\omega) = 0.
	\end{align}
	\end{subequations}
	The functions $R_\tf$ and $R_\etf$ depend upon the volatility function
	$\varphi_Y$ and are given element-wise by
	\begin{align*}
		R_{\tf,i}
		\left[ \tau, B_\tf; \omega\right]
		&= 
		\frac{\sigma_{v,i}^2}{2} B_{\tf,i}^2
		+
	\mathrm{i}\omega 
	\rho_i\sigma_{v,i}\varphi_{Y,i}(\tau)
	B_{\tf,i}
	- \frac{\omega^2}{2}
	\varphi_{Y,i}^2(\tau), \\
		R_{\etf,i}
		\left[ \tau, B_\tf, B_\etf; \omega\right]
		&=
		\sigma_{v,i}^2
		B_{\tf,i} B_{\etf,i}
	+
	\left[ 
		\mathrm{i}\omega B_{\etf,i}
	+
	B_{\tf,i}
	\right]
	\rho_i
	\sigma_{v,i}
	\varphi_{Y,i}(\tau)
		+
		\mathrm{i} \omega
		\varphi_{Y,i}^2(\tau),
	\end{align*}
	for $i = 1,\dots,n$.
	Note that $(A_\etf, B_\etf)$
	depends on $(A_\tf, B_\tf)$ through $R_\etf$, 
	but not vice versa.
\end{prop}
\begin{proof}
	See Appendix \ref{sub:proof2}.
\end{proof}
\begin{remark}
To price a European call option, 
we can simply price the equivalent put option and use put-call parity,
such that the price of a call option is given by $\mathcal{P}(t,T,K) + P(t,T)(Y(t) - K)$.
The above formula can then be used to price 
a $T$-maturity European option on a futures contract
with maturity $T_1 \geq T$ by setting $\varphi_Y(\tau) = \varphi(\Delta T + \tau)$
with $\Delta T \triangleq T_1 - T$.
Equivalently, options on the calendar spread $F(t,T_1) - F(t,T_2)$, where $T_2 > T_1 \geq T$,
can be priced by using 
$\varphi_Y(\tau) = \left(\varphi(\Delta T_1 + \tau) - \varphi(\Delta T_2 + \tau) \right)$
where $\Delta T_1 \triangleq T_1 - T$ and $\Delta T_2 \triangleq T_2 - T$. 
 \end{remark}
 \begin{remark}
The delta and vegas of the put are also available in semi-closed form. Since
$\etf = -\ii\,\partial\tf/\partial\omega$, an integration by parts in $\omega$
removes $\etf$ from the vegas. This gives
\begin{align*}
	\Delta_\mathcal{P}(t)
	&=
	\frac{\partial \mathcal{P}(t,T,K)}{\partial Y(t)}
	=
	P(t,T)
	\left[
	-\frac{1}{2}
	+
	\int_0^\infty
	\frac{\Im \left\{
		e^{-\ii\omega K + \tf(\omega)}
	\right\}}{\pi\omega}
	\mathrm{d}\omega
	\right],
	\\
	\mathcal{V}_{\mathcal{P},i}(t)
	&=
	\frac{\partial \mathcal{P}(t,T,K)}{\partial v_i(t)}
	=
	-P(t,T)
	\int_0^\infty
	\frac{\Re \left\{
		B_{\tf,i}(\tau;\omega)\,
		e^{-\ii\omega K + \tf(\omega)}
	\right\}}{\pi\omega^2}
	\mathrm{d}\omega,
\end{align*}
for $i = 1,\dots,n$. For the call, $\Delta_\mathcal{C}(t) = \Delta_\mathcal{P}(t) + P(t,T)$,
and the vegas are the same.
\end{remark}

The property that both vanilla options and spread options can be priced using the same formula,
although with different coefficients in the ODEs \eqref{eq:tfA_ode}--\eqref{eq:etfB_ode},
is not shared by geometric models. When pricing spread options in geometric models one typically has to resort to approximation or to methods that are significantly more numerically burdensome than when pricing vanilla options. See \cite{Schneider2018} for a survey of pricing techniques in a geometric framework.
\subsection{Moment formulas}%
\label{sub:moments}
The centered moments
\begin{equation*}
\varkappa_m(t,T)\triangleq\mathbb{E}^{\mathbb{Q}}_t\{(Y(T)-Y(t))^m\}
\end{equation*}
of the distribution of an underlying asset play a significant part in our empirical analysis. They can be
calculated from the model as well as directly from option prices. In the model, they follow from the
characteristic function of Proposition~\ref{prop:optionprice} by differentiation at $\omega=0$, and are given
by its Riccati equations expanded in powers of $\ii\omega$.
\begin{prop}\label{prop:modelmom}
Let $Y$ have the dynamics \eqref{eq:prop2Y}--\eqref{eq:prop2v} and let
$\tau=T-t$. For $m=2,3$,
\begin{equation}\label{eq:affine}
\varkappa_m(t,T)=A_m(\tau)+B_m(\tau)\cdot v(t),
\end{equation}
where $A_m$ and $B_m$, the coefficients of $(\ii\omega)^m/m!$ in $A_\tf$ and $B_\tf$, solve the $(n+1)$-dimensional
linear system
\begin{equation}\label{eq:momode}
\frac{\mathrm{d}A_m(\tau)}{\mathrm{d}\tau}=-(\Gamma\theta^\mathbb{Q})\cdot B_m(\tau),\quad A_m(0)=0;\qquad
\frac{\mathrm{d}B_m(\tau)}{\mathrm{d}\tau}=\Gamma^\top B_m(\tau)+R_m(\tau),\quad B_m(0)=0,
\end{equation}
with $R_m$ given element-wise by
\begin{equation}\label{eq:forcing}
R_{2,i}(\tau)=\varphi_{Y,i}^2(\tau),\qquad
R_{3,i}(\tau)=3\rho_i\sigma_{v,i}\varphi_{Y,i}(\tau)B_{2,i}(\tau),\qquad i=1,\dots,n .
\end{equation}
Equivalently, $B_m(\tau)=\int_0^\tau e^{\Gamma^\top u}R_m(\tau-u)\,\mathrm{d}u$
and $A_m(\tau)=-\int_0^\tau (\Gamma\theta^\mathbb{Q})\cdot B_m(u)\,\mathrm{d}u$. Note that $B_3$
depends on $B_2$ through $R_3$, but not vice versa.
\end{prop}
\begin{proof}
See Appendix~\ref{sub:proof3}.
\end{proof}
Alternatively, a model-free estimation of the moments of the increment $Y(T)-Y(t)$ can be performed directly from observed
option prices using the spanning formula of \cite{carr2001}. Write
$\bar{\mathcal{P}}(t,T,K)\triangleq\mathbb{E}^{\mathbb{Q}}_t\{(K-Y(T))^+\}$ and
$\bar{\mathcal{C}}(t,T,K)\triangleq\mathbb{E}^{\mathbb{Q}}_t\{(Y(T)-K)^+\}$ for the undiscounted put and call values on $Y(T)$. Applying
the spanning formula to $f(y)=(y-Y(t))^m$ gives
\begin{equation}\label{eq:moments}
\varkappa_m(t,T)=m(m-1)\left[\int_{-\infty}^{Y(t)}\!\!(K-Y(t))^{m-2}\bar{\mathcal{P}}(t,T,K)\,\mathrm{d}K
+\int_{Y(t)}^{\infty}\!(K-Y(t))^{m-2}\bar{\mathcal{C}}(t,T,K)\,\mathrm{d}K\right]
\end{equation}
for $m\geq2$. 

From the second and third moments we construct the model-free volatility and
the skewness of $Y(T)-Y(t)$, defined as follows
\begin{equation*}
\mathcal{M}_2 \triangleq \sqrt{\frac{\varkappa_2}{\tau}}
\qquad\text{and}\qquad
\mathcal{M}_3 \triangleq \frac{\varkappa_3}{\varkappa_2^{3/2}},
\end{equation*}
which can be computed from observed option prices and from the model alike. In the model, the two
are closely coupled: by Proposition~\ref{prop:modelmom}, $\varkappa_2$ and $\varkappa_3$ are affine in
the same volatility factors $v(t)$, and since $R_3$ is built from $B_2$, each factor enters
$\varkappa_3$ only through its contribution to $\varkappa_2$, weighted by its leverage correlation
$\rho_i$ and its volatility of volatility $\sigma_{v,i}$. $\mathcal{M}_2$ does not depend on the
$\rho_i$, while $\mathcal{M}_3$ is linear in them, so the skew can move only as the factors that
carry the volatility term structure move.

\subsection{Measure change and time-homogeneity}%
For the purpose of estimation, we further need to specify the 
dynamics of the state vector under the physical measure $\mathbb{P}$.
We use the \emph{completely} affine specification of \cite{Duffee2002}, detailed in
Appendix~\ref{sec:Measurechanges}, in which the Brownian motions are related by
\begin{align}
	\mathrm{d}W^\mathbb{P}(t)
	&= 
	\mathrm{d}W^\mathbb{Q}(t)
	- 
	D_v(v(t)) 
	\lambda_{x} 
	\mathrm{d}t\label{eq:measurechange1}\\
	\mathrm{d}Z^\mathbb{P}(t)
	&= 
	\mathrm{d}Z^\mathbb{Q}(t)
	- 
	D_v(v(t)) 
	\lambda_{v} 
	\mathrm{d}t,\label{eq:measurechange2}
\end{align}
where $\lambda_x$ and $\lambda_v$ are $n$-dimensional vectors.
For the statistical estimation we also need to ensure that a stationary distribution for the volatility process 
exists under the physical measure, and thus
in parallel to Assumption~\ref{assumption3}
we assume that
$\Gamma^\mathbb{P} \triangleq \Gamma + \Diag(\sigma_v)\Diag(\lambda_v)$ 
has strictly negative eigenvalues. Since the measure change only adds a term
linear in $v(t)$ to the drift, the constant $\vartheta$ is the same under both
measures, and the dynamics of $v(t)$ under $\mathbb{P}$ are
\begin{equation*}
	\mathrm{d}v(t) = \Gamma^\mathbb{P}\left( v(t) - \theta \right)\mathrm{d}t
	+ \Diag(\sigma_v) D_v(v(t)) \mathrm{d}Z^\mathbb{P}(t),
\end{equation*}
where $\theta$ is the stationary mean of $v(t)$ under $\mathbb{P}$ and
$\Gamma\theta^\mathbb{Q} = \Gamma^\mathbb{P}\theta = -\vartheta$.

The model as described above by construction fits 
the initial curve $F(0,T)$ for $T>0$, but for estimation we prefer a
time-homogeneous version. For simplicity we set the initial curve
to a constant $F_0$, which we treat as an additional parameter of the model.
Thus we assume that the infinite-maturity futures contract price is given
by some constant $F_0$, i.e.,
$\lim_{T\rightarrow \infty}F(t,T) = F_0$.
Finally, as per \cite{Dai2000},
the model above is not identifiable in its current form,
but we can achieve identification by setting $\theta^\mathbb{Q}_i= 1$ for all $i$,
such that the stationary mean of each volatility state variable under
$\mathbb{Q}$ equals one. The mean level under $\mathbb{P}$,
$\theta = (\Gamma^\mathbb{P})^{-1}\Gamma\theta^\mathbb{Q}$, then follows from the
estimates.

\subsection{Pricing other commodity derivatives}
As highlighted in \autoref{tab:aritvsgeom}, the arithmetic framework allows for semi-closed-form pricing of a wide variety of options depending on multiple futures prices at the same time. In this section we explain how this is done for the most common examples. 
\subsubsection{Pricing cross-commodity spread options}
The arithmetic framework can easily be extended to modeling multiple futures curves of different commodity prices. This is useful in particular for pricing cross-commodity spread options like options on the crack spread (crude oil vs. refined products like gasoline and heating oil), the crush spread (soybeans vs. soybean meal and oil), the spark spread (natural gas vs. electricity), or inter-location spreads, which involve the same commodity, same delivery month, but different exchanges or delivery locations (e.g., WTI in Cushing, Oklahoma vs. Brent in the North Sea).

To illustrate, let $F_A(t,T_1)$ and $F_B(t,T_2)$ be the futures prices of two different commodities, with arithmetic dynamics 
\begin{align}
		\mathrm{d}F_A(t,T_1) &= 
	\varphi_A^\top(T_1-t)
	D_v(v(t))\mathrm{d}W^\mathbb{Q}(t),\quad  \mathrm{d}F_B(t,T_2) = 
	\varphi_B^\top(T_2-t)
	D_v(v(t)) 
	\mathrm{d}W^\mathbb{Q}(t),
	\end{align}
where the dynamics of $v(t)$ are given in \eqref{eq:cir_dynamics}. By specifying $\varphi_A$ and $\varphi_B$ and the parameters of $v$, the model can thus have common and idiosyncratic factors with common and idiosyncratic stochastic volatility. A European option on the spread between these two futures prices can be priced using Proposition 2, by noting that if we set the spread $Y(t)\triangleq F_A(t,T_1)-F_B(t,T_2)$ we get arithmetic dynamics for the spread
\begin{align}	\mathrm{d}Y(t) &= \varphi_Y^\top(T-t)
	D_v(v(t))\mathrm{d}W^\mathbb{Q}(t)\\
\varphi_Y(T-t) &\triangleq \varphi_A(T_1-t)-\varphi_B(T_2-t),\quad 
T=\min(T_1,T_2).
\end{align}
\subsubsection{Commodity futures with multiple delivery periods}
Some commodity futures do not have a single discrete time of delivery, but instead a delivery period over which the commodity is delivered as a flow. For non-storable commodities like electricity this is the only form of delivery, but this type of futures also exists for a storable commodity like natural gas. The arithmetic framework presented in this section can be easily modified to accommodate this style of futures contract.

If we associate the futures price in \eqref{eq:fut_price_dynamics} with the shortest possible delivery period $\delta$, say a day or an hour\footnote{A different approach to achieve consistency between futures prices with different delivery periods can be found in \cite{BenthPiccirilliVargiolu2019}, where futures prices with an infinitesimally small delivery period are modeled.}, it is easy to see that the futures price expiring at time $T$ with an associated $N\times \delta$ delivery period, $F(t,T,N)$ for $N\in \mathbb{N}$, is given by
\begin{align}
    F(t,T,N)= \frac{1}{N}\sum_{k=1}^N F(t,T+(k-1)\delta),\quad N \in \mathbb{N},
    \label{eq:multideliveryfuturesprice}
\end{align}
with arithmetic dynamics
\begin{align}
\mathrm{d}F(t,T,N) = \sigma(t,T,N) \cdot \mathrm{d}W^\mathbb{Q}(t),\quad \sigma(t,T,N)=D_v (v(t))\frac{1}{N}\sum_{k=1}^N\fvol(T+(k-1)\delta-t).\label{eq:multiperiodfuturesdynamics}
\end{align}
Note that the decomposition in \eqref{eq:multideliveryfuturesprice} implies that options on multi-period delivery futures are simply another form of basket option and are thus priced directly by Proposition \ref{prop:optionprice}.

The fact that multiple-period futures prices retain the same structure, and thus remain tractable for any $N\in \mathbb{N}$ in \eqref{eq:multiperiodfuturesdynamics}, is again specific to the arithmetic framework and presents another advantage over the traditional geometric modeling framework.

\subsubsection{Average price options}
 
Average price options (APOs) are widely traded in several commodity markets. The
payoff is determined by the arithmetic average of a sequence of futures settlement prices rather than by a single
terminal price. Let $t_1<t_2<\dots<t_M=T$ denote the fixing dates, let $w_1,\dots,w_M$ be weights with
$\sum_{m=1}^Mw_m=1$, and let $T_m\ge t_m$ be the maturity of the futures contract observed at the $m$th fixing.
The average is defined as
\begin{equation}\label{eq:A(T)}
  A(T) \triangleq \sum_{m=1}^M w_m F(t_m,T_m),
\end{equation}
and an APO put with strike $K$ pays $[K-A(T)]^+$ at time $T$. For the most common contracts, all fixings are on
the same contract, so that $T_m=T_F$ for all $m$, the fixings are the daily settlement prices over a calendar
month, and the weights are $w_m=1/M$. Contracts whose fixings roll from one prompt contract to the next are
covered by letting $T_m$ vary with $m$. Define the running average process
\begin{equation}\label{eq:A(t)}
  A(t) \triangleq \sum_{m\,:\,t_m\le t} w_mF(t_m,T_m) + \sum_{m\,:\,t_m>t} w_mF(t,T_m),
\end{equation}
which collects the fixings already observed at time $t$. Inserting \eqref{eq:Fdynamics} and \eqref{eq:A(t)} in \eqref{eq:A(T)} we get
\begin{align*}
 A(T) ={}& \sum_{m\,:\,t_m\le t}w_mF(t_m,T_m) + \sum_{m\,:\,t_m>t}w_mF(t_m,T_m)\\
  ={}& \sum_{m\,:\,t_m\le t}w_mF(t_m,T_m) + \sum_{m\,:\,t_m>t}w_mF(t,T_m)\\
   &+\sum_{m\,:\,t_m>t}w_m\int_t^{t_m}\varphi^\top(T_m-u)D_v(v(u))\dd W^\Q(u)\\
   ={}&A(t) + \int_t^T\varphi_{\mathrm{APO}}^\top(T-u)D_v(v(u))\dd W^\Q(u),
\end{align*}
where we have used that $t_m>u$ is equivalent to
$T-t_m<T-u$, and $T_m-u=\Delta T_m+(T-u)$, and the definition 
\begin{equation}
  \varphi_{\mathrm{APO}}(\tau) \triangleq \sum_{m\,:\,T-t_m<\tau} w_m\,\varphi(\Delta T_m+\tau),
  \qquad \Delta T_m \triangleq T_m-T .
\end{equation}
$\varphi_{\mathrm{APO}}$ is then piecewise continuous, with jumps at the $M-1$ points $\tau=T-t_m$, $m<M$. It follows that we can price APO puts (and calls)  using Proposition 2 with $Y$ replaced by $A$ and $\varphi_Y$ replaced by $\varphi_{\mathrm{APO}}$.

%% file: data.tex
\section{Data description and estimation methodology}
\label{sec:data}
\subsection{Data}
The models are estimated on a comprehensive dataset of 342{,}637 prices of
crude oil futures, vanilla options and calendar spread options traded at the
NYMEX,\footnote{
The futures data was retrieved from Refinitiv, while both the vanilla and 
calendar spread options data was acquired from the CME.} from 2 January 2013
to 29 February 2024, a total of 2{,}810 business days.
For all three, we use the daily settlement prices set by the exchange.
For discounting, we use the Treasury yield curve of
\cite{gsw2007}.\footnote{The dataset is
	available at\newline
\url{https://www.federalreserve.gov/econres/feds/the-us-treasury-yield-curve-1961-to-the-present.htm}.}

\subsubsection{Futures}%
West Texas Intermediate (WTI)
crude oil futures traded on NYMEX are among the most liquid
commodity futures contracts.
They are physically settled in Cushing, Oklahoma, with delivery in the month
after expiry.\footnote{
Trading terminates
three business days before the 25th calendar day of the month 
preceding the delivery month, unless the 25th is not a business day,
in which case it is four business days before the 25th.
}

Monthly contracts are listed about ten years ahead, but liquidity is
concentrated in particular contracts, depending on their maturity and delivery
month. The contracts with delivery in the next seven months have a large open
interest whatever the delivery month, and their open interest peaks a few weeks
before expiry. We therefore select the first seven contracts with expiry more
than twelve business days ahead, M1 to M7.
\begin{figure}[htb]
	\centering
	\includegraphics[width=6in]{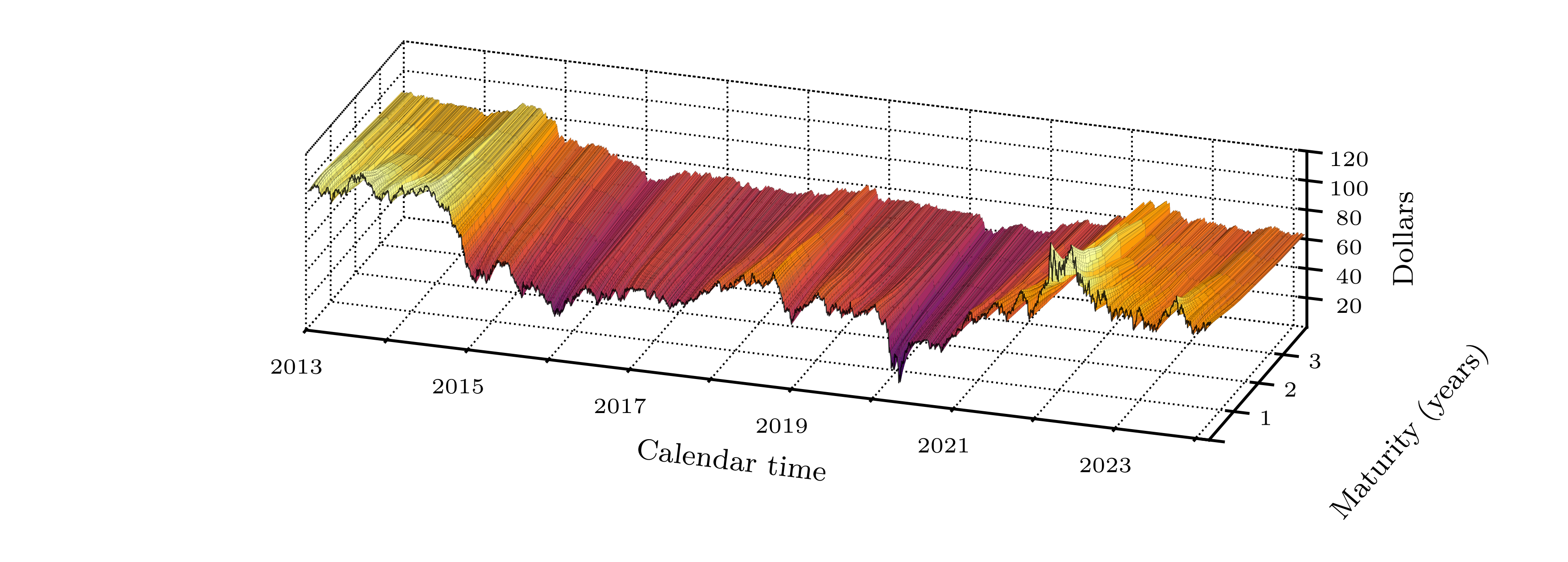}
	\caption{\textbf{Term structure of NYMEX WTI futures prices}\\ Settlement prices, in \$/bbl, of the M1--M7, Q1--Q3 and Y1--Y3 contracts, plotted at constant maturities equal to their sample averages; the black line traces M1. Daily data from 2 January 2013 to 29 February 2024.}
	\label{fig:wti_fut_ts}
\end{figure}

Beyond these, liquidity is concentrated first in the quarterly contracts and
then in the December contracts. We therefore add the next three contracts
for delivery in March, June, September or December, Q1 to Q3, and the next three
December contracts, Y1 to Y3. This gives 13 contracts on each date and 36{,}530
futures prices in all. \autoref{fig:wti_fut_ts} shows the term structure of
futures prices. Notable events include the drop in prices in 2014, the 2020
plunge during the Covid-19 pandemic and the spike in 2022 after the Russian
invasion of Ukraine.
Notice that the short-maturity end of  the term structure is more volatile than the far end, in agreement with the Samuelson effect.

\subsubsection{Vanilla options}%
Call and put options on futures contracts (or \emph{vanilla options}) are American-style with expiry three
business days before the underlying futures contract. The market is highly liquid, with options for a wide range of strikes and expiries actively traded each day. We select the options for the estimation only among those with an open
interest above 100 contracts and a settlement price above \$0.10.\footnote{Additionally, we screen the option prices on each contract each day
for static arbitrage.}

We only consider options on contracts M1--M7 and Q1--Q3, for two reasons:
First, our model assumes that interest rates are independent of futures
prices and futures volatilities, and \cite{Cheng2018} find that stochastic
interest rates matter mainly for pricing longer-dated options. Second, our
estimation procedure is only tractable for European-style options, requiring us
to \emph{de-americanize} the American-style options, an approximation whose
error grows with maturity. Following
\cite{trolle2009}, we compute the implied volatility of each American option
and use the corresponding closed-form formula to compute its European
counterpart, although in the \cite{Bachelier1900} rather than the
\cite{Black1976} model, since both futures prices and strikes can be negative.
Additional details are given in Appendix~\ref{sub:data_construction}.

To select a representative collection of options for each contract and date, we compute the implied Bachelier volatility and (undiscounted)
delta of each option. Calls and puts with an absolute delta between 0.45 and 0.55
are sorted into the middle bucket 0, representing options that are approximately
at-the-money (ATM). We then discard the remaining in-the-money options and sort
the out-of-the-money (OTM) puts and calls into four buckets each by the absolute
value of their delta, as shown in \autoref{tab:vo_dbuckets}.
	\begin{table}[!tb]
			\centering\small
\setlength{\aboverulesep}{0pt}\setlength{\belowrulesep}{0pt}%
\renewcommand{\arraystretch}{1.15}%
\begin{tabular}{@{}l|cccc|c|cccc@{}}
\toprule
Delta buckets & $-4$ & $-3$ & $-2$ & $-1$ & 0 & 1 & 2 & 3 & 4\\
\midrule
Type & \multicolumn{4}{c|}{OTM puts} & Both & \multicolumn{4}{c}{OTM calls}\\
$|\Delta| \pm 0.05$ & 0.1 & 0.2 & 0.3 & 0.4 & 0.5 & 0.4 & 0.3 & 0.2 & 0.1\\
\bottomrule
\end{tabular}
\caption{\textbf{Vanilla option delta buckets}\\ Each out-of-the-money option is assigned by its undiscounted delta to the bucket whose central absolute delta lies within 0.05 of its own. Calls and puts with absolute delta between 0.45 and 0.55 form the at-the-money bucket 0, and in-the-money options are discarded.}
			\label{tab:vo_dbuckets}
		\end{table}
For our measurements, we then select the option in each bucket closest to the middle of the bucket. This gives up to nine
options for each of the ten contracts on each date,
totaling 235{,}277 options across the full panel.

\begin{figure}[!htb]
	\centering
	\includegraphics[width=\textwidth]{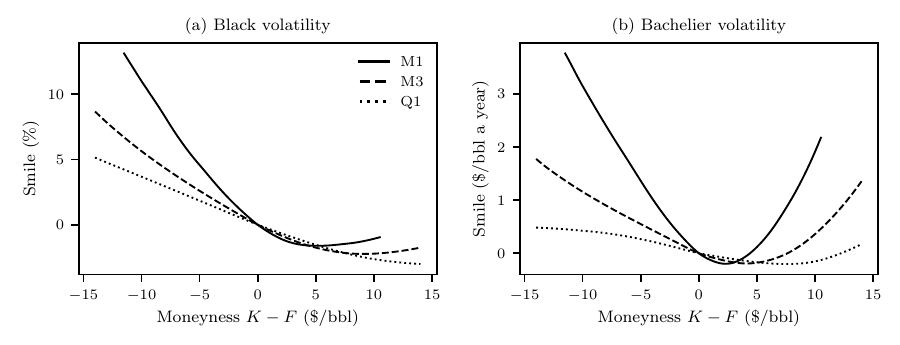}
	\caption{\textbf{Average NYMEX WTI implied smile in Black and Bachelier volatilities}\\ For M1, M3 and Q1, the implied volatility of the liquid out-of-the-money options less that at the money, against the moneyness $K-F$, averaged over the dates on which the moneyness is quoted and lightly smoothed. (a)~Black volatility, in percent. (b)~Bachelier volatility, in \$/bbl a year. Daily data from 2 January 2013 to 29 February 2024.}
	\label{fig:black_bach_smile}
\end{figure}

\autoref{fig:black_bach_smile} shows
the volatility smile averaged over all dates, measured as the implied volatility
of options with moneyness $K-F$ less that of the ATM option on the same contract. 
We plot the smile with respect to both the implied Black and implied Bachelier volatility.\footnote{
	See \cite{ChoiKwakTeeWang2022} for a survey and comparison
of the two frameworks.}
The Black model
assumes that the futures price follows a driftless geometric Brownian motion,
such that it is log-normal and hence positive. Positivity became
problematic in April 2020, when both negative futures prices and options with
negative strike prices were traded. As a consequence, CME switched to the Bachelier model for
option pricing purposes as this model assumes that the
futures price follows a driftless Brownian motion, thus implying Gaussian
distributions for prices.
While the Gaussian distribution
is symmetric, the log-normal distribution is not, and the latter will assign a lower probability 
to large downward moves in the price.
In implied Black and Bachelier volatility alike, the smile is most pronounced for M1 and flattens
with maturity. In implied Black
volatility, it rises steeply toward the puts but not toward the calls at each of the three
maturities, indicating that the market considers large downward movements in the price more probable
than log-normal dynamics would imply. In implied Bachelier volatility, the smile is much closer to
symmetric, with a median 20-delta risk reversal relative to the at-the-money volatility of $-0.05$,
against $-0.23$ in implied Black volatility.\footnote{The risk reversal is the implied volatility of the
20-delta call less that of the 20-delta put, with the strikes chosen by Bachelier delta. Choosing them
by Black delta does not change the result.} Since a smile measures how far market prices deviate from those
of the model behind the implied volatility, the near symmetry of the Bachelier smile suggests that the
risk-neutral distribution of the futures price is close to symmetric in the price level, as
\citet{Bouchouev2023} also concludes for WTI options.

\subsubsection{Calendar spread options}%
Alongside futures and vanilla options, calendar spread options are
a third source of information on the distribution of futures prices. The price of a vanilla option depends only on the risk-neutral marginal
distribution of a single futures price, while the price of a calendar spread
option depends on the joint risk-neutral distribution of two adjacent futures
prices, that is, on their marginal distributions and on the dependence structure
between them. This dependence structure is therefore not determined by the prices of
vanilla options, and it is only weakly identified by the time series of futures
prices.

NYMEX crude oil calendar spread options are the most liquid exotic oil
options, although they are much less liquid than the vanilla options. Across all
quoted options on the contracts we use, their volume is about three percent, and
their open interest about eight percent, of that of the vanilla
options.\footnote{On the median day, their volume is less than one percent of
that of the vanilla options, since their trading is concentrated on relatively
few days.}
They are European-style, so they need no de-americanization,
and expire on the first business day before the expiry of the first-to-expire 
underlying futures contract.
They exist in two varieties, physically and financially settled, depending on whether exercise entitles the option holder to the underlying futures spread or to an equivalent cash payment.
Furthermore, calendar spread options exist for spreads between contracts one to twelve months apart,
but most of the liquidity is concentrated in the one-month spread.
We only use the financial one-month calendar spread options,
and denote the contracts by the first-to-expire futures contract of the
underlying spread, such that an M1 calendar spread option is written on the
spread M1--M2.
Because the liquidity quickly decreases for longer-dated spreads, we select only the options on the first six spreads, M1 to M6.
Since calendar spread options are much cheaper than vanilla options, we screen
them with a lower
price floor, discarding those with no open interest or a settlement price of
\$0.02 or less. We
then apply the same selection by delta bucket as for the vanilla options, with five coarser buckets, shown in
\autoref{tab:cso_dbuckets}.
\begin{table}[!htb]
			\centering\small
\setlength{\aboverulesep}{0pt}\setlength{\belowrulesep}{0pt}%
\renewcommand{\arraystretch}{1.15}%
\begin{tabular}{@{}l|cc|c|cc@{}}
\toprule
Delta buckets & $-2$ & $-1$ & 0 & 1 & 2\\
\midrule
Type & \multicolumn{2}{c|}{OTM puts} & Both & \multicolumn{2}{c}{OTM calls}\\
$|\Delta| \pm 0.1$ & 0.1 & 0.3 & 0.5 & 0.3 & 0.1\\
\bottomrule
\end{tabular}
\caption{\textbf{Calendar spread option delta buckets}\\ As for the vanilla options in \autoref{tab:vo_dbuckets}, with each bucket spanning 0.1 on either side of its central absolute delta.}
			\label{tab:cso_dbuckets}
		\end{table}
This leaves 70{,}830 calendar spread options, with at most 30 on a date.

\begin{figure}[!htb]
	\centering
	\includegraphics[width=\textwidth]{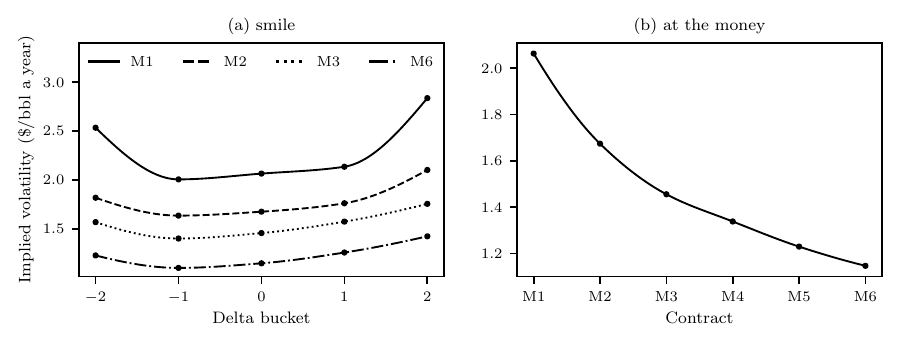}
	\caption{\textbf{Implied volatility of NYMEX WTI one-month calendar spread options}\\ (a)~The average implied Bachelier volatility, in \$/bbl a year, in each delta bucket (markers) for the calendar spread options on M1--M3 and M6. (b)~The average at-the-money implied Bachelier volatility for M1--M6. Daily data from 2 January 2013 to 29 February 2024.}
	\label{fig:cso_iv}
\end{figure}

Calendar spread options are usually quoted in terms of their implied Bachelier
volatilities, since both the underlying spread and the strike prices are often
negative. \autoref{fig:cso_iv} shows that the average smile rises in both wings with a moderate
tilt toward the calls, and that the at-the-money implied volatility decays rapidly with maturity.

The implied volatility of a calendar spread option summarizes the risk-neutral
distribution of the spread, which depends on the marginal distributions of its
two legs and on the dependence structure between them. Since the vanilla options
give the marginal distributions, the implied volatility of a calendar spread
option can be translated into a measure of the dependence. A simple and
interpretable such measure is the implied correlation of the bivariate Bachelier
model. If both legs follow Bachelier dynamics, with
volatilities $s_1$ and $s_2$ and correlation $\rho$, the volatility of the spread
is $s_{CS} = \sqrt{s_1^2 + s_2^2 - 2\rho s_1 s_2}$. We take $s_{CS}$ and $s_1$ as
the at-the-money implied Bachelier volatilities of the calendar spread option and
of the vanilla option on its first leg. The volatility $s_2$ of the second leg
until the calendar spread option expires is not observed.\footnote{Let the two legs be the futures contracts expiring at $T$ and
$T+\Delta$. The calendar spread option expires one business day before $T$,
while the vanilla option on the second leg expires three business days before
$T+\Delta$, about a month later. Its implied volatility therefore covers a
different and longer period than the life of the calendar spread option.}
As an approximation that needs no model, we therefore set $s_2 = s_1$, which
gives the implied one-month decorrelation, our measure of the
\emph{curve decorrelation} between adjacent contracts,
\begin{equation*}
	1-\rho = \frac{s_{CS}^2}{2 s_1^2}.
\end{equation*}
Up to a small error, the approximation is an upper bound on the
decorrelation.\footnote{Like any implied correlation, this measure depends on the
distribution assumed for the two legs \citep{Tavin2020}. If the two legs are
jointly Gaussian with correlation $\tilde\rho$, then
\begin{equation*}
	\frac{s_{CS}^2}{2 s_1^2} = \frac{1-\tilde\rho^2}{2} + \frac{(s_2-\tilde\rho s_1)^2}{2 s_1^2}.
\end{equation*}
Since the last term is non-negative, the approximation falls below the true
decorrelation $1-\tilde\rho$ by at most a fraction $(1-\tilde\rho)/2$ of it, about 0.1 percent on a
typical day, whatever the value of $s_2$.}
See \cite{Bouchouev2023} for a more
extensive discussion of calendar spread options and their use in oil markets.

\subsubsection{Option-implied volatility, skewness and decorrelation}%
\label{sub:data_moments}
We summarize the option prices by three option-implied quantities: the
model-free volatility $\mathcal M_2$ and skewness $\mathcal M_3$ of the vanilla
options, and the implied decorrelation of the calendar spread options. For each
contract and date, we compute $\mathcal{M}_2$ and $\mathcal{M}_3$ from the full cross-section of
out-of-the-money vanilla options. The spanning formula requires
option prices at every strike, but options are listed only at discrete strikes
and stop short of the wings. We therefore fit a smooth risk-neutral density to
each day's quotes and compute the moments from it. Beyond $\pm2$ at-the-money
standard deviations, we replace the density on each side by an exponential tail
with the same mass and first moment, so that $\mathcal M_2$ and $\mathcal M_3$
are moments of the whole distribution.\footnote{The fitted density reprices the quoted options only up to a
small error, part of which is the rounding of the settlement prices to the cent.
We discard a fit when its error is several times larger than this rounding can
explain, which happens for 10 of the 27{,}765 contract-dates.
Further details are given in Appendix~\ref{sub:data_construction}.}

\autoref{fig:ts3d_moments}(a) and (b) show the term structures of the
model-free volatility and skewness. The model-free volatility spikes
during the 2020 plunge and the sharp rise in 2022. The volatility of
soon-to-expire options fluctuates more than that of longer-dated options,
indicating mean reversion in the volatility. The skewness is positive
on most dates and increases with maturity, although its sign at the short end
varies across regimes, being predominantly negative during the 2020 plunge and
almost uniformly positive in 2022.

\autoref{fig:ts3d_moments}(c) shows the term structure of the implied one-month
decorrelation, for first legs maturing in two to five months. It is small, about
0.002 at two months on a typical day, so that adjacent contracts are almost
perfectly correlated. It falls with maturity on most days and varies most at the
front. It is highest when the futures curve is steep, in backwardation from mid-2013 to the autumn of
2014 and in contango in April 2020.\footnote{Its rank correlation is 0.56 with the absolute slope of the
curve and close to zero with the size of daily price moves.}
\begin{figure}[p]
	\centering
	\includegraphics[width=6in]{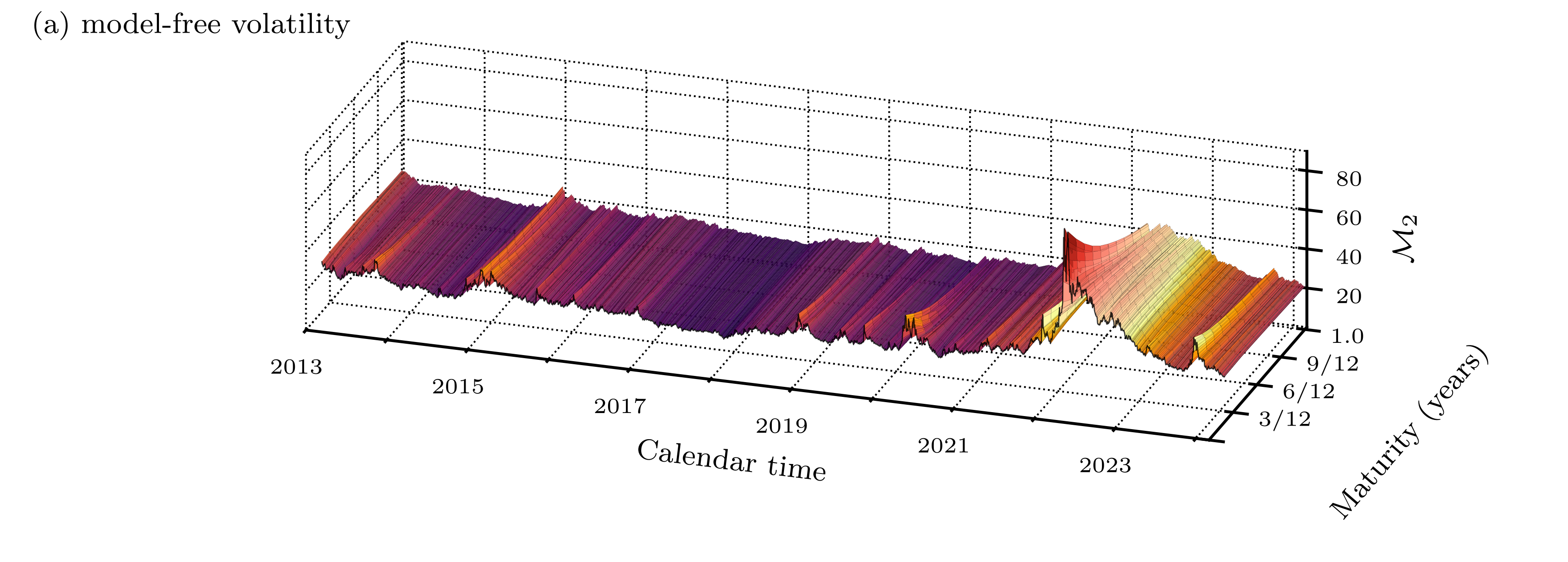}\\
	\includegraphics[width=6in]{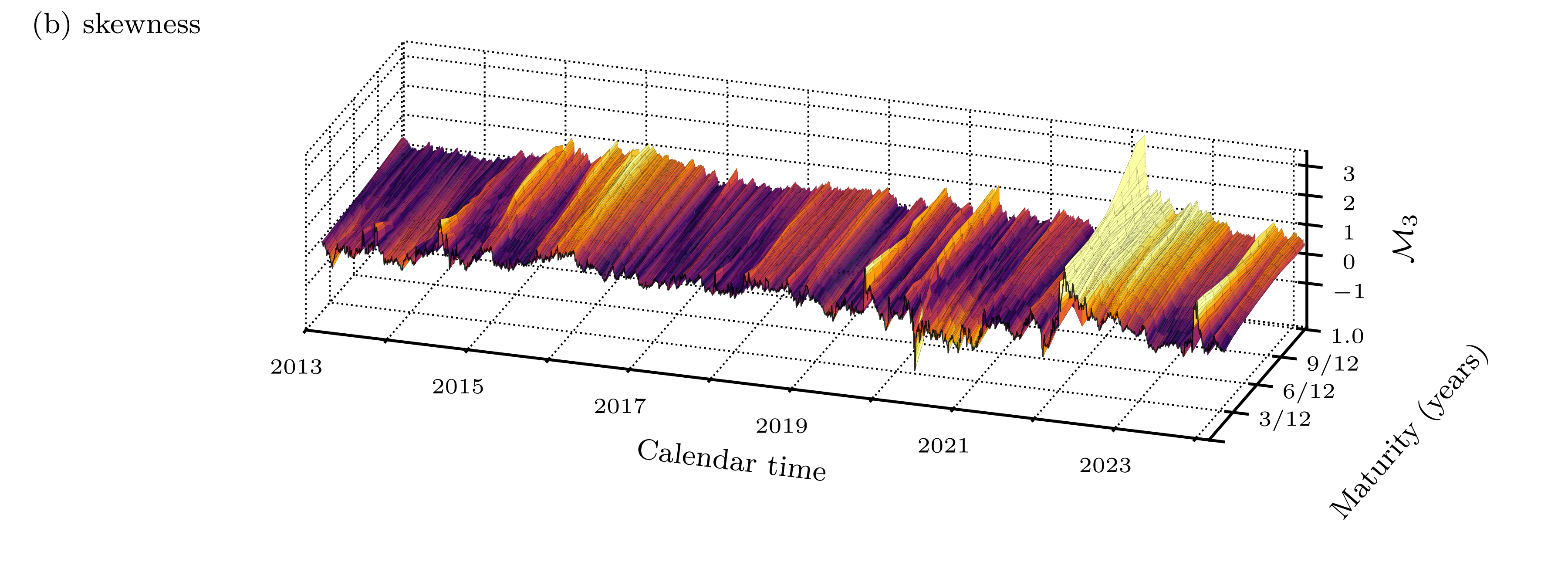}\\
	\includegraphics[width=6in]{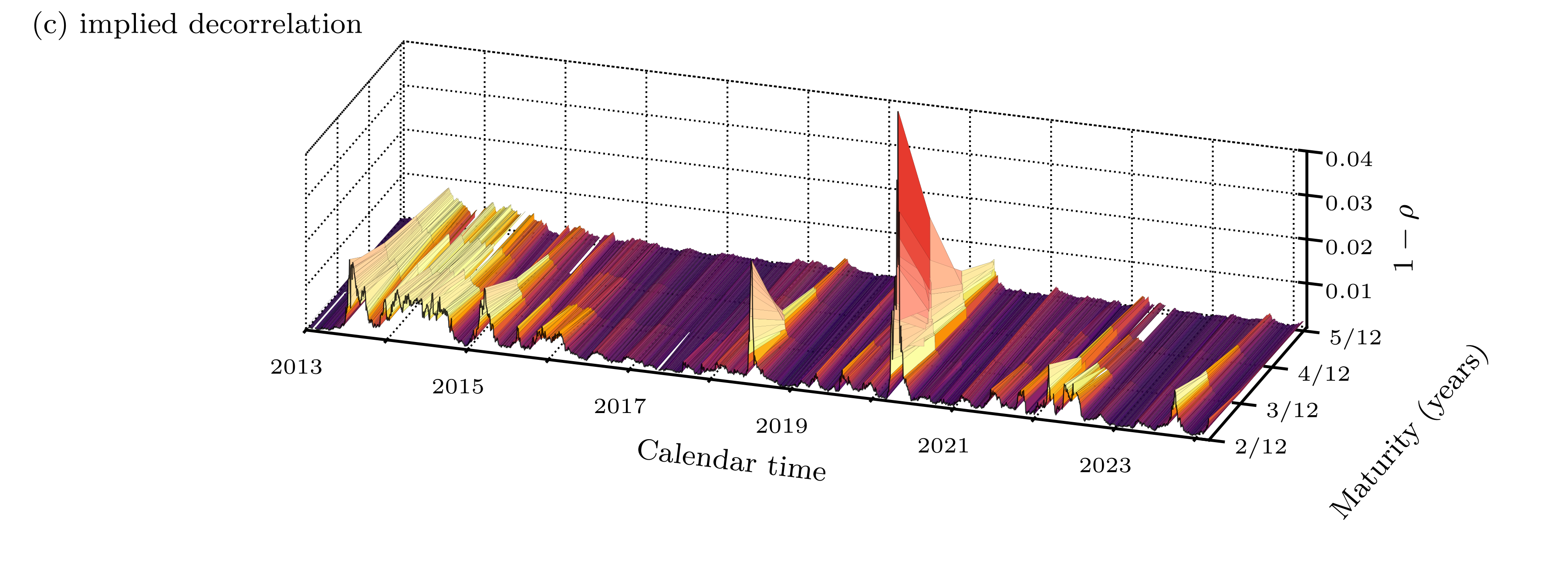}
	\caption{\textbf{Term structures of NYMEX WTI model-free volatility, model-free skewness and implied decorrelation}\\ (a)~The model-free volatility $\mathcal{M}_2$, in \$/bbl a year, and (b)~the skewness $\mathcal{M}_3$ of the vanilla options, at constant maturities from two months to one year. (c)~The implied one-month decorrelation $1-\rho$ from the calendar spread options, at constant maturities of the first leg from two to five months; the April 2020 peak, 0.066, rises above the axis. The black line traces the shortest maturity. Daily data from 2 January 2013 to 29 February 2024.}
	\label{fig:ts3d_moments}
\end{figure}

\subsubsection{Dimension of the option-implied quantities}%
\label{sub:data_dimension}
Whether a model can fit the volatility term structure, the skew and the curve decorrelation with the same
factors depends on whether they move together. We therefore study the daily changes of $\mathcal{M}_2^2$,
$\mathcal{M}_3$ and $1-\rho$ at constant maturities, and how much of the changes in each is explained by the others.\footnote{We use
$\mathcal{M}_2^2=\varkappa_2/\tau$ rather than $\mathcal{M}_2$, since it is affine in the volatility state
variables, so that its principal components indicate how many factors the volatility term structure needs.
$\mathcal{M}_3$ and $1-\rho$ are not affine, and their affine counterparts, $\varkappa_3/\tau$ and
$s_{CS}^2$, serve as a robustness check. The maturities are two to twelve months for $\mathcal{M}_2^2$, and
two to five months, the range of the calendar spread options, wherever the skewness or the decorrelation
enters. Further details are given in Appendix~\ref{sub:dimension}.}

The volatility term structure is essentially two-dimensional, since a level and a slope explain 96 percent
of the daily changes in $\mathcal{M}_2^2$. The skewness moves mostly as one curve, with its first principal
component explaining 80 percent of its daily changes. Regressing these changes on the level and slope of
$\mathcal{M}_2^2$, however, gives an $R^2$ of only 10 percent, indicating that most movements in the skew
are unrelated to movements in the volatility term structure. The same regression for the implied
decorrelation gives an $R^2$ of 17 percent, and adding the skewness to the regressors barely raises
it,\footnote{By 0.5 percentage points. The partial rank correlation of the skewness and the decorrelation,
0.07, falls to zero on weekly and monthly changes.} indicating that the decorrelation moves largely
independently of both. These shares remain low over weekly and monthly changes, over which noise in the
quotes weighs less, and without 2020.\footnote{Over monthly changes, the level and slope explain at most 30
percent of the skewness and the decorrelation, also in terms of $\varkappa_3/\tau$ and $s_{CS}^2$.}

These findings are model-free, and they suggest that the volatility term structure, the skew and the curve
decorrelation are largely separate features of the data. A model whose volatility factors are fitted to the
volatility term structure may therefore capture little of the movements in the skew and the curve
decorrelation, which the vanilla smiles and the calendar spread options carry.

%% file: estimation.tex
\subsection{Estimation procedure}
\label{sub:estimation}
We estimate the model using the
\emph{unscented} Kalman filter
\citep[see][]{Julier2004,Wan2004}
in conjunction with quasi-maximum likelihood.
We cast the model into a discrete-time system with
equidistant timesteps $\Delta$ of one business day,
at the time points $t_0, t_1, \dots, t_L$.
The transition equation for
the $3n$-dimensional state vector
$X(t) \triangleq \left( x(t), z(t), v(t) \right)^\top$
can be written as
\begin{equation*}
	X(t_{l+1}) = A X(t_l) + b + \omega_l,
\end{equation*}
where $\omega_l$ is assumed
Gaussian\footnote{The true distribution of the noise is not Gaussian, which is one of
the reasons this is only \emph{quasi}-maximum likelihood. Its conditional mean
and covariance are exact, but the volatility state variables $v(t)$ have no
closed-form transition for a general $\Gamma$, and $(x(t), z(t))$ is Gaussian
only conditional on the path of $v(t)$.}
with a covariance matrix that is affine in the volatility
state variables $v(t_l)$,
\begin{equation*}
	\Omega( v(t_l) ) = \Omega_0
	+ \sum^{n}_{i=1} \Omega_i v_i(t_l).
\end{equation*}
Additional details on the state-space form and the filter are given in
Appendix~\ref{sec:appendix_estimation}, and a simulation study of the estimation in Appendix~\ref{app:sim}.

We order the factors by decreasing $\kappa_i$, restrict $\Gamma$ to be upper
triangular and estimate the infinite-maturity futures price $F_0$ along with
the other parameters. We normalize $\beta_{0,i} > 0$ and require
$\beta_{1,i} \geq 0$, so that each loading $\fvol_i$ keeps its sign along the curve and is either a pure decay or a single hump.
For $n \geq 2$ we set $\beta_{1,n}=0$. When $\kappa_n$ is small, the loading $\fvol_n$
is nearly linear over the observed maturities,
$\fvol_n(\tau)\approx\beta_{0,n}+(\beta_{1,n}-\kappa_n\beta_{0,n})\tau$, so the data
identify only its slope $\beta_{1,n}-\kappa_n\beta_{0,n}$, and not $\beta_{1,n}$ and
$\kappa_n$ separately.

The level of volatility changes severalfold over the sample. So that the size
of the measurement errors does not change with it, we divide each observed and
model-implied price by a constant computed from the observed prices of its
date. We divide each option price by its implied vega, as \cite{trolle2009} do, and further by its implied volatility, so that its measurement error is,
to first order, the relative error in implied volatility. Since the model is
arithmetic, we use the Bachelier vega and volatility, for the vanilla and
calendar spread options alike. We divide each futures price by the typical
daily price move of its date, the average implied
Bachelier volatility of the at-the-money vanilla options times $\sqrt{1/252}$.
Let $y_l$ be the vector of the scaled prices observed at $t_l$. The measurement equation, nonlinear in the
state through the option prices, is
\begin{equation}
	\label{eq:nonlinear_meq}
	   y_l = h_l( X(t_l)) + \epsilon_l,
\end{equation}
with $h_l$ depending on the date through the contracts quoted on it. The option
prices are computed from the observed futures prices, so that an imperfect fit
to the futures curve does not enter them and they depend on the state only
through $v(t)$. The measurement noise $\epsilon_l$ is assumed Gaussian and
uncorrelated both serially and cross-sectionally. Its standard deviation is a
constant $\sigma_F$ for the futures and $\sigma_{CS}$ for the calendar spread
options. For the vanilla options, it is $\sigma_V$ at the money and grows towards
the wings, to $(1+|a_V|)\,\sigma_V$ for the far out-of-the-money options, since
the options at the money are the most liquid. Additional details on the
measurement model are given in Appendix~\ref{sub:measurements}.

Section~\ref{sub:data_dimension} showed that the level and slope of the volatility term
structure explain only a small part of the movements in the skewness. We
therefore remove the variation of the skew over time, to first order, from the
vanilla measurements, a device we call the \emph{skew forcing}, and
leave the average skew to the leverage correlations $\rho_i$. We also report
estimates without the forcing for comparison. Let $u_l$ be the
average skewness $\mathcal M_3$ of the contracts M1--M7 on date $t_l$, $\bar u$
its sample average, and $s_l$ the vector of the vanilla options' delta buckets divided by four,
from $-1$ for the far out-of-the-money puts to $+1$ for the far
out-of-the-money calls, with zero entries for the futures and calendar spread
options. The measurements $y_l$ in \eqref{eq:nonlinear_meq} are then
replaced by $y_l - \chi_V (u_l - \bar u) s_l$, which tilts each day's vanilla
smile in proportion to the day's skewness less its sample average, with an
estimated coefficient $\chi_V$. The forcing enters only the estimation:
every price, pricing error and hedge in Section~\ref{sec:results} is computed
without it.

The option prices are not defined for negative $v$ and are not smooth in $v$
close to zero. The filter therefore evaluates them at
$\max(v(t_l), \underline v)$, element-wise, for a small floor $\underline v$.
Since the Gaussian update can
place part of the distribution of $v_i(t_l)$ below zero, the filtered mean and
variance of each $v_i(t_l)$ are replaced after each update by those of the
posterior that combines the information in the update with a gamma approximation of the predicted distribution, which puts no mass below zero.

We complement quasi-maximum likelihood with a few requirements that keep the
model within reason, and take as our estimate the parameters that maximize the
quasi-likelihood among those that meet them. The filtered paths must be
consistent with the model's own dynamics under $\mathbb{P}$, the model's futures variance must agree
with the realized variance of the futures, the prices of risk must be financially reasonable, and the
volatility factors may not revert under $\mathbb{P}$ much more slowly than the variance in the
data, since the sample says little about slower mean reversion.\footnote{The one-month prediction errors of the filtered paths may not show more bias and
persistence than the model's dynamics would produce in one sample out of ten thousand, which guards
against drifts under $\mathbb{P}$ that the one-day quasi-likelihood barely penalizes. The model's
futures variance is tested against the realized variance in its level and in its slope across the
curve, each at the 1 percent level. The annualized maximal Sharpe ratio is at most 2, and the
relaxation time $\operatorname{Var}(v_i)/(\sigma_{v,i}^2\,\mathrm E[v_i])$ of each volatility state variable,
with its stationary moments under $\mathbb P$, is at most 1.25 years. For the level factor, this is a half-life of at most 1.73 years, against 0.8 years for
the variance in the data. Further details are given in Appendix~\ref{sub:constrained_estimation}.}

We solve the constrained problem by an augmented Lagrangian method and compute quasi-maximum likelihood sandwich standard errors restricted to the directions
that leave the active constraints unchanged.

%% file: results.tex
\section{Results}\label{sec:results}
We first examine the parameter estimates and the roles of the individual factors, then evaluate the models' pricing and hedging performance. We subsequently examine how the models capture the volatility term structure, curve decorrelation and calendar spread option prices, and variation in skew. Finally, we compare the pricing and hedging performance of a two-factor arithmetic model with that of its geometric counterpart.

\subsection{Parameter estimates and factor roles}\label{sec:res_est}
We compare the models by how they reproduce the three representative features
of the distribution of the futures curve implied by the options: the volatility
term structure, the curve decorrelation between adjacent contracts and the
skew. We measure them by the model-free volatility $\mathcal{M}_2$, the implied
one-month decorrelation and the model-free skewness $\mathcal{M}_3$,
respectively. Each factor of the model contributes to all three features, to
the volatility term structure through its contribution to the volatility of
each contract, to the curve decorrelation through how its loading differs along
the curve, and to the skew through its leverage correlation $\rho_i$ between
its shocks to the futures curve and the shocks to its volatility factor $v_i$.
Since the $\rho_i$ are constant, the skew can change over time only as the
volatility factors move, and fitting these changes requires volatility factors
that follow the skew as well. The skew forcing largely removes this need, since
it tilts each day's vanilla smile in proportion to the day's average skewness
and so captures the variation of the skew to first order. Estimating models
both with and without the forcing, we can therefore compare factors fitted
mainly to the volatility term structure and the curve decorrelation with
factors that must fit the skew as well.

We estimate five versions of the model, presented in \autoref{tab:names}. V denotes a
model estimated on the 36{,}530 futures prices and 235{,}277 vanilla options, VC a model also estimated on the 70{,}830 calendar spread options, and U a model without
the skew forcing. All five must reproduce the volatility term structure, and V2
and V3 show how many factors are needed. Without skew forcing, V2-U and V3-U must also
reproduce the skew through their factors. VC3 must also reproduce the
curve decorrelation. Since our focus is on the volatility term
structure and the curve decorrelation, V3 and VC3 are our main models, and we
compare them throughout.
\autoref{tab:res_est} reports the estimates of V2, V3 and VC3, and
\autoref{tab:app_est0} in the appendix those of V2-U and V3-U.

\begin{table}[!htbp]
\centering\small
\setlength{\aboverulesep}{0pt}\setlength{\belowrulesep}{0pt}%
\renewcommand{\arraystretch}{1.15}%
\begin{tabular}{@{}l|cc@{}}
\toprule
Data & With skew forcing & Without skew forcing\\
\midrule
Futures and vanilla options & V2, \textbf{V3} & V2-U, V3-U\\
Futures, vanilla and calendar spread options & \textbf{VC3} & \\
\bottomrule
\end{tabular}
\caption[Model names]{\textbf{Model names}\\ The digit in each name is the number of factors $n$; the main models are in bold.}
\label{tab:names}
\end{table}

\begin{figure}[!htbp]
\centering
\includegraphics[width=\textwidth]{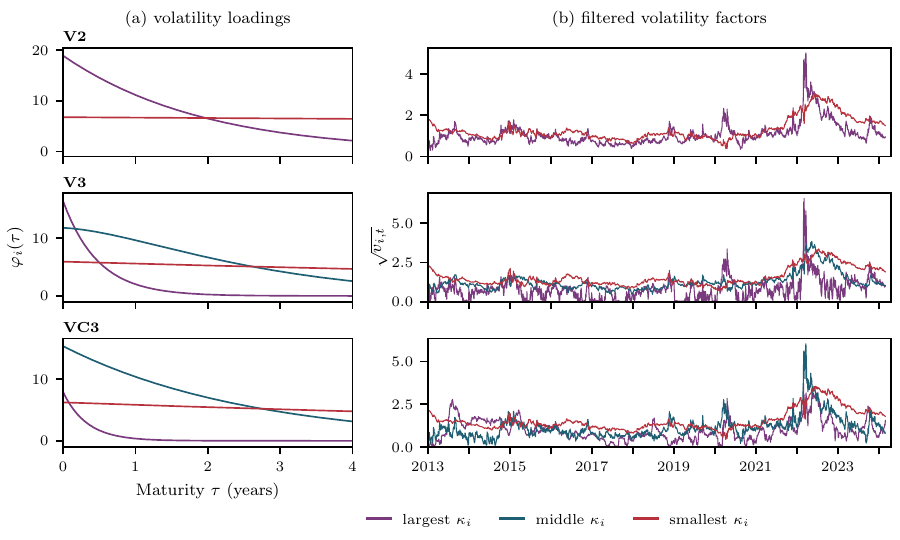}
\caption[Volatility loadings and filtered volatility state variables]{\textbf{Volatility loadings and filtered volatility state variables}\\ (a)~Volatility loadings $\varphi_i(\tau)$. (b)~Square roots of the filtered volatility state variables. Rows V2, V3 and VC3. Daily data from 2 January 2013 to 29 February 2024.}
\label{fig:res_factors}
\end{figure}

The volatility loadings of the four models estimated on futures and vanilla
options, shown in \autoref{fig:res_factors}(a) and, for V2-U and V3-U, in
\autoref{fig:app_factors_chi0}, share a common structure. Each has a level
factor, whose loading changes slowly along the curve, and a middle factor, whose
loading falls off over the first few years. The three-factor models add a
transitory factor, whose loading is concentrated on the front contracts and
halves within a few months. The two-factor models have no such factor, and their
middle factor covers the front contracts as well. VC3, estimated also on the
calendar spread options, shares this structure, although both its transitory and its
middle factor decay faster along the curve than those of V3.

\begin{table}[!htbp]
\centering\small
\setlength{\aboverulesep}{0pt}\setlength{\belowrulesep}{0pt}%
\renewcommand{\arraystretch}{1.15}%
\begin{tabular}{@{}l cc| ccc| ccc@{}}
\toprule
 & \multicolumn{2}{c|}{V2} & \multicolumn{3}{c|}{V3} & \multicolumn{3}{c}{VC3}\\
\cmidrule(lr){2-3}\cmidrule(lr){4-6}\cmidrule(l){7-9}
factor $i$ & 1 & 2 & 1 & 2 & 3 & 1 & 2 & 3\\
\midrule
$\kappa_i$ & 0.659 & 0.013 & 2.237 & 0.702 & 0.060 & 3.268 & 0.457 & 0.066\\
 & {\scriptsize (0.0409)} & {\scriptsize ( --- )} & {\scriptsize (0.0521)} & {\scriptsize (0.0137)} & {\scriptsize (0.0020)} & {\scriptsize (0.0387)} & {\scriptsize (0.0053)} & {\scriptsize (0.0030)}\\[2pt]
$\beta_{0,i}$ & 18.931 & 6.759 & 16.476 & 11.814 & 5.946 & 7.944 & 15.369 & 6.243\\
 & {\scriptsize (1.3889)} & {\scriptsize (0.3309)} & {\scriptsize (0.3873)} & {\scriptsize (0.1365)} & {\scriptsize (0.1580)} & {\scriptsize (0.2038)} & {\scriptsize (0.1657)} & {\scriptsize (0.1391)}\\[2pt]
$\beta_{1,i}$ & 2.648 & --- & 2.578 & 7.685 & --- & 1.633 & 1.061 & ---\\
 & {\scriptsize (0.8773)} &  & {\scriptsize (0.3406)} & {\scriptsize (0.4215)} &  & {\scriptsize (0.2582)} & {\scriptsize (0.0915)} & \\[2pt]
\midrule
$\sigma_{v,i}$ & 3.449 & 0.774 & 14.794 & 1.027 & 0.901 & 4.348 & 4.386 & 0.862\\
 & {\scriptsize (0.3961)} & {\scriptsize (0.0242)} & {\scriptsize (0.5464)} & {\scriptsize (0.0484)} & {\scriptsize (0.0339)} & {\scriptsize (0.2354)} & {\scriptsize (0.1875)} & {\scriptsize (0.0233)}\\[2pt]
$\rho_i$ & $-$0.128 & 0.286 & $-$0.071 & $-$0.274 & 0.268 & 0.319 & $-$0.150 & 0.272\\
 & {\scriptsize (0.0036)} & {\scriptsize (0.0384)} & {\scriptsize (0.0123)} & {\scriptsize (0.0260)} & {\scriptsize (0.0224)} & {\scriptsize (0.0189)} & {\scriptsize (0.0058)} & {\scriptsize (0.0204)}\\[2pt]
$\Gamma_{1,i}$ & $-$5.770 & 4.043 & $-$5.544 & 3.800 & 0.351 & $-$1.801 & 0.000 & 0.000\\
 & {\scriptsize (0.2874)} & {\scriptsize ( --- )} & {\scriptsize (0.4121)} & {\scriptsize ( --- )} & {\scriptsize ( --- )} & {\scriptsize (0.1605)} & {\scriptsize ( --- )} & {\scriptsize ( --- )}\\[2pt]
$\Gamma_{2,i}$ & --- & $-$0.394 & --- & $-$2.585 & 1.931 & --- & $-$5.941 & 4.398\\
 &  & {\scriptsize (0.0557)} &  & {\scriptsize (0.0721)} & {\scriptsize ( --- )} &  & {\scriptsize (0.2897)} & {\scriptsize ( --- )}\\[2pt]
$\Gamma_{3,i}$ & --- & --- & --- & --- & $-$0.554 & --- & --- & $-$0.559\\
 &  &  &  &  & {\scriptsize (0.0168)} &  &  & {\scriptsize (0.0162)}\\[2pt]
$\Gamma^{\mathbb P}_{i,i}$ & $-$7.378 & $-$0.400 & $-$6.601 & $-$2.411 & $-$0.400 & $-$1.354 & $-$6.687 & $-$0.400\\
 & {\scriptsize (1.6636)} & {\scriptsize ( --- )} & {\scriptsize (4.8574)} & {\scriptsize (0.1798)} & {\scriptsize ( --- )} & {\scriptsize (0.4895)} & {\scriptsize (0.8313)} & {\scriptsize ( --- )}\\[2pt]
$\lambda_{x,i}$ & 1.430 & $-$0.175 & $-$0.449 & $-$0.188 & $-$0.311 & $-$0.602 & $-$0.241 & $-$0.356\\
 & {\scriptsize (0.2693)} & {\scriptsize (0.1889)} & {\scriptsize (0.2206)} & {\scriptsize (0.0634)} & {\scriptsize (0.0549)} & {\scriptsize (0.1299)} & {\scriptsize (0.0997)} & {\scriptsize (0.0637)}\\[2pt]
\midrule
$F_0$ & \multicolumn{2}{c|}{26.963} & \multicolumn{3}{c|}{68.575} & \multicolumn{3}{c}{58.640}\\
 & \multicolumn{2}{c|}{{\scriptsize (2.7650)}} & \multicolumn{3}{c|}{{\scriptsize (0.6721)}} & \multicolumn{3}{c}{{\scriptsize (0.6553)}}\\[2pt]
$\log\sigma_F$ & \multicolumn{2}{c|}{$-$0.800} & \multicolumn{3}{c|}{$-$1.669} & \multicolumn{3}{c}{$-$1.751}\\
 & \multicolumn{2}{c|}{{\scriptsize (0.0126)}} & \multicolumn{3}{c|}{{\scriptsize (0.0156)}} & \multicolumn{3}{c}{{\scriptsize (0.0155)}}\\[2pt]
$\sigma_V$ & \multicolumn{2}{c|}{0.0170} & \multicolumn{3}{c|}{0.0091} & \multicolumn{3}{c}{0.0171}\\
 & \multicolumn{2}{c|}{{\scriptsize (0.0004)}} & \multicolumn{3}{c|}{{\scriptsize (0.0002)}} & \multicolumn{3}{c}{{\scriptsize (0.0004)}}\\[2pt]
$a_V$ & \multicolumn{2}{c|}{2.049} & \multicolumn{3}{c|}{4.614} & \multicolumn{3}{c}{2.081}\\
 & \multicolumn{2}{c|}{{\scriptsize (0.0737)}} & \multicolumn{3}{c|}{{\scriptsize (0.0945)}} & \multicolumn{3}{c}{{\scriptsize (0.0733)}}\\[2pt]
$\sigma_{CS}$ & \multicolumn{2}{c|}{---} & \multicolumn{3}{c|}{---} & \multicolumn{3}{c}{0.1700}\\
 & \multicolumn{2}{c|}{} & \multicolumn{3}{c|}{} & \multicolumn{3}{c}{{\scriptsize (0.0012)}}\\[2pt]
$\chi_V$ & \multicolumn{2}{c|}{0.123} & \multicolumn{3}{c|}{0.123} & \multicolumn{3}{c}{0.124}\\
 & \multicolumn{2}{c|}{{\scriptsize (0.0013)}} & \multicolumn{3}{c|}{{\scriptsize (0.0015)}} & \multicolumn{3}{c}{{\scriptsize (0.0014)}}\\[2pt]
$\log\mathcal L$ & \multicolumn{2}{c|}{418{,}221.09} & \multicolumn{3}{c|}{481{,}426.46} & \multicolumn{3}{c}{467{,}553.61}\\
\bottomrule
\end{tabular}
\caption[Parameter estimates]{\textbf{Parameter estimates}\\ Models with the skew forcing, sandwich standard errors in parentheses. VC3 is estimated also on the calendar spread options, and its log-likelihood includes them. Dashes mark parameters fixed at zero or absent from a model. An em dash in place of a standard error marks a parameter held by an active constraint. The other standard errors are conditional on these.}
\label{tab:res_est}
\end{table}

To see how the factors share the sample's volatility shocks, consider the
filtered volatility factors in \autoref{fig:res_factors}(b). In V3, the Covid-19
shock in spring 2020 appears as a sharp spike in $v_1$ alone, consistent with the
transitory factor's by far highest volatility of volatility, and the spike has
faded within months. The disruptions after the Russian invasion of Ukraine in
2022, by contrast, are a shock to the level volatility factor $v_3$, which stays
high to the end of the sample. The volatility factors $v_1$ and $v_2$ rise
with it, consistent with the spillover through the off-diagonal entries of $\Gamma$. This spillover is shared by every three-factor
model, also when V3 is estimated on either half of the sample alone.\footnote{In each, the spillover provides at least 78 percent of the stationary
mean of $v_1$ or $v_2$ under $\mathbb P$. The estimates on each half are given in
Appendix~\ref{app:oos}.} The filtered volatility factors thus suggest that the 2020 shock was predominantly transitory, while the 2022 shock had a more persistent effect on the volatility term structure.

Including the calendar spread options in the estimation hardly alters the level
factor, but it reassigns the transitory and middle factors. In VC3, the middle factor now carries
the day-to-day volatility of the front contracts. Its volatility factor $v_2$ moves almost
exactly with $v_1$ of V3 from day to day, and it receives the entire spillover from the level
volatility factor. The transitory factor, instead, appears to be a \emph{curve
decorrelation factor}. Its volatility factor $v_1$ moves closely with the
implied decorrelation at two months, while no volatility factor of V3 does.

\begin{figure}[!htbp]
\centering
\includegraphics[width=\textwidth]{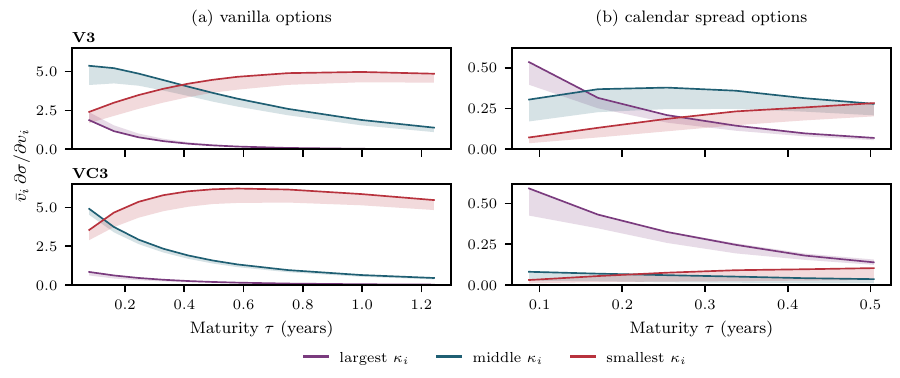}
\caption[Sensitivity of implied volatility to the volatility state variables]{\textbf{Sensitivity of implied volatility to the volatility state variables}\\ Response of the at-the-money implied Bachelier volatility to a proportional move in each $v_i$, $\bar v_i\,\partial\sigma/\partial v_i$, at the median filtered state $\bar v$, against maturity; the bands span the delta buckets. Rows V3 and VC3; (a)~vanilla options, (b)~calendar spread options. }
\label{fig:res_div}
\end{figure}

\autoref{fig:res_div} shows these roles in the option prices, as the response of the
at-the-money implied volatility to a proportional move in each volatility
factor, with bands spanning the delta buckets. It also shows what including the calendar spread
options does to the factors. In V3, the transitory factor carries part of the
volatility of the front vanilla options, in panel (a), and most of the front
calendar spread volatility, in panel (b), so a single factor sets both. VC3
separates the two: its transitory factor carries little of the vanilla volatility
but most of the front calendar spread volatility, while the front vanilla
volatility rests on the middle factor instead.

Finally, we turn to the skew and the leverage correlations $\rho_i$. With the
forcing, which carries the variation of the skew, the estimated $\rho_i$ are
modest in every model, so each volatility factor moves the level of the vanilla
smile far more than its tilt. In V3 and VC3, the
level factor's $\rho_i$ is positive, so a rise in prices tends to come with a rise in the
volatility of the whole curve. Without the forcing, the factors must reproduce
the skew as well. In V3-U, the transitory and middle factors then have large
$\rho_i$ of opposite sign, $\rho_1=0.89$ and $\rho_2=-0.70$, one producing a call skew and the
other a put skew. Since the $\rho_i$ are constant, the skew can then change only through the
balance between these two volatility factors, and the transitory volatility factor of V3-U in
\autoref{fig:app_factors_chi0} follows the skew rather than the volatility of the curve. The
leverage correlations of V2-U, in contrast, stay close to those of V2, so its two factors, which
also carry the volatility term structure, do not take on the skew.

\subsection{Pricing performance and hedging: futures and vanilla options}\label{sec:res_pricing}

We now evaluate the models' pricing and hedging performance, starting with futures prices, which provide a direct test of the factor loadings. \autoref{tab:res_futrmse} reports the
\emph{root mean squared errors} (RMSE), in percent of the observed prices. The
errors of V3 are smaller than those of V2 along the entire curve, with the most important
improvement at M1 and M2, where V2 lacks the transitory factor. The errors of VC3 are smaller
still, mainly at M1. Those of V3-U are close to those of V3, apart from a larger error at
M1.\footnote{The RMSE is dominated by a few extreme days, above all April 2020. In the
Diebold--Mariano tests on the daily errors in \autoref{tab:app_dm}, the errors of V3 are smaller
than those of V2 at every contract, and those of VC3 are smaller than those of V3 over the curve
as a whole. Those of V3-U do not differ significantly from those of V3, at M1 or over the curve.}

\begin{table}[!htbp]
\centering\small
\setlength{\aboverulesep}{0pt}\setlength{\belowrulesep}{0pt}%
\renewcommand{\arraystretch}{1.15}%
\begin{tabular}{@{}lrrrrrrrrrrrrr|r@{}}
\toprule
Model & M1 & M2 & M3 & M4 & M5 & M6 & M7 & Q1 & Q2 & Q3 & Y1 & Y2 & Y3 & All\\
\midrule
V2 & 3.12 & 1.06 & 0.44 & 0.51 & 0.68 & 0.80 & 0.87 & 0.88 & 0.87 & 0.84 & 0.77 & 0.73 & 1.38 & 1.19\\
V3 & 1.09 & 0.26 & 0.41 & 0.41 & 0.35 & 0.29 & 0.23 & 0.20 & 0.29 & 0.40 & 0.49 & 0.31 & 0.55 & 0.46\\
VC3 & 0.80 & 0.26 & 0.37 & 0.33 & 0.25 & 0.21 & 0.19 & 0.20 & 0.28 & 0.37 & 0.42 & 0.25 & 0.55 & 0.38\\
V3-U & 1.44 & 0.27 & 0.41 & 0.44 & 0.37 & 0.31 & 0.25 & 0.24 & 0.32 & 0.41 & 0.44 & 0.32 & 0.55 & 0.54\\
\bottomrule
\end{tabular}
\caption[Futures pricing errors]{\textbf{Futures pricing errors}\\ Root mean squared error (RMSE) of the relative futures pricing errors, (model $-$ data)/data, in percent, with model prices at each date's filtered state, over the 36{,}530 futures prices from 2 January 2013 to 29 February 2024. The column ``All'' aggregates every contract.}
\label{tab:res_futrmse}
\end{table}

We next consider how well the models price the vanilla options, which depend on
the volatility factors as well as on the loadings. \autoref{tab:res_vanmae} shows
the \emph{mean absolute errors} (MAE) of their implied Bachelier volatilities, in dollars per barrel a year, by contract and
moneyness, for V2 and V3 in addition to VC3, whose
factors must also reproduce the curve decorrelation, and V3-U, whose factors must also carry the
skew. We stress that V2, V3 and VC3 were
estimated using the skew forcing, but that their pricing errors, like every price
in this paper, are computed without it. The errors of all four models therefore
measure what the factors and their $\rho_i$ alone price. For all models, the pricing error is largest at M1
and in the wings. Against V2, the third factor reduces the
error substantially at the money but only slightly in the wings, since with
modest $\rho_i$ a further factor adds flexibility to the level of volatility
rather than to the shape of the smile.

Including the calendar spread options in the estimation undoes the gain of the third factor at the money, above all at M1. Leaving out the skew forcing instead improves the shape of the smile. Except at M1, the errors of V3-U are much smaller than those of V3 in the wings, where its
factors now carry the skew. At the money, however, they are larger at M1 and beyond M3. Its
at-the-money prices are on average too high at the front and too low at the longer contracts,
so that its at-the-money term structure is too steep. This indicates that the estimated factors change with what they
must reproduce, and that requiring them also to capture curve decorrelation or skew comes at some cost to the fit of the at-the-money volatility term structure.\footnote{These patterns also hold in the Diebold--Mariano tests on the daily
errors in \autoref{tab:app_dm}.}

\begin{table}[!htbp]
\centering\small
\setlength{\aboverulesep}{0pt}\setlength{\belowrulesep}{0pt}%
\renewcommand{\arraystretch}{1.15}%
\begin{tabular}{@{}lcrrrrrrrrrr|r@{}}
\toprule
Model & Moneyness & M1 & M2 & M3 & M4 & M5 & M6 & M7 & Q1 & Q2 & Q3 & All\\
\midrule
V2 & OTM puts & 1.42 & 1.06 & 0.96 & 0.91 & 0.90 & 0.88 & 0.86 & 0.81 & 0.75 & 0.76 & 0.94\\
V3 & OTM puts & 1.23 & 1.04 & 0.96 & 0.91 & 0.89 & 0.88 & 0.87 & 0.84 & 0.77 & 0.74 & 0.92\\
VC3 & OTM puts & 1.50 & 1.12 & 1.00 & 0.95 & 0.92 & 0.90 & 0.87 & 0.83 & 0.76 & 0.76 & 0.97\\
V3-U & OTM puts & 1.23 & 0.75 & 0.67 & 0.68 & 0.70 & 0.71 & 0.71 & 0.75 & 0.76 & 0.79 & 0.78\\
\midrule
V2 & ATM & 0.73 & 0.36 & 0.35 & 0.37 & 0.38 & 0.36 & 0.34 & 0.30 & 0.31 & 0.39 & 0.39\\
V3 & ATM & 0.48 & 0.36 & 0.32 & 0.27 & 0.26 & 0.25 & 0.27 & 0.26 & 0.23 & 0.23 & 0.30\\
VC3 & ATM & 0.73 & 0.37 & 0.36 & 0.38 & 0.38 & 0.36 & 0.34 & 0.30 & 0.31 & 0.39 & 0.39\\
V3-U & ATM & 0.73 & 0.33 & 0.27 & 0.29 & 0.31 & 0.30 & 0.28 & 0.26 & 0.28 & 0.36 & 0.34\\
\midrule
V2 & OTM calls & 1.32 & 1.22 & 1.25 & 1.28 & 1.29 & 1.27 & 1.22 & 1.16 & 1.06 & 1.07 & 1.22\\
V3 & OTM calls & 1.19 & 1.20 & 1.21 & 1.21 & 1.22 & 1.22 & 1.18 & 1.15 & 1.05 & 1.02 & 1.17\\
VC3 & OTM calls & 1.32 & 1.23 & 1.26 & 1.29 & 1.30 & 1.28 & 1.23 & 1.17 & 1.08 & 1.09 & 1.23\\
V3-U & OTM calls & 1.16 & 0.77 & 0.61 & 0.55 & 0.56 & 0.54 & 0.52 & 0.54 & 0.61 & 0.70 & 0.66\\
\midrule\midrule
V2 & All & 1.16 & 0.88 & 0.85 & 0.85 & 0.86 & 0.84 & 0.81 & 0.75 & 0.70 & 0.73 & 0.85\\
V3 & All & 0.96 & 0.87 & 0.83 & 0.80 & 0.79 & 0.79 & 0.78 & 0.75 & 0.67 & 0.65 & 0.79\\
VC3 & All & 1.19 & 0.91 & 0.87 & 0.87 & 0.87 & 0.85 & 0.82 & 0.76 & 0.71 & 0.74 & 0.86\\
V3-U & All & 1.04 & 0.62 & 0.52 & 0.51 & 0.52 & 0.52 & 0.51 & 0.51 & 0.54 & 0.60 & 0.59\\
\bottomrule
\end{tabular}
\caption[Vanilla pricing errors]{\textbf{Vanilla pricing errors}\\ Mean absolute error (MAE) of the differences between model and observed implied Bachelier volatilities, in \$/bbl a year, with model prices at each date's filtered state, over the 235{,}277 vanilla options from 2 January 2013 to 29 February 2024, by contract and moneyness: out-of-the-money puts (delta buckets $-4$ to $-2$), at the money ($-1$ to $1$) and out-of-the-money calls ($2$ to $4$). The column ``All'' aggregates the contracts, and the rows ``All'' the moneyness groups.}
\label{tab:res_vanmae}
\end{table}

\begin{figure}[!htbp]
\centering
\includegraphics[width=\textwidth]{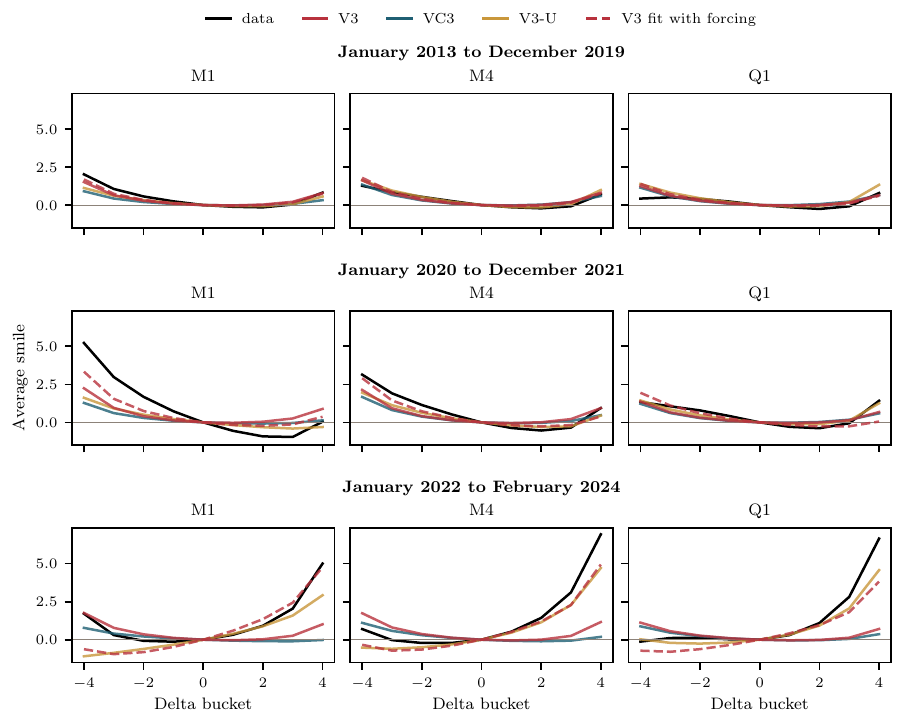}
\caption[Average vanilla smiles by subperiod]{\textbf{Average vanilla smiles by subperiod}\\ Implied Bachelier volatility in each delta bucket less that at the money, in \$/bbl a year, averaged within each of the subperiods 2013--2019, 2020--2021 and 2022--2024, for M1, M4 and Q1: the data, V3, VC3 and V3-U, and the fitted smile of V3 in the estimation, which includes the skew forcing (dashed).}
\label{fig:res_smile}
\end{figure}

The pricing errors above aggregate the whole sample. To see how the smile changes over
time, we instead split the sample into three periods, keeping the models estimated
on the whole sample: the calmer early period 2013--2019, the Covid-19 shock in
2020--2021 and the disruptions after the Russian invasion of Ukraine in
2022--2024. \autoref{fig:res_smile} shows the average smiles of M1, M4 and
Q1 in each period, as the implied volatility of each delta bucket with the
at-the-money implied volatility subtracted. The skew changes sign over the sample,
with puts at a premium in 2020--2021 and calls in 2022--2024. The smiles that V3 and VC3 price, which again never include the skew forcing, are
close to the data in 2013--2019, follow the put skew of 2020--2021 only partly, and hardly follow the call skew of 2022--2024, as their modest correlations imply. The fitted smile of V3 in
the estimation, which includes the skew forcing and is shown dashed, follows both
skews much more closely. In the estimation of the models with the skew forcing, the changes of the skew over time are
thus carried by the skew forcing rather than by the factors. In
V3-U the factors carry them instead. They follow the call skew of 2022--2024 about as well as the fitted smile with the skew
forcing, but the put skew of 2020--2021 no better than the factors of
V3.

Beyond the fit on each date, a model of option prices must describe how the
volatility surface changes from one day to the next. The pricing errors do not
measure these changes, but a hedge tests the model's
description of them. If the factors of a model are
the risks that move the volatility surface, a hedge with futures and as
many options as the model has factors removes almost all of the daily risk of an
option portfolio. We therefore hedge the portfolio of every vanilla option
used in the estimation over one business day at a time. Each option is delta-hedged in its own futures
contract, and we add at-the-money options on M4, Q2 and M1 one at a time, up to $n$. The
positions minimize the model's one-day variance of each hedged option.\footnote{Further details on the hedge construction are given in
Appendix~\ref{app:hedge}.} We
report three measures: the \emph{hedging effectiveness} of \cite{ederington1979},
$1-\operatorname{Var}(\Pi)/\operatorname{Var}(\Pi^u)$ with $\Pi$ and $\Pi^u$ the
hedged and unhedged P\&L per unit of the portfolio's vega, its version based on the median
absolute deviation, and the mean absolute daily P\&L per option.\footnote{Weighting
the options by their open interest changes none of the conclusions, as
\autoref{tab:app_hedge_oi} shows.}

\begin{table}[!htbp]
\centering\small
\setlength{\aboverulesep}{0pt}\setlength{\belowrulesep}{0pt}%
\renewcommand{\arraystretch}{1.15}%
\begin{tabular}{@{}l|c|c|ccc@{}}
\toprule
 & & \multicolumn{4}{c}{hedge options $k$}\\
\cmidrule(l){3-6}
model & unhedged & 0 & 1 & 2 & 3\\
\midrule
\multicolumn{6}{@{}l}{\emph{A. Hedging effectiveness}}\\
V2 & --- & 0.09\textsuperscript{n.s.} & 0.94 & \textbf{0.95} & ---\\
 &  & (0.12) & (0.92) & (0.96) & \\[2pt]
V3 & --- & 0.09\textsuperscript{n.s.} & 0.94 & 0.96 & \textbf{0.97}\\
 &  & (0.12) & (0.93) & (0.95) & (0.96)\\[2pt]
VC3 & --- & 0.09\textsuperscript{n.s.} & 0.94 & 0.95 & \textbf{0.96}\textsuperscript{\dag}\\
 &  & (0.13) & (0.93) & (0.96) & (0.96)\\[2pt]
V2-U & --- & 0.09\textsuperscript{n.s.} & 0.94 & \textbf{0.95} & ---\\
 &  & (0.12) & (0.93) & (0.96) & \\[2pt]
V3-U & --- & 0.03\textsuperscript{n.s.} & 0.96 & 0.97 & \textbf{0.83}\textsuperscript{n.s.}\\
 &  & (0.08) & (0.94) & (0.97) & (0.94)\\[2pt]
\midrule
\multicolumn{6}{@{}l}{\emph{B. Mean absolute daily P\&L per option, cents}}\\
V2 & 25.38 & 7.85 & 3.94 & \textbf{3.26} & ---\\[2pt]
V3 & 25.38 & 7.83 & 4.54 & 3.31 & \textbf{2.93}\\[2pt]
VC3 & 25.38 & 7.85 & 3.92 & 3.27 & \textbf{3.21}\\[2pt]
V2-U & 25.38 & 7.85 & 3.93 & \textbf{3.25} & ---\\[2pt]
V3-U & 25.38 & 7.97 & 3.84 & 3.21 & \textbf{5.27}\\[2pt]
\bottomrule
\end{tabular}
\caption[Hedging the vanilla options]{\textbf{Hedging the vanilla options}\\ The portfolio of every vanilla option used in the estimation, hedged over one business day with $k$ at-the-money hedge options (M4, Q2, M1) and each option's own future, over 2{,}809 dates. Positions minimize the model's one-day variance of each hedged option. (A)~Hedging effectiveness per unit of portfolio vega, with the version based on the median absolute deviation in parentheses. (B)~Mean absolute daily P\&L per option, in cents per barrel. Bold where $k=n$. Each change from the column to the left is tested with the Newey--West $t$-statistic (50 lags). Unless marked, every change is significant at the 1 percent level; \textsuperscript{\dag} marks significance at the 5 percent level only and \textsuperscript{n.s.} no significance at the 5 percent level.}
\label{tab:res_hedge}
\end{table}

\autoref{tab:res_hedge} reports the hedges. In every model, futures alone remove
less than 10 percent of the variance of the portfolio, while a single at-the-money option on
M4 removes most of the rest.\footnote{Since the portfolio and $\mathcal M_2$ both
aggregate the vanilla options across maturities and moneyness, they are closely
related. After the futures hedge, the daily change in the level of
$\mathcal M_2$ explains 78 percent of the variance of the portfolio's P\&L.} This indicates that most of the daily risk of an
option portfolio is volatility risk that the futures do not hedge, consistent with the
unspanned stochastic volatility that \cite{trolle2009} document for crude oil
options. For the models estimated using the skew forcing, each further option then
reduces the risk again, until a hedge with as many options as the model has
factors removes almost all of the variance of the portfolio. This is consistent with
the factors of these models being the risks that move the volatility
surface. The comparison between the models points the same way once each has as many hedge
options as factors. The hedges of V3 then have a lower mean absolute daily P\&L per option than
those of V2, 2.93 against 3.26 cents, although with two hedge options each they do slightly
worse.\footnote{The paired difference in mean absolute daily P\&L per option at two hedge options
is $+0.05$ cents ($t=3.17$).}

Including the calendar spread options in the estimation costs VC3 accuracy in the
level of the vanilla prices, and somewhat in their dynamics. Although its factors are
assigned differently, the hedges of VC3 do at least as well as those of V3 with one or two
hedge options. With all three, however, their mean absolute P\&L per option is higher, by a
difference that is significant at the 1 percent level.\footnote{The paired difference in mean
absolute daily P\&L per option is $+0.28$ cents ($t=12.54$).}

V3-U shows the reverse of VC3, smaller pricing errors but worse dynamics. With one
or two hedge options, the hedges of V3-U do at least as well as those of V3, but with the
third at-the-money option their mean absolute P\&L rises sharply and their
effectiveness drops to 0.83.\footnote{The fall in effectiveness is not
significant, since a few extreme days dominate the variance, while the rise in the
P\&L is. In 2020--2021, the effectiveness is 0.61.} This failure is likely related to the skew,
since to produce it the transitory and middle factors of V3-U take leverage correlations of
opposite sign, which makes them nearly degenerate at the money, where the hedge options are.
V2-U, whose two factors are both needed for the volatility term structure and do not follow the
skew, prices and hedges the vanilla options comparably to V2.\footnote{Aggregated over all options,
the mean absolute error is 0.85 for both.} The better prices of V3-U thus seem to come from fitting
each day's skew at the expense of describing how the volatility surface moves.

The pricing and hedging results above are in sample. To test them out of sample, we estimate V3
on each half of the sample alone, from January 2013 to July 2018 and from August 2018 to February 2024,
and evaluate each estimate on the other half. Both estimates keep the factors, leverage
correlations and spillover of the full-sample estimate. Out of sample, each hedges the vanilla
options nearly as well as the estimate on that half itself, and its vanilla pricing errors are
not significantly larger. Its futures pricing errors, by contrast, grow
out of sample in both halves. The hedges and the vanilla prices are thus robust out of sample,
while the fit to the futures curve depends more on the period of estimation.
The estimates, pricing errors and hedges on each half are given in Appendix~\ref{app:oos}.

\subsection{Volatility}\label{sec:res_mfm}

\begin{figure}[!htbp]
\centering
\includegraphics[width=\textwidth]{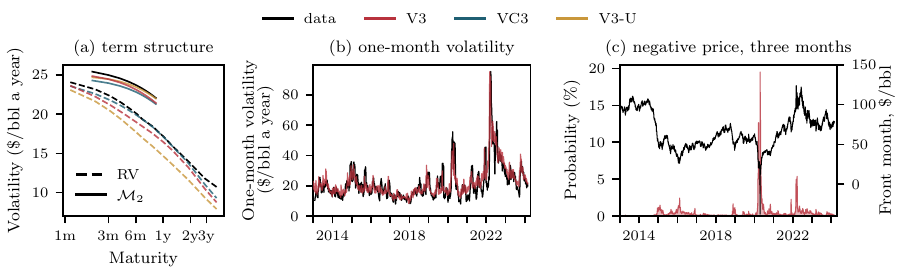}
\caption[Futures volatility]{\textbf{Futures volatility}\\ (a)~Volatility against maturity, averaged over 2013--2024, in \$/bbl a year: for the data, the realized volatility of each futures contract (dashed) and the model-free volatility $\mathcal M_2$ at constant maturities (solid); for the models, their model-implied counterparts. (b)~The one-month volatility of V3 and the 21-day rolling realized volatility of the M1 contract. (c)~The probability under $\mathbb P$ that the spot price is at or below zero three months ahead, from V3 at each date's filtered state, and the settlement price of the futures contract nearest to expiry (right axis). Daily data from 2 January 2013 to 29 February 2024.}
\label{fig:res_futvol}
\end{figure}

We now compare the volatility the models imply with the volatility in the data.
\autoref{fig:res_futvol}(a) shows the term structure of the realized volatility of
the futures and of the implied volatility $\mathcal M_2$, together with their
model-implied counterparts for V3, VC3 and V3-U. In the data, $\mathcal M_2$ lies above the realized volatility at every maturity,
and all three models reproduce both term structures well. VC3 is slightly closer
to the realized volatility, while V3 and V3-U lie a little below it. In panel (b), the one-month volatility of V3 moves over time with the 21-day realized volatility of the
M1 contract, but falls short of the peak of the Covid-19 shock in 2020. Since an arithmetic model
assigns a positive probability to negative prices, panel (c) shows this probability for V3, under
$\mathbb P$ and three months ahead. It is close to zero on most dates and peaks in April 2020, when
the front-month futures price turned negative. The probabilities that the model assigns to negative
prices thus appear reasonable, practically zero in normal times and large only when negative prices
were observed.

\begin{figure}[!htbp]
\centering
\includegraphics[width=\textwidth]{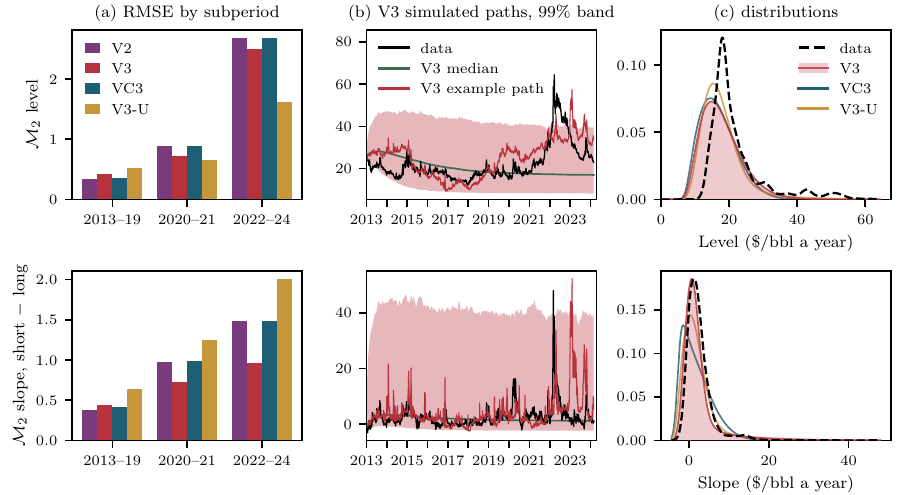}
\caption[Model-free volatility]{\textbf{Model-free volatility}\\ Level (mean over maturities from 42 to 210 business days, top) and slope (42-day less 210-day value, bottom) of $\mathcal M_2$, in \$/bbl a year. (a)~Root mean squared error (RMSE) of the model-implied level and slope for V2, V3, VC3 and V3-U in each of the subperiods 2013--2019, 2020--2021 and 2022--2024. (b)~Data against the median and the 99 percent band of 10{,}000 paths of V3 simulated under $\mathbb P$ from 2 January 2013, and one simulated path. (c)~Distribution of the data (dashed) and stationary distributions of V3, VC3 and V3-U, from 10{,}000 paths simulated under $\mathbb P$ with a ten-year burn-in.}
\label{fig:res_cm2}
\end{figure}

To assess how well the models reproduce $\mathcal M_2$ over time, we compare its
level and slope in the data with their model-implied counterparts. Since
$\mathcal M_2$ does not depend on the leverage correlations $\rho_i$, it tests the
volatility term structure of the factors alone. \autoref{fig:res_cm2}(a) shows the
RMSE of the difference in each subperiod for V2, V3, VC3 and V3-U. For every
model, the errors are largest in 2022--2024, around the 2022 peak. In the calmer early period,
V2, V3 and VC3 miss the slope by about the same, but from 2020 on, when the volatility term
structure moves most, V3 misses it by far the least. Although the volatility term structure in
the data is essentially two-dimensional, it thus takes three factors to reproduce its slope
through turbulent periods. Over the whole sample, V3 fits the slope of $\mathcal M_2$ better than
VC3 and V3-U, whose factors must also reproduce the curve decorrelation and the skew,
respectively. Both models also price the at-the-money vanilla options worse than V3, which
suggests that fitting either feature comes at the expense of the volatility term
structure.\footnote{The RMSEs of the level and the slope
of $\mathcal M_2$, in dollars per barrel a year, are 1.27 and 0.83 for V2, 1.19 and 0.63 for V3,
1.27 and 0.84 for VC3, and 0.87 and 1.15 for V3-U.}

We next ask whether the dynamics of the models produce paths of $\mathcal M_2$
like the observed one. \autoref{fig:res_cm2}(b) shows the observed path against
the 99 percent band of 10{,}000 paths of V3, simulated under $\mathbb P$ from the
start of the sample. The observed slope stays within the band on 99 percent of the
dates and the level on 91 percent. The level leaves the band in essentially one
episode, the 2022 peak, where it lies above the band from February to December
2022. The simulated paths of VC3 cover the slope on nearly the same share and the
level on a smaller one, 89 percent,
and panel (c) shows that the stationary distributions of VC3 and V3-U are close to
that of V3. This suggests that the dynamics of
the volatility factors capture the evolution of $\mathcal M_2$ reasonably well,
except for the magnitude of the 2022 shock.

\subsection{Calendar spread options and curve decorrelation}\label{sec:res_cso}

We now turn to the calendar spread options, which carry information on the
dependence structure of futures prices that neither the futures nor the vanilla
options provide. V3 is estimated on futures and vanilla options alone, while VC3
is estimated on the calendar spread options as well. We compare the two in how they price the calendar spread options, in the
implied one-month decorrelation and in how they hedge them. The comparison shows how much of the dependence structure
can be inferred from futures and vanilla options alone, and what including the
calendar spread options in the estimation changes in the factors.

\begin{table}[!htbp]
\centering\small
\setlength{\aboverulesep}{0pt}\setlength{\belowrulesep}{0pt}%
\renewcommand{\arraystretch}{1.15}%
\begin{tabular}{@{}l|ccc|ccc|cc@{}}
\toprule
 & \multicolumn{3}{c|}{V3} & \multicolumn{3}{c|}{VC3} & \multicolumn{2}{c}{VC3 $-$ V3}\\
contract & MAE & mean error & MAE (\%) & MAE & mean error & MAE (\%) & RMSE & $t$\\\midrule
M1 & 0.744 & $-0.152$ & 36.7 & 0.568 & $-0.462$ & 18.0 & $-$0.17 & ($-$3.19)\\
M2 & 0.524 & $+0.137$ & 33.2 & 0.334 & $-0.251$ & 14.4 & $-$0.18 & ($-$4.67)\\
M3 & 0.475 & $+0.279$ & 34.7 & 0.245 & $-0.166$ & 13.5 & $-$0.26 & ($-$5.84)\\
M4 & 0.473 & $+0.350$ & 37.3 & 0.223 & $-0.145$ & 13.5 & $-$0.29 & ($-$5.87)\\
M5 & 0.481 & $+0.406$ & 40.6 & 0.206 & $-0.121$ & 13.8 & $-$0.30 & ($-$6.31)\\
M6 & 0.491 & $+0.436$ & 43.7 & 0.192 & $-0.106$ & 13.9 & $-$0.32 & ($-$6.45)\\\midrule
at the money & 0.535 & $+0.125$ & 37.7 & 0.237 & $-0.187$ & 12.0 & $-$0.31 & ($-$9.62)\\
all & 0.529 & $+0.244$ & 37.6 & 0.293 & $-0.207$ & 14.5 & $-$0.22 & ($-$7.16)\\
\bottomrule\end{tabular}
\caption[Calendar spread option pricing errors]{\textbf{Calendar spread option pricing errors}\\ Mean absolute error (MAE) and mean error (model less data) of the implied Bachelier volatilities, in \$/bbl a year, and the MAE relative to the observed implied volatility, in percent, by contract, over the 70{,}654 calendar spread options valued in the data and by both models from 2 January 2013 to 29 February 2024. V3 prices them at its state filtered from the futures and vanilla options alone, VC3 at its state filtered also from the calendar spread options. The last two columns are the mean difference between the models' daily RMSEs, VC3 less V3, and its Newey--West $t$-statistic (50 lags). Every difference is significant at the 1 percent level.}
\label{tab:res_cso_err}
\end{table}

We price the calendar spread options with both models at their filtered
states.\footnote{On each date, the state of V3 is filtered from the futures and
vanilla options up to that date, so the calendar spread options enter neither
its estimation nor its filter.} \autoref{tab:res_cso_err} reports the pricing
errors of both models in implied Bachelier volatilities. V3 overprices the
calendar spread options at every contract but M1, increasingly so with the maturity,
and its implied volatilities miss the observed ones by 38 percent on average. VC3 prices them with a small error that varies little across
contracts and is slightly too low, most at M1. The gap between the models widens over
the sample, as the errors of V3 grow from before 2020 to after the 2022 shock,
while those of VC3 stay small in every subperiod.\footnote{The median daily RMSE
of the relative errors of V3 rises from 32 percent before 2020 to 64 percent
after the 2022 shock, while that of VC3 stays below 22 percent in every
subperiod.} Since V3 prices the vanilla options well, and with them the volatility of each
contract, this points to the dependence structure implied by V3 as the source of
its errors.

\begin{figure}[!htbp]
\centering
\includegraphics[width=\textwidth]{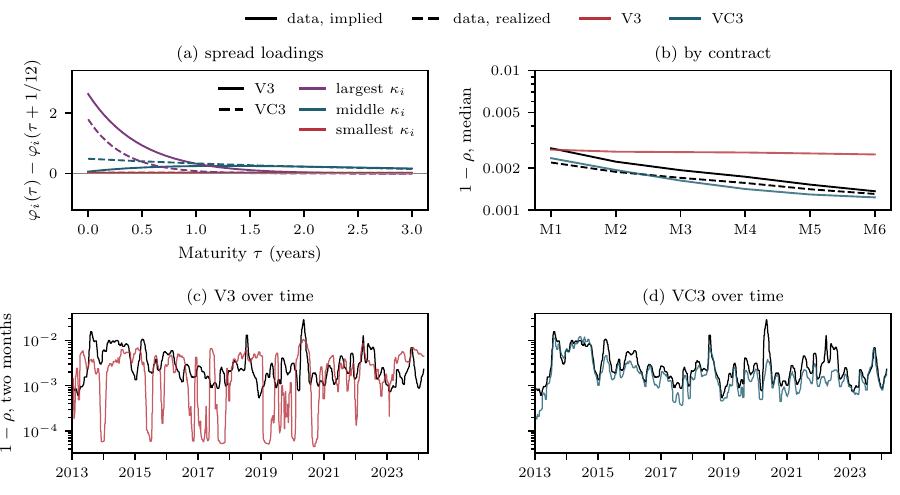}
\caption[Spread loadings and implied decorrelation]{\textbf{Spread loadings and implied decorrelation}\\ (a)~One-month spread loadings $\varphi_i(\tau)-\varphi_i(\tau+1/12)$ of the transitory, middle and level factors; V3 solid, VC3 dashed. (b)~Median implied decorrelation by contract, in the data and at the prices of V3 and VC3, and its counterpart from the realized variances of the futures spread and its first leg over the life of each option (dashed). (c), (d)~The implied decorrelation at a constant two months in the data, against the instantaneous decorrelation at the filtered state of V3 in (c) and of VC3 in (d), as 21-day rolling medians on a common scale.}
\label{fig:res_cso_corr}
\end{figure}

To check this, consider the implied one-month decorrelation, whose median by contract
\autoref{fig:res_cso_corr}(b) shows. V3 overstates it at every contract but M1, by a
factor rising from 1.2 at M2 to 1.8 at M6, since its decorrelation hardly falls with maturity,
while VC3 lies slightly below the data. For a given volatility of the
first leg, the implied volatility of a calendar spread option scales with the
square root of the decorrelation, so the overstatement accounts for most of the
overpricing of V3 and for its growth with maturity.\footnote{Such an overstatement
implies an overpricing of 9 percent at M2, rising to 35 percent at M6, in line with the mean
errors of V3 in \autoref{tab:res_cso_err}. At M1, where V3 does not overstate the
decorrelation, its errors have mixed signs.}

We next ask whether the models also follow the decorrelation over time.
\autoref{fig:res_cso_corr}(c) and (d) show it at a constant maturity of two months,
for V3 and for VC3. The
decorrelation of VC3 moves closely with that of the data, while that of V3
hardly does.\footnote{The rank correlation of the model's decorrelation at two months with
that of the data is 0.13 for V3, against 0.92 for VC3.}
That V3
fits the level and slope of $\mathcal M_2$ but not the curve decorrelation is
consistent with the two moving largely independently in the data. In the model, the instantaneous decorrelation\footnote{For consistency with the implied decorrelation of the data, the ratio
uses the variance of the first leg for both legs.} of the contracts with maturities
$\tau$ and $\tau+1/12$ is
\begin{equation*}
\frac{\sum_i\left(\varphi_i(\tau)-\varphi_i(\tau+1/12)\right)^2 v_i}
{2\sum_i\varphi_i(\tau)^2 v_i}.
\end{equation*}
The numerator contains each factor's \emph{spread loading}
$\varphi_i(\tau)-\varphi_i(\tau+1/12)$, shown in \autoref{fig:res_cso_corr}(a).
The decorrelation thus depends on how the volatility is divided between factors
whose loadings change over one month and factors whose loadings do not. The
vanilla options depend only on the denominator, the volatility of each contract,
while the calendar spread options depend on the fraction, so models that
divide the volatility differently can price the vanilla options alike but the
calendar spread options differently.

In V3, the loading of the transitory factor falls steeply at the front of the curve, so it has
a large spread loading there. It also carries much of the volatility of the front contracts, so the
decorrelation of V3 moves with that volatility.\footnote{The rank
correlation of the decorrelation of V3 at two months with its two-month volatility is 0.41,
against $-0.19$ for VC3. The at-the-money calendar spread implied volatility of V3 also responds
to $v_2$ and $v_3$, although their spread loadings are small, consistent with the spillover
through which $v_3$ raises $v_2$ and $v_2$ raises $v_1$ ($\Gamma_{2,3}=1.93$, $\Gamma_{1,2}=3.80$).
In VC3 there is no spillover to $v_1$, and it goes only to $v_2$ ($\Gamma_{2,3}=4.40$), whose
spread loading is small.} Including the calendar spread options in the estimation changes this. In VC3,
the volatility of the calendar spread options rests mainly on the transitory
factor, the curve decorrelation factor, which carries little of the volatility of
the vanilla options, as \autoref{fig:res_div} showed. Its volatility factor $v_1$
tracks the at-the-money variance of the calendar spread options, so the
decorrelation of VC3 moves with $v_1$ rather than with the volatility of the
front contracts. The middle factor of VC3, in turn, has a spread loading of a
constant 3.2 percent of its loading, so it moves adjacent contracts almost one
for one.

\begin{figure}[!htbp]
\centering
\includegraphics[width=\textwidth]{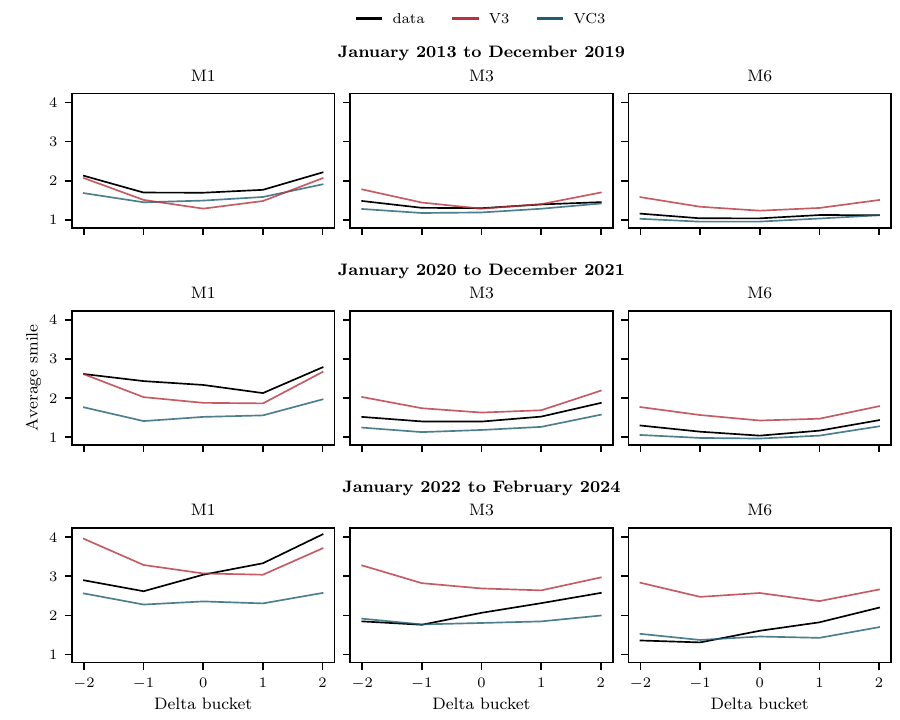}
\caption[Average calendar spread option smiles by subperiod]{\textbf{Average calendar spread option smiles by subperiod}\\ Implied Bachelier volatility in each delta bucket, in \$/bbl a year, averaged within each of the subperiods 2013--2019, 2020--2021 and 2022--2024, for the calendar spread options on M1, M3 and M6 valued in the data and by V3 and VC3.}
\label{fig:res_cso_smile}
\end{figure}

\autoref{fig:res_cso_smile} shows the average smiles of the calendar spread
options in the three subperiods, and they agree with the pricing errors. Before 2022, the
smiles in the data are close to symmetric, and VC3 follows their shape, below their level,
while V3 lies above them beyond M1. After the 2022 shock, the smiles in the data
tilt toward the calls at every contract, as do the
smiles of the vanilla options, and neither model reproduces this tilt.

\begin{table}[!htbp]
\centering\small
\setlength{\aboverulesep}{0pt}\setlength{\belowrulesep}{0pt}%
\renewcommand{\arraystretch}{1.15}%
\begin{tabular}{@{}lc|c|c|cc@{}}
\toprule
 & & & \multicolumn{3}{c}{hedge options $k$}\\
\cmidrule(l){4-6}
window & model & unhedged & 0 & 1 & 2\\
\midrule
\multicolumn{6}{@{}l}{\emph{A. Hedging effectiveness}}\\
1 day & V3 & --- & 0.26\textsuperscript{n.s.} & 0.26\textsuperscript{n.s.} & 0.68\textsuperscript{n.s.}\\
 &  &  & (0.55) & (0.03) & (0.29)\\
 & VC3 & --- & 0.44\textsuperscript{\dag} & 0.33\textsuperscript{n.s.} & 0.64\textsuperscript{n.s.}\\
 &  &  & (0.51) & (0.13) & (0.43)\\
 & $t$ &  & $-$1.11 & $-$0.84 & 0.72\\[2pt]
\midrule
\multicolumn{6}{@{}l}{\emph{B. Mean absolute daily P\&L per option, cents}}\\
1 day & V3 & 1.59 & 0.99 & 1.03 & 0.95\\
 & VC3 & 1.59 & 0.92 & 0.97 & 0.89\\
 & $t$ &  & $-$4.59 & $-$3.85 & $-$3.83\\[2pt]
\bottomrule
\end{tabular}
\caption[Hedging the calendar spread options]{\textbf{Hedging the calendar spread options}\\ The portfolio of every calendar spread option used in the estimation, hedged over one business day with the six futures spreads ($k=0$) and $k$ at-the-money calendar spread options (M2; M2 and M4), over 2{,}806 dates. Positions minimize the model's one-day variance of each hedged option. (A)~Hedging effectiveness per unit of portfolio vega, with the version based on the median absolute deviation in parentheses. (B)~Mean absolute daily P\&L per option, in cents per barrel. Each change from the column to the left is tested with the Newey--West $t$-statistic (50 lags). Unless marked, every change is significant at the 1 percent level; \textsuperscript{\dag} marks significance at the 5 percent level only and \textsuperscript{n.s.} no significance at the 5 percent level. The rows $t$ are the Newey--West $t$-statistics of the paired difference between the models, VC3 less V3, in the variance of the P\&L per unit of portfolio vega (A) or in the absolute P\&L per option (B).}
\label{tab:res_cso_hedge}
\end{table}

We also hedge the portfolio of all calendar spread options used in the estimation.
The hedge is built and evaluated as for the vanilla portfolio, over one business day,
with positions that minimize the model's one-day variance of each hedged option and
with the same three measures. Only the hedge instruments differ. Since the
calendar spread options are written on futures spreads, the hedge without options
($k=0$) uses the six spreads they are written on rather than single futures, and the hedge options
are at-the-money calendar spread options, on M2 ($k=1$), whose option is quoted on the most dates,
and on M2 and M4 ($k=2$). A
few extreme dates dominate the variance of this portfolio, so we judge a typical day
by the effectiveness based on the median absolute deviation, in parentheses in
panel A of \autoref{tab:res_cso_hedge}.\footnote{The top 1 percent of the dates
carry 64 percent of the variance of the unhedged portfolio.} By this measure, the spreads alone
remove more than 50 percent of the risk of this portfolio in both models, against about 12 percent
for the futures in the vanilla portfolio. This is partly because, unlike the vanilla
portfolio, this portfolio is far from delta-neutral, since it holds fewer options, more
often on one side of the money than the other.\footnote{On the median date, the
portfolio holds 26 calendar spread options against 84 vanilla options, and its net
delta is 25 percent of its gross delta against 7 percent for the vanilla portfolio.} The hedges of VC3 have a slightly but significantly lower
mean absolute daily P\&L per option in panel B,\footnote{The Newey--West
$t$-statistics of the paired difference in absolute P\&L per option, VC3 less
V3, are $-3.85$ with one hedge option and $-3.83$ with two.} but by the median-based effectiveness neither
model's hedge options improve on the spreads.

The calendar spread options are less liquid than the vanilla options, so their settlements are
noisier. We therefore also evaluate the hedges over ten business days, over which the settlement
errors largely cancel. Over this horizon, the hedge options improve clearly on the spreads in both
models, and with two hedge options the hedges of VC3 are better in a typical ten-day period in every subperiod, with a significantly lower
mean absolute P\&L per option over the full sample.\footnote{\autoref{tab:app_cso_hedge_periods} in the appendix reports
the ten-day hedges by subperiod. The Newey--West $t$-statistics of the paired difference in
absolute P\&L per option with two hedge options, VC3 less V3, are $-2.24$ over the full sample and
$-2.86$, $0.89$ and $-1.22$ in the three subperiods.} In the variance, which a few extreme days
dominate, the differences between the models are mostly not significant. Given the illiquidity of
the calendar spread options, this evidence is weaker than that of their prices, but it points the
same way.

\subsection{Skewness}\label{sec:res_skew}

\begin{figure}[!htbp]\centering
\includegraphics[width=\textwidth]{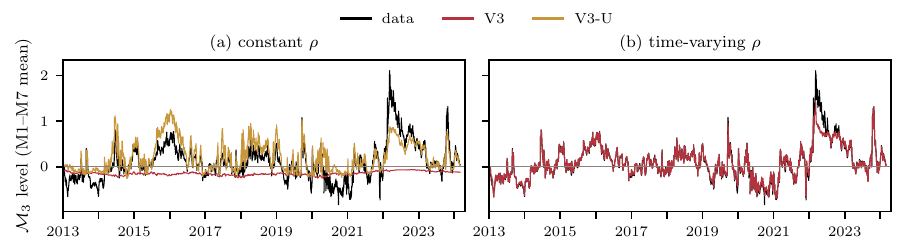}
\caption[Model-free skewness]{\textbf{Model-free skewness}\\ Level of $\mathcal M_3$ (mean over M1--M7). (a)~Data, and the own skew of V3 and V3-U, the $\mathcal M_3$ that their factors and constant leverage correlations $\rho_i$ produce. (b)~Data, and V3 with one common shift of all its leverage correlations each day, fitted to the day's term structure of $\mathcal M_3$ as described in Appendix~\ref{app:cm}. Daily data from 2 January 2013 to 29 February 2024.}
\label{fig:res_cm3}
\end{figure}

Unlike $\mathcal M_2$, the skew $\mathcal M_3$ depends on the leverage
correlations $\rho_i$. With constant $\rho_i$, it can therefore move over time only
as the volatility factors move, the same factors that carry the volatility term
structure. \autoref{fig:res_cm3}(a) shows the level of $\mathcal M_3$ and each
model's \emph{own skew}, the $\mathcal M_3$ that the factors and their $\rho_i$
produce, in V3 and V3-U. As \autoref{fig:res_smile} showed, the skew changes sign over the sample, and $\mathcal M_3$ shows a call skew already
in 2014--2016. The own skew of V3 is nearly flat and explains none of this
variation, as intended, since in the estimation the skew forcing carries
it.\footnote{The same holds for VC3.} V3-U, whose factors must carry the skew
themselves, follows it more closely and explains 55 percent of its variation, but misses the put
skew of 2020--2021. This indicates that, with constant $\rho_i$, the factors can produce part of
the skew's movement only at the cost of a worse fit to the slope of the volatility term structure, consistent with the skew moving largely independently of the volatility term
structure in the data.

We next ask what it would take for a model whose factors fit the volatility term structure to also
fit the skew. Since the skew in the data moves mostly as one curve across
maturities, we keep V3 as estimated and fit, each day, one common shift of all its leverage
correlations to the observed skew, as described in Appendix~\ref{app:cm}. With this shift, V3 fits
the level of $\mathcal M_3$ closely in \autoref{fig:res_cm3}(b).\footnote{The close fit is expected,
since the shift is fitted to the skew each day, and the $R^2$ is 0.94. What the exercise shows is how
large and how persistent the shift must be for V3 to fit the skew.} The shift, which traces how the leverage would have to move, is small before 2022 and large in 2022, and it is
persistent, with a half-life of about four months. This suggests that most of the movement in the
skew could be accounted for by a persistent stochastic leverage driven by its own factor, leaving the volatility factors free to
fit the volatility term structure.

The data thus do not support coupling the skew to the
factors of the volatility term structure through constant leverage correlations, since factors that also carry the skew fit the slope of the
volatility term structure worse and hedge the vanilla options less well. A model
of the full volatility surface would therefore need a
stochastic leverage separate from these factors. In this paper, the skew forcing
keeps the skew out of the factors in the estimation, and a model with
stochastic leverage decoupled from the volatility term structure is left for
future work.

\subsection{Arithmetic and geometric models}\label{sec:res_geom}

To see whether the data give a reason to prefer the arithmetic structure to the traditional geometric one, we compare the two-factor arithmetic model with its geometric counterpart. The geometric model is that of \cite{chiarella2013}, in which the same loadings and volatility factors drive the relative changes of the futures prices, $\mathrm{d}F/F$, instead of the changes $\mathrm{d}F$. In \cite{chiarella2013} the volatility factors evolve independently, and we couple them through the upper-triangular $\Gamma$ of our model and impose the same parameter restrictions as in the arithmetic model. We estimate both models without the skew forcing on the same futures and vanilla options, with the same filter and constraints, so that only the price-proportionality differs. The geometric model cannot price options whose futures price or strike is not positive, as in April 2020, so we leave these options out for both models.\footnote{This removes 7 of the 235{,}277 vanilla options.} The calendar spread options are left out, since the geometric model has no semi-closed-form price for them.

\autoref{tab:res_geom} reports the mean absolute errors in implied Bachelier and implied Black volatility. The errors of the geometric model are 19 percent larger at the money and 45 percent larger in the out-of-the-money puts, at 1.29 against 0.89~\$/bbl a year. Only in the out-of-the-money calls are they slightly smaller. Over all options they are 16 percent larger in implied Bachelier and 22 percent larger in implied Black volatility. With Newey--West standard errors over 50 lags, the Diebold--Mariano $t$-statistic of the difference in daily mean absolute error is 2.58. The difference is significant at the 1 percent level, though not by a wide margin.\footnote{Without March--June 2020 the $t$-statistic is 2.24, and the arithmetic model has the smaller mean absolute error in seven of the twelve years.}

\begin{table}[!htbp]
\centering\small
\setlength{\aboverulesep}{0pt}\setlength{\belowrulesep}{0pt}%
\renewcommand{\arraystretch}{1.15}%
\begin{tabular}{@{}l|rr|rr@{}}
\toprule
 & \multicolumn{2}{c|}{Bachelier} & \multicolumn{2}{c}{Black}\\
Moneyness & Arithmetic & Geometric & Arithmetic & Geometric\\
\midrule
OTM puts & 0.89 & 1.29 & 1.66 & 2.46\\
ATM & 0.48 & 0.57 & 0.79 & 0.92\\
OTM calls & 1.33 & 1.26 & 1.76 & 1.74\\
\midrule
All & 0.90 & 1.04 & 1.40 & 1.70\\
Diebold--Mariano $t$ & \multicolumn{2}{c|}{2.58} & \multicolumn{2}{c}{2.71}\\
\bottomrule
\end{tabular}
\caption[Arithmetic and geometric models]{\textbf{Arithmetic and geometric models}\\ Mean absolute error (MAE) of the implied Bachelier volatilities, in \$/bbl a year, and of the implied Black volatilities, in percentage points, for the two-factor arithmetic model and its geometric counterpart, both estimated without the skew forcing on the same 235{,}270 vanilla options: out-of-the-money puts (delta buckets $-4$ to $-2$), at the money ($-1$ to $1$) and out-of-the-money calls ($2$ to $4$). The last row is the Diebold--Mariano $t$-statistic of the difference in daily MAE, geometric less arithmetic, with Newey--West standard errors (50 lags).}
\label{tab:res_geom}
\end{table}

This pattern is consistent with the average smiles of \autoref{fig:black_bach_smile}, which
rise steeply toward the puts in implied Black volatility but are close to
symmetric in implied Bachelier volatility. Without leverage, the smile of the
geometric model is symmetric in implied Black volatility and that of the
arithmetic model in implied Bachelier volatility. In implied Bachelier
volatility, the baseline of the geometric model therefore tilts toward the
calls, since its volatility in dollars rises and falls with the price, which
amplifies rises and damps falls. Its leverage correlations must first offset this tilt and
then produce the put skew in the data, which on average takes a skew about five
times as large as in the arithmetic model.\footnote{Measured by the median
20-delta risk reversal relative to the at-the-money volatility, which is $-0.05$
in implied Bachelier volatility in the data, while a symmetric Black smile
corresponds to $+0.17$.
The leverage correlations of the arithmetic model are $-0.12$ and $+0.23$.}
Accordingly, both leverage correlations of the geometric model are negative, at
$-0.67$ and $-0.41$, while those of the arithmetic model differ in sign.
The errors of the geometric model nevertheless remain largest in the puts, so constant leverage
correlations apparently cannot fully offset the tilt of its baseline. In the
arithmetic model, which has symmetry as its baseline, the leverage correlations
are left for the skew in the data.

The arithmetic model also fits the futures better, with a root mean squared error of \$0.58 against
\$0.68 per barrel. With futures and one or two options in the hedge, the hedges of the arithmetic
model, reported in \autoref{tab:app_geom_hedge}, have a significantly lower mean absolute daily P\&L per
option, while their effectiveness does not differ significantly. The arithmetic model thus fits the
futures and vanilla options better than its geometric counterpart, and its hedges of the vanilla
options are slightly better per option.

\begin{table}[!htbp]
\centering\small
\setlength{\aboverulesep}{0pt}\setlength{\belowrulesep}{0pt}%
\renewcommand{\arraystretch}{1.15}%
\begin{tabular}{@{}l|c|cc@{}}
\toprule
 & & \multicolumn{2}{c}{hedge options $k$}\\
\cmidrule(l){3-4}
model & unhedged & 1 & 2\\
\midrule
\multicolumn{4}{@{}l}{\emph{A. Hedging effectiveness}}\\
Arithmetic & --- & 0.95 & 0.96\\
 &  & (0.92) & (0.94)\\[2pt]
Geometric & --- & 0.95 & 0.96\\
 &  & (0.93) & (0.95)\\[2pt]
\midrule
Difference & --- & +0.00 & +0.00\\
 &  & (0.72)\textsuperscript{n.s.} & (0.41)\textsuperscript{n.s.}\\[2pt]
\midrule
\multicolumn{4}{@{}l}{\emph{B. Mean absolute daily P\&L per option, cents}}\\
Arithmetic & 25.38 & 4.29 & 3.55\\[2pt]
Geometric & 25.38 & 4.48 & 3.68\\[2pt]
\midrule
Difference & --- & +0.19 & +0.13\\
 &  & (2.26)\textsuperscript{\dag} & (2.94)\\[2pt]
\bottomrule
\end{tabular}
\caption[Hedging with the arithmetic and geometric models]{\textbf{Hedging with the arithmetic and geometric models}\\ The portfolio of every vanilla option in the comparison, hedged over one business day with $k$ hedge options (M4, then M4 and Q2) and each option's own future, over 2{,}809 dates, for the two-factor arithmetic model and its geometric counterpart. Neither model is estimated with the skew forcing. (A)~Hedging effectiveness per unit of portfolio vega, with the version based on the median absolute deviation in parentheses. (B)~Mean absolute daily P\&L per option, in cents per barrel. Each change from the column to the left is tested with the Newey--West $t$-statistic (50 lags), and the last row of each panel is the difference, geometric less arithmetic, with its $t$-statistic in parentheses. Unless marked, every change is significant at the 1 percent level; \textsuperscript{\dag} marks significance at the 5 percent level only and \textsuperscript{n.s.} no significance at the 5 percent level.}
\label{tab:app_geom_hedge}
\end{table}

%% file: conclusion.tex
\section{Conclusion}
\label{sec:conclusion}

This paper develops an arithmetic multifactor stochastic volatility framework for the commodity term structure. Futures are priced in closed form, and vanilla options and basket options, in particular calendar spread options, in semi-closed form, while comparable geometric models rely on approximations for the latter. This tractability allows us to include calendar spread options directly in the estimation, alongside futures and vanilla options. Furthermore, the model also has a Markov representation with fewer state variables than its geometric counterpart.

We estimate the model on an extensive panel of 342{,}637 daily settlement prices of WTI crude oil futures, vanilla options and calendar spread options from 2013 to 2024. We find that three factors are required to reproduce both the level and slope of the volatility term structure, and that hedges built on these factors remove almost all of the daily risk of a portfolio of vanilla options. Compared with its geometric counterpart, estimated with two factors in the same way, the arithmetic model fits the prices of futures and vanilla options better and hedges the vanilla options slightly better.

A key result is that calendar spread options carry information on the dependence structure of futures prices that neither futures nor vanilla options provide. Estimated on futures and vanilla options alone, the model overprices calendar spread options beyond the front month, because it overstates the decorrelation between adjacent futures contracts, the curve decorrelation. Estimated on calendar spread options too, the model prices them well and follows the curve decorrelation over time. This hinges on how the volatility of each contract is divided between factors whose loadings change from one contract to the next and factors whose loadings do not.
Including calendar spread options reorganizes the factors so that one of them carries the curve decorrelation. The hedges of calendar spread options improve, at only a small loss of accuracy in the pricing and hedging of vanilla options.

The model can be extended in several directions. For instance, stochastic interest rates correlated with futures prices would make it better suited to longer-dated options. Furthermore, with leverage correlations that are constant over time, the factors generate only part of the time variation in the skew of implied volatilities. When they must also account for the skew, they fit the slope of the volatility term structure less well, and a third hedge option no longer improves the hedges of the vanilla options. This, along with the model-free analysis in Section~\ref{sec:data}, suggests the model could be improved by extending it to include a stochastic leverage component, separate from the factors driving the volatility term structure. We leave this possible extension for future work.


Finally, the empirical analysis could also be extended to other basket options, such as cross-commodity spread options, average price options and options on futures with delivery periods. The arithmetic framework is well suited to such an extension, since these options can also be priced in semi-closed form.

%% file: appendix_estimation.tex
\section{Estimation details}%
\label{sec:appendix_estimation}

\subsection{State space representation}%
Time is measured in years of 252 business days. Under the physical measure, the state variables follow
\begin{align*}
	\mathrm{d}x(t) &= \left[ - K x(t) + \Diag(\lambda_x) v(t) \right] \mathrm{d}t
	+ D_v(v(t)) \mathrm{d}W^\mathbb{P}(t),\\
	\mathrm{d}z(t) &= \left[ x(t) - K z(t) \right]\mathrm{d}t,\\
	\mathrm{d}v(t) &= \left[ \vartheta + \Gamma^\mathbb{P} v(t) \right]\mathrm{d}t
	+ \Diag(\sigma_v) D_v(v(t)) \mathrm{d}Z^\mathbb{P}(t),
\end{align*}
where $\vartheta = -\Gamma \mathbf{1}$, since the stationary mean of every volatility factor under
$\mathbb{Q}$ is one, and $\mu = -(\Gamma^\mathbb{P})^{-1}\vartheta$ is the stationary mean under
$\mathbb{P}$. The combined system $X(t) \triangleq (x(t), z(t), v(t))^\top$ has the affine form
\begin{equation}
	\label{eq:big_sde}
	\mathrm{d}X(t) = \left( K_0 + K_1 X(t) \right)\mathrm{d}t
	+ \Sigma_X S(X(t)) \mathrm{d} B^\mathbb{P}(t),
\end{equation}
where $B^\mathbb{P}(t) \triangleq \left( W^\mathbb{P}(t), Z^\mathbb{P}(t) \right)^\top$ is a
$2n$-dimensional Brownian motion with correlation matrix
\begin{equation*}
	R =
	\begin{bmatrix}
		I_{n\times n} & \Diag(\rho)\\
		\Diag(\rho) & I_{n\times n}
	\end{bmatrix},
\end{equation*}
and
\begin{gather*}
	K_0 =
	\begin{bmatrix}
		0_{n\times 1} \\
		0_{n\times 1} \\
		\vartheta
	\end{bmatrix},
	\quad
	K_1 =
	\begin{bmatrix}
		-K & 0_{n\times n} & \Diag(\lambda_x) \\
		I_{n\times n} & - K & 0_{n\times n} \\
		0_{n\times n} & 0_{n\times n} & \Gamma^\mathbb{P}
	\end{bmatrix},
	\\[4pt]
	\Sigma_X =
	\begin{bmatrix}
		I_{n\times n} & 0_{n\times n} \\
		0_{n\times n} & 0_{n\times n} \\
		0_{n\times n} & \Diag(\sigma_v)
	\end{bmatrix},
	\quad
	S(X(t)) =
	\begin{bmatrix}
		D_v(v(t)) & 0_{n\times n} \\
		0_{n\times n}  & D_v(v(t))
	\end{bmatrix}.
\end{gather*}
Applying Itô's lemma to $e^{-K_1 t}X(t)$ gives the $\mathcal{F}_t$-conditional mean
\begin{equation}
	\label{eq:cond_mean}
	\mathbb{E}^\mathbb{P}_t\left\{ X(T) \right\}
	= e^{K_1 (T-t)} X(t) + \left( e^{K_1 (T-t)} - I \right) K_1^{-1} K_0,
\end{equation}
and the $\mathcal{F}_t$-conditional covariance
\begin{equation*}
	\mathbb{V}^\mathbb{P}_t\left\{ X(T) \right\}
	= \int_t^T e^{K_1 (T-u)} \Sigma_X
	S(\mathbb{E}^\mathbb{P}_t\{X(u)\}) R\, S(\mathbb{E}^\mathbb{P}_t\{X(u)\})
	\Sigma_X^\top e^{K_1^\top (T-u)} \mathrm{d}u .
\end{equation*}
Since $R$ correlates each $W_i$ only with $Z_i$, $S(X) R\, S(X)$ is linear in $v$, so the conditional
covariance is affine in $v(t)$,
\begin{equation}
	\label{eq:cond_cov}
	\mathbb{V}^\mathbb{P}_t\left\{ X(T) \right\}
	= \Omega_0 + \sum^{n}_{i=1} \Omega_i v_i(t),
\end{equation}
with $\Omega_0, \dots, \Omega_n$ depending on $T-t$ alone, which we evaluate by Gauss--Legendre
quadrature. With $\Delta$ one business day, the transition of Section~\ref{sub:estimation} has
$A = e^{K_1 \Delta}$, $b = ( e^{K_1 \Delta} - I ) K_1^{-1} K_0$, and the covariance of $\omega_l$ is
\eqref{eq:cond_cov} with $T-t=\Delta$.

\subsection{Measurements and measurement noise}%
\label{sub:measurements}
\autoref{tab:measurements} gives the measurement, its scaling constant and the standard deviation of its
noise for each instrument. The observed and the model-implied prices are divided by the same constants,
which are computed from observed prices and do not depend on the parameters. The covariance matrix
$R_\epsilon$ of the measurement noise is diagonal.
\begin{table}[!htb]
	\centering
	\footnotesize
	\setlength{\tabcolsep}{4pt}
	\begin{tabular}{lllll}
		\toprule
		Instrument & Measurement & Scaling constant & Noise & Parameters \\
		\midrule
		Futures & $F_j / c^F_l$ & typical daily move & $\sigma_F$ & $\sigma_F$ \\
		Vanilla options & $P_j / c_j$ & vega $\times$ implied volatility & $\sigma_V \sqrt{\bar c_D / c_j}\,(1 + |a_V|\,|s_j|)$ & $\sigma_V$, $a_V$ \\
		Calendar spread options & $P_j / c_j$ & vega $\times$ implied volatility & $\sigma_{CS}$ & $\sigma_{CS}$ \\
		\bottomrule
	\end{tabular}
	\caption[Measurements and measurement noise]{\textbf{Measurements and measurement noise}\\
	The noise is given by its standard deviation. $F_j$ is the settlement price of futures contract $j$
	on date $t_l$ and $c^F_l$ the typical daily price move of the date. $P_j$ is the undiscounted European
	price of option $j$, $c_j$ its implied Bachelier vega times its implied Bachelier volatility,
	$\bar c_D$ the median of $c_j$ in its delta bucket $D$ over the sample, and $s_j = D/4$.}
	\label{tab:measurements}
\end{table}

The typical daily price move $c^F_l$ is the average of the observed implied Bachelier volatilities of the
at-the-money vanilla measurements of date $t_l$, times $\sqrt{1/252}$, so a scaled futures price is a
number of typical daily moves. The scaling constant $c_j$ of an option is the product of its discounted
implied Bachelier vega and its implied Bachelier volatility, both computed from its observed price,
de-americanized for a vanilla option and with the spread between its two legs as the underlying for a
calendar spread option. A change $\mathrm d \sigma$ in the implied volatility $\sigma$ moves $P_j/c_j$ by
$\mathrm d \sigma / \sigma$ divided by the discount factor, so the measurement error of an option is, to
first order, its relative error in implied volatility.

The noise of a vanilla option falls with $c_j$, and $\sigma_V$ is the noise of a typical at-the-money
option of its bucket. The factor $(1 + |a_V|\,|s_j|)$ lets the noise grow towards the wings, to
$(1+|a_V|)\,\sigma_V$ for the far out-of-the-money options. The noise of the futures, in typical daily
moves, and of the calendar spread options, in relative implied volatility, is constant.

The skewness $u_l$ of the skew forcing is the average of $\mathcal M_3$ over the contracts M1--M7,
carried forward from the previous date on the few dates on which one of them is missing, and $\bar u$ is
its average over the dates on which all seven are quoted. In the filter, the forcing
$\chi_V (u_l - \bar u) s_l$ is added to the predicted measurements, which is equivalent to subtracting it
from $y_l$. The floor in the option prices is $\underline v = 0.001$.

\subsection{The filter}%
\label{sub:filter_details}
The filter is an unscented Kalman filter \citep{Julier2004,Wan2004} adapted to two features of the model.
The futures prices are affine in the curve state variables and enter the update exactly, so only the
option prices, which depend on the state through the volatility factors alone, need the unscented
transform. The volatility factors are non-negative, while the Gaussian distributions of the filter are
not. The option prices, which are not defined for negative $v$, are therefore evaluated at
$w = \max(v, \underline v)$, with the moments of $w$ computed under the gamma distribution with the
predicted mean and variance of $v$. After each update, the distribution of each $v_i$ is recombined with
the predicted gamma distribution, since truncating the Gaussian update at zero would raise the filtered
mean of a factor near zero.

The filter carries the mean $m_{l|l}$ and covariance $P_{l|l}$ of the state given the measurements up to
$t_l$. For $n \geq 2$, $\beta_{1,n}=0$ leaves $z_n$ out of every measurement, so the filter carries the
remaining $3n-1$ state variables. The first date only initializes the filter. The start of $(x,z)$ is the
ridge regression of the first date's futures curve on its loadings, with the stationary covariance of
$(x_i, \kappa_i z_i)$ at $v_i = 3$ as prior, and a covariance in which a direction the curve barely
determines starts loose. The volatility factors start at the maximum a posteriori estimate given the first
date's squared model-free volatilities, which are affine in $v$, with measurement errors of 5 percent and
independent exponential priors with mean one, the stationary mean under $\mathbb{Q}$.

On each later date $t_{l+1}$, the filter computes the following from $(m_{l|l}, P_{l|l})$. Below,
$m^-_i$ and $p^-_i$ are the predicted mean and variance of $v_i$, $P^-_{vv}$ the predicted covariance of
$v$, and $P^-_{\cdot v}$ the columns of $P_{l+1|l}$ that belong to $v$.
\begin{enumerate}
	\item Prediction:
	\begin{equation*}
		m_{l+1|l} = A m_{l|l} + b,
		\qquad
		P_{l+1|l} = A P_{l|l} A^\top + \Omega_0 + \sum^{n}_{i=1}\Omega_i v_{i,l|l},
	\end{equation*}
	with $v_{i,l|l} \geq 0$ the filtered mean of $v_i$.

	\item Moments of the floored volatility factors. For each $i$, with $a_i = (m^-_i)^2/p^-_i$,
	$r_i = m^-_i/p^-_i$, $T_{i,k}$ the probability that a gamma variable of shape $a_i+k$ and rate $r_i$
	falls below $\underline v$, and $M_i = (m^-_i)^2 + p^-_i$,
	\begin{align*}
		\bar w_i &= \underline v\, T_{i,0} + m^-_i (1 - T_{i,1}),
		\qquad
		s^2_{w,i} = \underline v^2\, T_{i,0} + M_i (1 - T_{i,2}) - \bar w_i^2,\\
		\Lambda_i &= \frac{\underline v\, m^-_i T_{i,1} + M_i (1 - T_{i,2}) - m^-_i \bar w_i}{p^-_i},
	\end{align*}
	the mean and variance of $w_i$ and its regression coefficient on $v_i$. The covariance $P^w$ of $w$ is
	$\Diag(\Lambda) P^-_{vv} \Diag(\Lambda)$ with its diagonal replaced by $s^2_{w,i}$.

	\item Unscented transform of the option prices. With $\lambda_U = \alpha_U^2 n - n$, $\alpha_U = 0.1$,
	$\beta_U = 2$ and $\left[\sqrt{P^w}\right]_i$ the $i$th column of the Cholesky factor of $P^w$,
	\begin{align*}
		\mathcal{W}^{(0)} &= \bar w,
		\qquad
		\mathcal{W}^{(\pm i)} = \bar w \pm \sqrt{n + \lambda_U} \left[ \sqrt{P^w} \right]_i,
		\qquad
		\mathcal Y^{(j)} = h^O(\mathcal W^{(j)}),\\
		m^O &= \sum_j W^{(m)}_j \mathcal{Y}^{(j)},
		\qquad
		P^O = \sum_j W^{(c)}_j \left( \mathcal{Y}^{(j)} - m^O \right)
		\left( \mathcal{Y}^{(j)} - m^O \right)^\top,\\
		P^{Ow} &= \sum_j W^{(c)}_j \left( \mathcal{Y}^{(j)} - m^O \right)
		\left( \mathcal{W}^{(j)} - \bar w \right)^\top,
		\qquad
		G = P^{Ow} (P^w)^{-1} \Diag(\Lambda),
	\end{align*}
	where $h^O$ gives the scaled option prices at the floored volatility factors, and
	$W^{(m)}_0 = \lambda_U/(n+\lambda_U)$, $W^{(c)}_0 = W^{(m)}_0 + 1 - \alpha_U^2 + \beta_U$ and
	$W^{(m)}_{\pm i} = W^{(c)}_{\pm i} = 1/(2(n+\lambda_U))$. $G$ is the response of the option prices to
	$v$.

	\item Kalman update. With $H^F X + d^F$ the scaled futures prices, affine in the state,
	\begin{align*}
		m^\mathcal{Y} &=
		\begin{bmatrix}
			H^F m_{l+1|l} + d^F\\
			m^O + \chi_V (u_{l+1} - \bar u) s_{l+1}
		\end{bmatrix},
		\qquad
		P^{\mathcal{X} \mathcal{Y}} =
		\begin{bmatrix}
			P_{l+1|l} (H^F)^\top &
			P^-_{\cdot v} G^\top
		\end{bmatrix},\\
		P^\mathcal{Y}_{l+1} &=
		\begin{bmatrix}
			H^F P_{l+1|l} (H^F)^\top &
			H^F P^-_{\cdot v} G^\top\\
			G (P^-_{\cdot v})^\top (H^F)^\top &
			P^O
		\end{bmatrix}
		+ R_\epsilon,
		\qquad
		\nu_{l+1} = y_{l+1} - m^\mathcal{Y},\\
		K &= P^{\mathcal{X} \mathcal{Y}} \left( P^\mathcal{Y}_{l+1} \right)^{-1},
		\qquad
		m^+ = m_{l+1|l} + K \nu_{l+1},
		\qquad
		P^+ = P_{l+1|l} - K P^\mathcal{Y}_{l+1} K^\top .
	\end{align*}

	\item Gamma posterior. For each $i$, with $m^+_i$ and $p^+_i$ the updated mean and variance of $v_i$,
	\begin{equation*}
		\pi_i(v) \propto v^{a_i - 1}
		\exp\left( b_i v - \tfrac12 \varpi_i v^2 \right),
		\quad v \geq 0,
		\qquad
		\varpi_i = \frac{1}{p^+_i} - \frac{1}{p^-_i},
		\qquad
		b_i = \frac{m^+_i}{p^+_i} - \frac{2 m^-_i}{p^-_i},
	\end{equation*}
	the predicted gamma distribution times the Gaussian likelihood in $v_i$ implied by the update. Its mean
	$\mu^\pi_i$ and variance $s^{2}_{\pi,i}$ are computed by tanh-sinh quadrature \citep{Takahasi1974} on
	either side of its mode, with the singularity at zero subtracted analytically when $a_i < 1$. Then
	\begin{gather*}
		m_{l+1|l+1} = m^+ + S_v \left( \mu^\pi - m^+_v \right),
		\qquad
		P_{l+1|l+1} = D P^+ D,\\
		D = I + S_v \left( \Diag\!\left( s_{\pi,i}/\sqrt{p^+_i} \right) - I \right) S_v^\top,
	\end{gather*}
	where $S_v$ selects the rows of $v$ in the state and $m^+_v$ is the updated mean of $v$.
\end{enumerate}

The quasi-log-likelihood follows from the prediction error decomposition,
\begin{equation*}
	\log \mathcal{L}
	= - \frac{1}{2} \sum^{L}_{l=1}
	\left\{ N_l \log(2\pi) + \log \left| P^\mathcal{Y}_{l} \right|
	+ \nu_{l}^\top \left( P^\mathcal{Y}_{l} \right)^{-1} \nu_{l} \right\},
\end{equation*}
with $N_l$ the number of measurements on $t_l$. \cite{Sarkka2013} treats the unscented Kalman filter and
\cite{Harvey1991} Kalman filtering in general.

\subsection{Numerical evaluation of the option prices}\label{sub:numerics}
The prices of Proposition~\ref{prop:optionprice} require, for every option on every date, the integral
over the frequency $\omega$ in \eqref{eq:put_option_price}, whose integrand requires the solutions of the
Riccati system \eqref{eq:tfA_ode}--\eqref{eq:etfB_ode}. These depend only on the maturity and the
frequency, so we solve the system once for each parameter vector at every maturity of the panel and every
node of the integral, and evaluate every option, date and sigma point of the filter from the stored
solutions. The loading $\varphi_Y$ of a vanilla option is that of the futures contract that expires three
business days after it, so one table serves every vanilla option. That of a calendar spread option depends
on the gap between the expiries of its two legs, and we solve one table per gap.

The integral is approximated by the double-exponential rule of \citet{Takahasi1974}. The substitution
$\omega=e^{2\sinh s}$ maps $(0,\infty)$ onto the real line, and the trapezoidal rule with step $h$ in $s$
gives the nodes $\omega_k = e^{2\sinh(kh)}$ and weights $w_k = 2h\cosh(kh)\,\omega_k$ for $k=-N,\dots,N$.
We take $N=50$ and the step $h=W(2\pi N)/N$ of \citet{AndersenLake2019}, where $W$ is the Lambert
function, and keep the nodes between $10^{-14}$ and $100$. Increasing $N$ to 400 does not change the
prices of the options in the panel. For each node, the real and imaginary parts of the Riccati system are
integrated with the Runge--Kutta pair of order 5(4) of \citet{Tsitouras2011}, with relative and absolute
tolerances of $10^{-7}$.

\subsection{Data construction}%
\label{sub:data_construction}
The vanilla options enter the panel at two levels. The measurements are chosen among the liquid quotes, with an
open interest above 100 contracts and a settlement above \$0.10, while the arbitrage screen and the
model-free moments use every quote.

On each date, contract and option type, the raw settlements must admit a convex and monotone
curve with bounded slope that passes within half a tick of each of them. Where a ladder does not, we remove
the smallest number of quotes that makes it admissible, found exactly as the solution of a mixed-integer
linear program. The screen runs on the raw settlements, where the half-tick bound is exact, since the
conditions hold for American options as well.

In the Bachelier model, the price of an American option on a futures contract can be written as
$V = s\,W(y;\varrho)$, with $s=\sigma\sqrt{\tau}$, $y=\pm(F-K)/s$ for a call and a put, and
$\varrho = r\tau$, where $W$ solves
\begin{equation*}
	\partial_u W = \tfrac12 \partial_{yy} W - \varrho W, \qquad
	W(y,0) = y^+, \qquad W \geq y^+ ,
\end{equation*}
over the fraction $u$ of the option's life remaining. We solve the European and the American problems on
the same grid by Crank--Nicolson with the early-exercise constraint imposed exactly, and find each
option's implied volatility by inverting the closed-form European price plus the early-exercise premium, their
difference. Against a finite-difference reference solution, the error of this price is at most
$6\cdot 10^{-5}$ dollars.

The risk-neutral density of the futures price at expiry is fitted for each contract and date in the
standardized strike $z = (K-F)/\xi$, with $\xi=\sigma_{ATM}\sqrt{\tau}$. It is a non-negative combination
of cubic B-splines inside the quoted strikes and of exponential tails beyond them, fitted to the
out-of-the-money option prices by least squares with a roughness penalty. Beyond $|z|=2$, each tail of the fitted density is replaced by an exponential tail with the same mass $D$
and the same first moment $M$ about $|z|=2$, so that $b=M/D$ is its mean distance beyond that point. The
second and third moments are then the integrals \eqref{eq:moments} over strikes with $|z|\le2$ plus, from
the upper tail, $2Mb$ and $6M(2b+b^2)$, and from the lower tail the same with the opposite sign in the third. A fit is discarded when the root mean square of its pricing
errors exceeds $5\max(1, \xi/\bar \xi)$ times the standard deviation of rounding to the cent,
$0.01/\sqrt{12}$, with $\bar \xi$ the median of $\xi$ over the sample.

\subsection{Dimension of the model-free moments}%
\label{sub:dimension}
Each moment is interpolated date by date to constant maturities between the two contracts that bracket
the maturity: monthly from two to twelve months for $\mathcal M_2^2$ alone, and from two to five months, the range of the calendar spread options, wherever $\mathcal M_3$ or $1-\rho$
enters. We replace the changes of each series by their ranks, so that a few days in April 2020 do not
dominate, and standardize them. The explained share of a group of series is $1-\mathrm{RSS}/\mathrm{TSS}$
of the group regressed on a set of regressors. The level and slope are the first two principal components
of $\mathcal M_2^2$, and the partial correlation is that of the first principal components of
$\mathcal M_3$ and of $1-\rho$ after both are regressed on the level and slope. The confidence intervals
are from a moving-block bootstrap.

Noise in the quotes adds the same variance to a change over any interval, while the persistent variation
of the moments accumulates with the interval, so we repeat the analysis on non-overlapping weekly and
monthly changes. In \autoref{tab:dimension}, the share of the curvature component of $\mathcal M_2^2$ falls as the interval lengthens, so part of it is noise. The share of the skew explained by the level and slope of
$\mathcal M_2^2$ rises to 19 percent on monthly changes, while the partial correlation of the skew and
the decorrelation falls to zero, so neither result is an artifact of noise in the daily changes, and neither changes when 2020 is excluded. The last two rows of \autoref{tab:dimension} replace $\mathcal M_3$ and
$1-\rho$ by $\varkappa_3/\tau$ and $s_{CS}^2$, which, like $\mathcal M_2^2 = \varkappa_2/\tau$, are affine
in the volatility state variables. In these coordinates, the level and slope of $\mathcal M_2^2$ explain at
most 28 percent of the skew, and the partial correlation of the skew and the decorrelation is at most
0.07.
\begin{table}[!htbp]
\centering\small
\setlength{\aboverulesep}{0pt}\setlength{\belowrulesep}{0pt}%
\renewcommand{\arraystretch}{1.15}%
\begin{tabular}{@{}lccc@{}}
\toprule
 & Daily & Weekly & Monthly\\
\midrule
$\mathcal M_2^2$: level and slope & 0.96 & 0.98 & 0.98\\
$\mathcal M_2^2$: curvature & 0.022 & 0.013 & 0.011\\
$\mathcal M_3$: level and slope of $\mathcal M_2^2$ & 0.10 [0.07, 0.13] & 0.13 [0.07, 0.18] & 0.19 [0.07, 0.33]\\
$1-\rho$: level and slope of $\mathcal M_2^2$ & 0.17 & 0.12 & 0.04\\
$1-\rho$: added by $\mathcal M_3$ & 0.005 & 0.005 & 0.007\\
Partial correlation of $\mathcal M_3$ and $1-\rho$ & 0.07 [0.02, 0.12] & 0.00 [$-$0.09, 0.11] & 0.00 [$-$0.18, 0.18]\\
\midrule
$\varkappa_3/\tau$: level and slope of $\mathcal M_2^2$ & 0.16 & 0.17 & 0.28\\
Partial correlation of $\varkappa_3/\tau$ and $s_{CS}^2$ & 0.07 & 0.05 & $-$0.02\\
\bottomrule
\end{tabular}
\caption[Explained shares of the changes in the model-free moments]{\textbf{Explained shares of the changes in the model-free moments}\\ Explained shares $1-\mathrm{RSS}/\mathrm{TSS}$ and partial correlations for the ranks of daily changes and of non-overlapping weekly and monthly changes, with 95 percent moving-block bootstrap intervals in brackets. The last two rows use $\varkappa_3/\tau$ and $s_{CS}^2$ in place of $\mathcal M_3$ and $1-\rho$.}
\label{tab:dimension}
\end{table}

\subsection{Constrained estimation}%
\label{sub:constrained_estimation}
Beyond fitting the prices, we require the estimate to be consistent, with filtered paths that its own
dynamics under $\mathbb{P}$ could produce and dynamics that match the data. We impose this through the requirements below and take as our estimate the parameters that maximize
the quasi-likelihood among those that meet them. Each test among them is at a level that a correctly
specified model passes with a probability of at least 99 percent. $\mu$ and $\Sigma^\infty$ denote the
stationary mean and covariance of $v$ under $\mathbb{P}$.
\begin{itemize}
	\item The non-overlapping 21-day prediction errors of the filtered $x$ and $v$, standardized by the model's
	conditional covariance, show no mean or first-order autocorrelation, tested against the 99.99 percent
	quantile of the $\chi^2(4n)$ distribution, so that only grossly wrong dynamics are rejected.
	\item The mean square of each standardized prediction-error series is at most 5.
	\item The sample mean of each filtered volatility factor lies between $\mu_i/1.25$ and $2\mu_i$. A
	path far from its stationary mean would describe a different market under $\mathbb{P}$ than the one
	filtered from the prices. The band is wider above, since a rare excursion such as that of 2022 need
	not set the stationary mean.
	\item The model's futures variance at the filtered state matches the realized futures variance in
	its level and in two slopes across the curve: from the contracts that expire within two months to
	those that expire later in the first year, and from these to the contracts that expire after a year.
	Each is tested at the 1 percent level, with equal-weighted cosine standard errors with 8 degrees of
	freedom \citep{Mueller2007,Lazarus2018}. The two-factor models test a single slope, from
	the front of the curve to the contracts that expire after a year.
	\item The shape $\mu_i^2/\Sigma^\infty_{ii}$ of each volatility factor, the shape parameter of the
	gamma distribution with its stationary mean and variance, is at least 0.05 under both measures,
	which rules out only a factor concentrated at zero. The level factor, which carries the overall level
	of volatility, has a shape of at least one, so that it never reaches zero.
	\item The one-day predicted shape of each volatility factor in the filter has a geometric mean of at
	least 0.5 over the sample.
	\item The relaxation time $\Sigma^\infty_{ii}/(\sigma_{v,i}^2\mu_i)$ of each volatility factor is at
	most 1.25 years.
	\item The annualized maximal Sharpe ratio implied by the measure change is at most 2, both at the
	stationary mean and at the sample mean of the filtered states.
\end{itemize}

Besides these constraints, the spillover from slower factors may supply at most 75 percent of each factor's
mean under $\mathbb{Q}$, so that each factor keeps a drift of its own, and the other parameters are kept
within wide bounds. The constrained problem is solved by the augmented Lagrangian method
\citep{Hestenes1969,Powell1969,Rockafellar1973}, and the estimate is the point at which a subsequent
simplex search \citep{Nelder1965} finds no further improvement. The standard errors are sandwich standard
errors restricted to the directions that leave the active constraints unchanged.

\subsection{The geometric comparison model}%
\label{sub:geometric}
The geometric model of Section~\ref{sec:res_geom} applies the loadings and volatility factors of the
arithmetic model to $\mathrm dF/F$ instead of $\mathrm dF$, as in \cite{chiarella2013}. Its log futures
price is affine in five curve state variables per factor, $x_i$ and $z_i$ as in the arithmetic model and
three convexity terms of finite variation driven by $v_i$. The filter measures $\ln F$, with noise of
constant standard deviation, and prices the vanilla options from the same Riccati system with a different
source term, integrated in log-price frequency with $N=150$. Calendar spread options cannot be priced
this way, since a spread of two log-normal prices has no such transform.

The two models of the comparison are estimated on the vanilla options that both can price, each measured
as its price over its Black vega at the observed Black volatility, with noise
$\sigma_V(1+|a_V|\,|s_j|)$. Neither uses the skew forcing. They meet the requirements of
Appendix~\ref{sub:constrained_estimation} except the comparison with the realized futures variance, which
is in dollars for one model and in log returns for the other.

%% file: appendix_simulation.tex
\section{Simulation study}\label{app:sim}

We verify the filter and the estimation on panels simulated from a known model.
We ask whether the filter infers the state variables at the true parameters, and
whether the estimation recovers the parameters.

\subsection{Synthetic panels}\label{app:sim_panels}

The true model is V3 at its estimate, with $\chi_V=0$. Each panel is
one path of the model under $\mathbb P$, simulated with four steps a day from one
year before the first date, starting from a flat futures curve at $F_0$ and the
volatility state variables at their stationary mean. The volatility factor without spillover from slower factors is simulated exactly, and the others to second order in
the step. The futures and vanilla options are priced on the
observed panel's dates, contracts and maturities, with each option's strike set
so that it lies in its delta bucket, and the measurement noise is drawn at V3's
estimated levels.

\subsection{Filtering at the true parameters}\label{app:sim_filter}

\begin{figure}[!htbp]
\centering
\includegraphics[width=\textwidth]{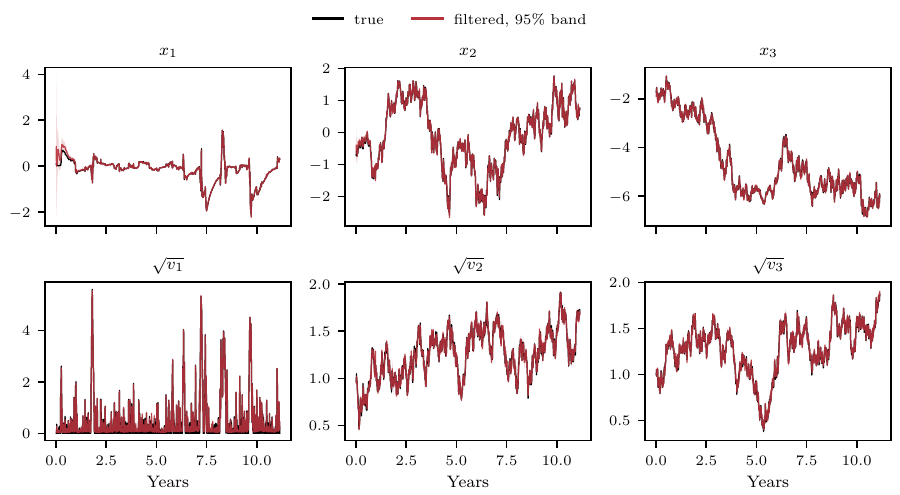}
\caption[True and filtered state variables]{\textbf{True and filtered state variables}\\
One simulated panel, filtered at the true parameters: the true value and the filtered mean, with the
filtered 95 percent band, of the curve state variables (above) and, as square roots, of the volatility state
variables (below).}
\label{fig:sim_states}
\end{figure}

We run the filter at the true parameters and compare the filtered state
variables with the simulated ones. \autoref{tab:sim_filter} shows that they are
essentially unbiased. For every state variable, the median of its mean error over the panels is below one percent of its
standard deviation, and the mean error is below seven percent on every panel. The state variable $x_1$ is inferred least
precisely, with a root mean squared error of 15 percent of its standard deviation.
\autoref{fig:sim_states} shows the true and filtered state variables on one panel.

\begin{table}[!htbp]
\centering\small
\setlength{\aboverulesep}{0pt}\setlength{\belowrulesep}{0pt}%
\renewcommand{\arraystretch}{1.15}%
\begin{tabular}{@{}lrcr@{}}
\toprule
 & \multicolumn{2}{c}{bias} & RMSE\\
state & median & range & \\
\midrule
$x_1$ & +0.005 & [$-$0.049, +0.066] & 0.153\\
$x_2$ & $-$0.000 & [$-$0.010, +0.007] & 0.035\\
$x_3$ & +0.000 & [$-$0.001, +0.008] & 0.024\\
$v_1$ & $-$0.000 & [$-$0.002, +0.001] & 0.026\\
$v_2$ & +0.000 & [$-$0.006, +0.004] & 0.069\\
$v_3$ & +0.000 & [$-$0.003, +0.003] & 0.037\\
\bottomrule
\end{tabular}
\caption[Filtering at the true parameters]{\textbf{Filtering at the true parameters}\\ The filtered state variables against the true ones on 20 simulated panels, relative to each state variable's standard deviation: the mean error, true less filtered, its median and range over the panels, and the median root mean squared error.}
\label{tab:sim_filter}
\end{table}

\subsection{Recovering the parameters}\label{app:sim_recovery}

\begin{figure}[!htbp]
\centering
\includegraphics[width=\textwidth,height=0.78\textheight,keepaspectratio]{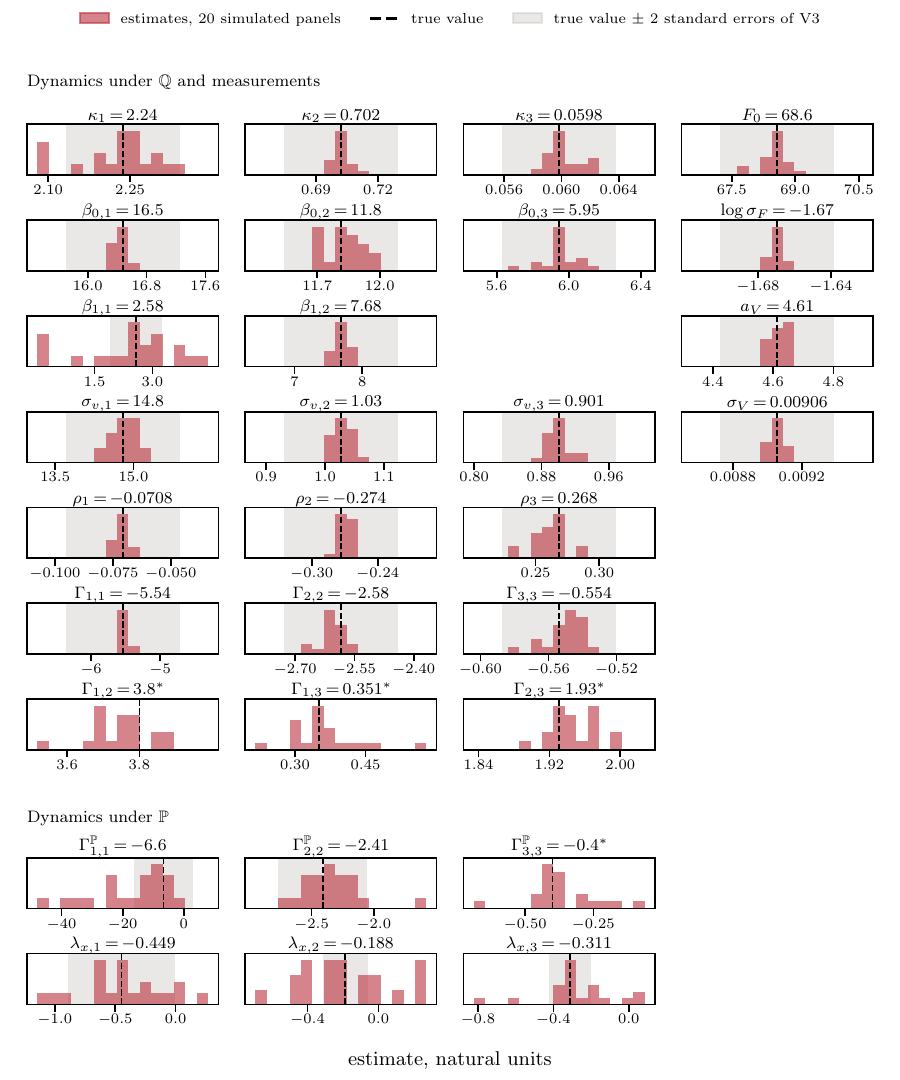}
\caption[Recovery of the parameters]{\textbf{Recovery of the parameters}\\
Histograms of the unconstrained estimates on 20 simulated panels. The dashed line is the true value,
and the grey band spans two of V3's sandwich standard errors on either side of it. Each title gives
the parameter's true value. Above, the parameters of the dynamics under $\mathbb Q$ and of the
measurements; below, those of the dynamics under $\mathbb P$. $^*$: V3's standard error is zero,
since an active constraint holds the parameter, so no band is shown, and the text measures the errors against 5 percent of the true value.}
\label{fig:sim_recovery}
\end{figure}

We estimate the model on each of 20 panels by maximizing the quasi-likelihood
without the constraints of Appendix~\ref{sub:constrained_estimation}, starting
from the true parameters. \autoref{fig:sim_recovery} shows the estimates against the true values and V3's standard errors.

The parameters of the dynamics under $\mathbb Q$ and of the measurements are
recovered. For each of them, the median error is within 0.3 standard errors, and the robust spread
of the errors is at most 1.2 standard errors with no panel beyond three, except for $\beta_{1,1}$
and $\Gamma_{1,3}$, whose estimates scatter more widely around a median close to the true value.
The parameters of the dynamics under $\mathbb P$ are recovered less precisely.
The median estimate of $\Gamma^{\mathbb P}_{1,1}$ is $-12.2$ against a true value of $-6.60$, and
the market prices of risk $\lambda_{x,i}$ are centered on their true values but widely spread. The
option prices thus determine the dynamics under $\mathbb Q$, while the dynamics under $\mathbb P$
rest on the drift of the state variables over the sample alone.

%% file: appendix_proofs.tex
\section{Proofs}
\label{sec:appendix_proofs}

\subsection{Measure change}
\label{sec:Measurechanges}
The completely affine measure change of \cite{Duffee2002} can be used in the context of our model in the following manner. Let 
$$B^\mathbb{Q}(t)\triangleq \begin{bmatrix}W^\mathbb{Q}(t)\\Z^{\mathbb{Q}}(t)
\end{bmatrix}$$
and $R_{i,j}\mathrm{d}t=\mathrm{d}\langle 
	B_{i}^\mathbb{Q},
	B_{j}^\mathbb{Q} 
	\rangle(t),\, i,j=1,\dots 2n.$
It follows from the definition of $W^\mathbb{Q}$ and $Z^{\mathbb{Q}}$  that 
\begin{align}
    R=\begin{bmatrix} I_{n\times n} & \Diag(\rho_1,\dots,\rho_n)\\ \Diag(\rho_1,\dots,\rho_n) & I_{n\times n}\end{bmatrix}.
\end{align}
If we define an adapted stochastic process $\phi$ living in $\mathbb{R}^{2n}$, as $\phi(t) =[D_v(v(t))\lambda_x, \,D_v(v(t))\lambda_v]^\top,\forall t\geq0$, where $\lambda_x,\lambda_v \in\mathbb{R}^n$, we can then define the physical measure $\mathbb {P}$ via the Radon--Nikodym derivative:
$$\frac{\mathcal{\mathrm{d}\mathbb{P}}}{{\mathrm{d}\mathbb{Q}}}\Big |\mathcal{F}_t\triangleq \exp\left(\int_0^t\phi(s)^\top R^{-1}\mathrm{d}B^{\mathbb{Q}}(s)-\frac{1}{2}\int_0^t\phi(s)^\top R^{-1}\phi(s)ds\right)\quad \forall t\geq 0$$ 
It follows from Girsanov's theorem that, under standard assumptions (e.g. Novikov's condition; see again \citealp{Duffee2002} for more details), $\mathbb{P}\sim \mathbb{Q}$,
$\frac{\mathcal{\mathrm{d}\mathbb{P}}}{{\mathrm{d}\mathbb{Q}}}\big |\mathcal{F}_t$ is a $\mathbb{Q}$-martingale and 
\begin{align}
B^\mathbb{P}(t)\triangleq B^\mathbb{Q}(t)-\int_0^t\phi(s) ds
\end{align}
is a (non-standard) Brownian motion under $\mathbb{P}$. If we decompose $B^\mathbb{P}(t)$ into two $n$-dimensional Brownian motions, $[W^\mathbb{P}(t), Z^\mathbb{P}(t)]^\top=B^\mathbb{P}(t)$, the result in \eqref{eq:measurechange1} and \eqref{eq:measurechange2} follows. Finally, we can apply \eqref{eq:measurechange1} and \eqref{eq:measurechange2} to \eqref{eq:Fdynamics} and \eqref{eq:cir_dynamics}, to get
 \begin{align}
\mathrm{d} F(t,T) &=\fvol^\top(T-t)\Diag(v(t))\lambda_xdt
+\fvol^\top(T-t) D_v(v(t)) \mathrm{d}
W^\mathbb{P
}(t),
\label{eq:Fdynamics P}
		\\
		\mathrm{d}v(t) 
		&= \Gamma^{\mathbb{P}}\left( v(t) - \theta \right)\mathrm{d}t
		+ \Diag(\sigma_v) D_v(v(t)) \mathrm{d}Z^\mathbb{P}(t),\label{eq:cir_dynamicsP}
	\end{align}
where $\Gamma^{\mathbb{P}}\triangleq\Gamma+\Diag(\sigma_v)\Diag(\lambda_v)$.

\subsection{Proof of Proposition 1}
\label{sub:proof1}
\begin{proof}
	Integrating the futures price dynamics \eqref{eq:fut_price_dynamics}
	and using the separability assumption yields
	\begin{align*}
		F(t,T) &=
		F(0,T) +
		\int_0^t 
		\sum^{n}_{i=1} 
		\left( \beta_{0,i} + \beta_{1,i}(T-u) \right)
		e^{-\kappa_i(T-u)} 
		\sqrt{v_i(u)}
		\mathrm{d}W_{i}^\mathbb{Q}(u)\\
		       &= F(0,T)+
		       \sum^{n}_{i=1} 
		       \left( 
		       e^{-\kappa_i(T-t)}
		       \left( \beta_{0,i}+ \beta_{1,i}(T-t) \right)
		       \right)
		       \int_0^t
		       e^{-\kappa_i(t-u)} 
		       \sqrt{v_i(u)}
		       \mathrm{d}W_{i}^\mathbb{Q}(u)\\
		       &+ 
		       \sum^{n}_{i=1} 
		       \left( 
		       e^{-\kappa_i(T-t)}
		       \beta_{1,i}
		       \right)
		       \int_0^t
		       (t-u)
		       e^{-\kappa_i(t-u)} 
		       \sqrt{v_i(u)}
		       \mathrm{d}W_{i}^\mathbb{Q}(u)
	\end{align*}
	from which we get the expressions for $x(t)$ and $z(t)$, 
	as well as $\fvol(\tau)$ and $\fvol_z(\tau)$. The dynamics can be found by
	applying the Leibniz rule.
\end{proof}

\subsection{Proof of Proposition 2 with model extension}
\label{sub:proof2}
In this section we give an example of how to extend the model to include stochastic interest rates. For completeness we prove Proposition 2 in Section~\ref{Sec:Option pricing} in this generalized setting.  We assume a simple extension in which the short rate is $r(t)=\zeta_r\cdot X^r(t)$ with $\zeta_r\in\mathbb R^{n_r}$, where
\begin{equation}\label{eq:10c}
  \dd X^r(t) = \bigl(\theta_r + \gamma_r X^r(t)\bigr)\dd t + \Sigma_r D_v(v(t))\,\dd W^\Q(t) + \sigma^r(X^r(t))\,\dd W^r(t).
\end{equation}
Here $\theta_r\in\mathbb R^{n_r}$, $\gamma_r\in\mathbb R^{n_r\times n_r}$ and $\Sigma_r\in\mathbb R^{n_r\times n}$.
$W^r$ is an $n_r$-dimensional Brownian motion independent of $(W^\Q,Z^\Q)$, and
$\sigma^r:\mathbb R^{n_r}\to\mathbb R^{n_r\times n_r}$ satisfies
\begin{equation*}
  \sigma^r(x)\sigma^r(x)^\top = H_0 + \sum_{k=1}^{n_r} x_k H_k
\end{equation*}
for symmetric matrices $H_0,H_1,\dots,H_{n_r}\in\mathbb R^{n_r\times n_r}$.
When $n_r=n$, the specification $\Sigma_r=\Diag(\sigma_r)$ gives
$\Diag(\sigma_r)D_v(v(t))\,\dd W^\Q(t)$. 
 
The dynamics of $Y$ and $v$ are given as in \eqref{eq:prop2Y} and \eqref{eq:prop2v}. We can easily see that  $X(t)\triangleq(Y(t),v(t),X^r(t))$ defines an affine process with affine drift and
the instantaneous covariations 
\begin{align*}
  \dd\langle X^r\rangle_t &= \Bigl(\Sigma_r\Diag(v(t))\Sigma_r^\top + H_0 + \textstyle\sum_k X_k^r(t)H_k\Bigr)\dd t, \\
  \dd\langle X^r, Y\rangle_t &= \Sigma_r\Diag(v(t))\,\varphi_Y(T-t)\,\dd t,\\
  \dd\langle X^r, v\rangle_t &= \Sigma_r\Diag(v(t))\Diag(\rho)\Diag(\sigma_v)\,\dd t.
\end{align*}
All of these are affine in $(v,X^r)$, so $X$ is a (time-inhomogeneous) affine diffusion, subject to standard admissibility conditions, see for example \cite{filipovic2005time}. Finally, if we set $\Sigma_r=0$, we recover the independent setting in Section~\ref{Sec:Option pricing}.
 
  \renewcommand{\theprop}{2}
\begin{prop}[With Stochastic Rates]
\label{prop:2pp}
Let $(Y,v,X^r)$ have dynamics described by \eqref{eq:prop2Y}, \eqref{eq:prop2v} and \eqref{eq:10c}, with $\varphi_Y$ an $n$-dimensional deterministic function,
and let $r=\zeta_r\cdot X^r$. The time-$t$ price of a European put option with maturity $T>t$ and strike $K$ on $Y$ is
\begin{align}
  \mathcal{P}(t,T,K)
  &= \E^\Q_t\Bigl\{ e^{-\int_t^T r(s)\dd s}\,\bigl(K-Y(T)\bigr)^+ \Bigr\} \nonumber\\
  &= P(t,T)\,\frac{K-\bar Y(t,T)}{2}
   - \int_0^\infty \frac{\Imag\bigl\{[K-\chi(\omega)]\,e^{-\ii\omega K+\psi(\omega)}\bigr\}}{\pi\omega}\,\dd\omega,
  \label{eq:11d}
\end{align}
where
\begin{align*}
  \psi[\omega;\tau,v(t),X^r(t),Y(t)] &= A_\psi(\tau;\omega) + B_\psi(\tau;\omega)\cdot v(t) + E_\psi(\tau)\cdot X^r(t) + \ii\omega Y(t),\\
  \chi[\omega;\tau,v(t),Y(t)] &= A_\chi(\tau;\omega) + B_\chi(\tau;\omega)\cdot v(t) + Y(t),
\end{align*}
with $\tau=T-t$. The bond price and the $T$-forward price of $Y(T)$ are
\begin{equation}\label{eq:11dd}
  P(t,T) = e^{\psi(0)}, \qquad
  \bar Y(t,T) \triangleq \frac{\E^\Q_t\bigl\{e^{-\int_t^T r(s)\dd s}\,Y(T)\bigr\}}{P(t,T)} = \chi(0).
\end{equation}
The real-valued, $\omega$-independent loading $E_\psi(\tau)\in\mathbb R^{n_r}$ solves the Riccati system
\begin{equation}\label{eq:12nod}
  \frac{\dd E_{\psi,k}(\tau)}{\dd\tau} = -\zeta_{r,k} + \bigl(\gamma_r^\top E_\psi(\tau)\bigr)_k + \tfrac12 E_\psi(\tau)^\top H_k E_\psi(\tau),
  \qquad E_\psi(0)=0,\qquad k=1,\dots,n_r.
\end{equation}
$(A_\psi,B_\psi)$ and $(A_\chi,B_\chi)$ solve
\begin{align}
  \frac{\dd A_\psi(\tau;\omega)}{\dd\tau} &= \vartheta\cdot B_\psi(\tau;\omega) + \theta_r\cdot E_\psi(\tau) + \tfrac12E_\psi(\tau)^\top H_0E_\psi(\tau),
     & A_\psi(0;\omega)&=0; \label{eq:12a}\\
  \frac{\dd B_\psi(\tau;\omega)}{\dd\tau} &= \Gamma^\top B_\psi(\tau;\omega) + \widetilde R_\psi\bigl[\tau,B_\psi(\tau;\omega);\omega\bigr],
     & B_\psi(0;\omega)&=0; \label{eq:12b}\\
  \frac{\dd A_\chi(\tau;\omega)}{\dd\tau} &= \vartheta\cdot B_\chi(\tau;\omega),
     & A_\chi(0;\omega)&=0; \label{eq:12c}\\
  \frac{\dd B_\chi(\tau;\omega)}{\dd\tau} &= \Gamma^\top B_\chi(\tau;\omega) + \widetilde R_\chi\bigl[\tau,B_\psi(\tau;\omega),B_\chi(\tau;\omega);\omega\bigr],
     & B_\chi(0;\omega)&=0. \label{eq:12d}
\end{align}
Define the real vector $s(\tau)\triangleq\Sigma_r^\top E_\psi(\tau)\in\mathbb R^n$ and the complex loadings
\begin{equation*}
  a_i(\tau;\omega) \triangleq \ii\omega\,\varphi_{Y,i}(\tau) + s_i(\tau), \qquad i=1,\dots,n.
\end{equation*}
Then, element-wise,
\begin{align*}
  \widetilde R_{\psi,i}[\tau,B_\psi;\omega]
   &= \frac{\sigma_{v,i}^2}{2}B_{\psi,i}^2 + \rho_i\sigma_{v,i}\,a_i(\tau;\omega)\,B_{\psi,i} + \frac12 a_i^2(\tau;\omega)\\
   &= \frac{\sigma_{v,i}^2}{2}B_{\psi,i}^2 + \ii\omega\rho_i\sigma_{v,i}\varphi_{Y,i}(\tau)B_{\psi,i} - \frac{\omega^2}{2}\varphi_{Y,i}^2(\tau)\\
   &\quad + s_i(\tau)\Bigl[\rho_i\sigma_{v,i}B_{\psi,i} + \ii\omega\varphi_{Y,i}(\tau) + \tfrac12 s_i(\tau)\Bigr],\\[4pt]
  \widetilde R_{\chi,i}[\tau,B_\psi,B_\chi;\omega]
   &= \sigma_{v,i}^2B_{\psi,i}B_{\chi,i} + \bigl[a_i(\tau;\omega)B_{\chi,i} + \varphi_{Y,i}(\tau)B_{\psi,i}\bigr]\rho_i\sigma_{v,i} + a_i(\tau;\omega)\varphi_{Y,i}(\tau)\\
   &= \sigma_{v,i}^2B_{\psi,i}B_{\chi,i} + \bigl[\ii\omega B_{\chi,i}+B_{\psi,i}\bigr]\rho_i\sigma_{v,i}\varphi_{Y,i}(\tau) + \ii\omega\varphi_{Y,i}^2(\tau)\\
   &\quad + s_i(\tau)\bigl[\rho_i\sigma_{v,i}B_{\chi,i} + \varphi_{Y,i}(\tau)\bigr].
\end{align*}
\end{prop}
\begin{remark}
If $\Sigma_r=0$, then $s\equiv0$ and $\widetilde R=R$, so $\chi$ and $B_\psi$
  coincide with the expressions of Proposition 2 in Section~\ref{Sec:Option pricing}. Furthermore $\bar Y(t,T)=Y(t)$, 
  $P(t,T)=\exp\{A_r(\tau)+E_\psi(\tau)\cdot X^r(t)\}$, where $\dd A_r/\dd\tau=\theta_r\cdot E_\psi+\frac12E_\psi^\top H_0E_\psi$, with $A_r(0)=0$. Hence \eqref{eq:11d} reduces to
  $P(t,T)\E^\Q_t\{(K-Y(T))^+\}$.
\end{remark}
 
\begin{proof}
Write $D(t,T)\triangleq e^{-\int_t^T r(s)\dd s}$. The price of a European put option on $Y(T)$ can be decomposed
into the two terms
\begin{align}
  \mathcal{P}(t,T,K) &= \E^\Q_t\Bigl\{D(t,T)\bigl(K-Y(T)\bigr)^+\Bigr\}\nonumber\\
   &= K\,\E^\Q_t\bigl\{D(t,T)\mathbf 1_{Y(T)\le K}\bigr\} - \E^\Q_t\bigl\{D(t,T)Y(T)\mathbf 1_{Y(T)\le K}\bigr\},
  \label{eq:33}
\end{align}
where we denote the two expectations above as 
\begin{align}
G(K)\triangleq\E^\Q_t\{D(t,T)\mathbf 1_{Y(T)\le K}\},\quad H(K)\triangleq\E^\Q_t\{D(t,T)Y(T)\mathbf 1_{Y(T)\le K}\}.
\end{align} Since the short rate is correlated with $Y$ through
$\Sigma_r$, the discount factor cannot be taken outside the expectations. We first consider $G(K)$, which is the distribution function of the finite measure $$\mu_t(A)=\E^\Q_t\{D(t,T)\mathbf 1_{Y(T)\in A}\}.$$ Define a function $\psi$ as the logarithm of the discounted conditional
characteristic function, such that
\begin{equation}\label{eq:34}
  \psi(\omega;t,T) \triangleq \log\E^\Q_t\Bigl\{D(t,T)\,e^{\ii\omega Y(T)}\Bigr\}.
\end{equation}
Knowledge of the discounted characteristic function allows us to compute $G(K)$ using the \cite{gilpelaez1951}
inversion theorem applied to $\mu_t/P(t,T)$, or equivalently Proposition 2 of \cite{Duffie2000}, where
\begin{equation}\label{eq:35}
  G(K) = \frac{e^{\psi(0)}}{2} - \int_0^\infty\frac{\Imag\bigl\{e^{\psi(\omega)-\ii\omega K}\bigr\}}{\pi\omega}\,\dd\omega .
\end{equation}
Since $X=(Y(t),v(t),X^r(t))$ is an affine process and the short rate is an affine function of $X$, the
$\mathcal F_t$-conditional expectation in \eqref{eq:34} is exponentially affine in $X$, or, equivalently, \eqref{eq:34} has an
affine form
\[
  \psi(\omega;t,T) = A_\psi(T-t;\omega) + B_\psi(T-t;\omega)\cdot v(t) + C_\psi(T-t;\omega)Y(t) + E_\psi(T-t;\omega)\cdot X^r(t).
\]
To derive the solution to $(A_\psi,B_\psi,C_\psi,E_\psi)$ for a fixed $\omega$, first note that \eqref{eq:34} implies that the
terminal values are given by $A_\psi(0;\omega)=B_{\psi,i}(0;\omega)=E_{\psi,k}(0;\omega)=0$ for all $i$ and $k$, and
$C_\psi(0;\omega)=\ii\omega$. Now define the complex-valued Itô process $M(t)$ as
\[
  M(t) = e^{-\int_0^t r(s)\dd s+A_\psi(T-t,\omega)+B_\psi(T-t,\omega)\cdot v(t)+C_\psi(T-t,\omega)Y(t)+E_\psi(T-t,\omega)\cdot X^r(t)} .
\]
Notice that by the law of iterated expectations with $t\le s\le T$, $M(t)=\E^\Q_t\{M(s)\}$ such that $M(t)$ is a
martingale. Applying Itô's formula we can derive the dynamics of $M(t)$ of the form
\begin{align*}
  \frac{\dd M(t)}{M(t)} ={}& \mu_M\bigl(t,X^r(t),v(t),Y(t)\bigr)\dd t
   + \bigl(C_\psi(\tau)\varphi_Y(\tau)+\Sigma_r^\top E_\psi(\tau)\bigr)\cdot D_v(v(t))\dd W^\Q(t)\\
   &+ \bigl(\Diag(\sigma_v)B_\psi(\tau)\bigr)\cdot D_v(v(t))\dd Z^\Q(t)
   + E_\psi(\tau)^\top\sigma^r(X^r(t))\dd W^r(t),
\end{align*}
where the drift function $\mu_M$ is given by
\begin{align*}
  \mu_M\bigl(t,X^r(t),v(t),Y(t)\bigr) ={}& -\zeta_r\cdot X^r(t)
    -\Bigl(\frac{\dd A_\psi(\tau)}{\dd\tau}+\sum_{i=1}^n\frac{\dd B_{\psi,i}(\tau)}{\dd\tau}v_i(t)
    +\frac{\dd C_\psi(\tau)}{\dd\tau}Y(t)\Bigr)\\
   &-\sum_{k=1}^{n_r}\frac{\dd E_{\psi,k}(\tau)}{\dd\tau}X_k^r(t)
   +\sum_{i=1}^nB_{\psi,i}(\tau)\vartheta_i+\sum_{i=1}^nv_i(t)\Bigl(\sum_{j=1}^n\Gamma_{ji}B_{\psi,j}(\tau)\Bigr)\\
   &+\sum_{k=1}^{n_r}E_{\psi,k}(\tau)\theta_{r,k}+\sum_{k=1}^{n_r}X_k^r(t)\Bigl(\sum_{l=1}^{n_r}(\gamma_r)_{lk}E_{\psi,l}(\tau)\Bigr)\\
   &+\sum_{i=1}^nv_i(t)\Bigl[\frac{\sigma_{v,i}^2}{2}B_{\psi,i}^2(\tau)+\rho_i\sigma_{v,i}a_i(\tau)B_{\psi,i}(\tau)
    +\frac{a_i^2(\tau)}{2}\Bigr]\\
   &+\frac12E_\psi(\tau)^\top H_0E_\psi(\tau)+\frac12\sum_{k=1}^{n_r}X_k^r(t)\,E_\psi(\tau)^\top H_kE_\psi(\tau),
\end{align*}
using $\tau=T-t$ and writing
$a_i(\tau)\triangleq C_\psi(\tau)\varphi_{Y,i}(\tau)+\bigl(\Sigma_r^\top E_\psi(\tau)\bigr)_i$. We have used that $W^r$ is independent of $(W^\Q,Z^\Q)$ and that
\[
  E_\psi^\top\sigma^r(x)\sigma^r(x)^\top E_\psi=E_\psi^\top H_0E_\psi+\sum_{k=1}^{n_r}x_kE_\psi^\top H_kE_\psi .
\]
Since $M(t)$ is a martingale, $\mu_M(t,X^r(t),v(t),Y(t))=0$. By separation of variables we can first collect the terms proportional to $X^r_k(t)$ and set them to zero. This shows
that $E_\psi$ solves the Riccati system \eqref{eq:12nod}. We can use that $C_\psi$ has the trivial solution
$C_\psi(\tau)=\ii\omega$ for all $\tau$, so that $a_i(\tau)=\ii\omega\varphi_{Y,i}(\tau)+s_i(\tau)$ with
$s=\Sigma_r^\top E_\psi$. By applying separation of variables to the remaining terms we get the system described by
\eqref{eq:12a} and \eqref{eq:12b}, where the second expression for $\widetilde R_{\psi,i}$ follows by expanding
$\frac12a_i^2=-\frac{\omega^2}{2}\varphi_{Y,i}^2+\ii\omega s_i\varphi_{Y,i}+\frac12s_i^2$. All coefficients are real
for $\omega=0$, in which case \eqref{eq:34} gives $e^{\psi(0)}=\E^\Q_t\{D(t,T)\}=P(t,T)$.  
To evaluate the second term $H(K)$ in \eqref{eq:33}, we note that its Fourier--Stieltjes transform $\mathcal H(\omega)$ is
given by
\begin{equation}\label{eq:36}
  \mathcal H(\omega) = \int_{\mathbb R}e^{\ii\omega y}\dd H(y) = \E^\Q_t\Bigl\{D(t,T)\,Y(T)e^{\ii\omega Y(T)}\Bigr\}.
\end{equation}
The extended Lévy inversion theorem of \citet{Duffie2000} can then be used to compute $H(K)$ from
$\mathcal H(\omega)$ as
\begin{equation}\label{eq:37}
  H(K) = \frac{\mathcal H(0)}{2} - \int_0^\infty\frac{\Imag\bigl\{\mathcal H(\omega)e^{-\ii\omega K}\bigr\}}{\pi\omega}\,\dd\omega,
\end{equation}
where $\mathcal H(0)=\E^\Q_t\{D(t,T)Y(T)\}$ is the discounted expectation of $Y(T)$, which differs from
$P(t,T)Y(t)$ because the discount factor and $Y(T)$ are correlated. Furthermore, \eqref{eq:36} is of the extended
transform type from \cite{Duffie2000}, where they show that the solution has the form
\[
  \mathcal H(\omega)=\chi(\omega)e^{\psi(\omega)},
\]
where $\chi(\omega)$ is again affine with
\[
  \chi(\omega) = A_\chi(T-t) + B_\chi(T-t)\cdot v(t) + C_\chi(T-t)Y(t) + E_\chi(T-t)\cdot X^r(t).
\]
Notice that by \eqref{eq:36}, we have that $A_\chi(0;\omega)=B_{\chi,i}(0;\omega)=E_{\chi,k}(0;\omega)=0$ for all $i$ and $k$,
and $C_\chi(0;\omega)=1$. Analogous to the previous case, the law of iterated expectations implies that the
discounted process $N(t)\triangleq e^{-\int_0^tr(s)\dd s}\mathcal H(\omega)$, which equals
$\chi(t;\omega,T)M(t;\omega,T)$, is a martingale, which we can use to derive the solution to
$(A_\chi,B_\chi,C_\chi,E_\chi)$ as a system of ODEs. Applying the Itô product rule to
$N(t)=\chi(t;\omega,T)M(t;\omega,T)$ for fixed $T$ and $\omega$ results in
\[
  \dd N(t) = \chi(t)\dd M(t) + M(t)\dd\chi(t) + \dd\langle\chi(t),M(t)\rangle,
\]
and since the previously derived conditions ensure that the first term above has no drift, we only need to
consider the last two terms. Straightforward calculations, writing
$c_i(\tau)\triangleq C_\chi(\tau)\varphi_{Y,i}(\tau)+(\Sigma_r^\top E_\chi(\tau))_i$ for the loading of $\chi$ on
$\sqrt{v_i(t)}\dd W_i^\Q(t)$, result in
\begin{align*}
  M(t)\dd\chi(t) ={}& M(t)\Bigl\{-\Bigl(\frac{\dd A_\chi(\tau)}{\dd\tau}+\sum_{i=1}^n\frac{\dd B_{\chi,i}(\tau)}{\dd\tau}v_i(t)
    +\frac{\dd C_\chi(\tau)}{\dd\tau}Y(t)+\sum_{k=1}^{n_r}\frac{\dd E_{\chi,k}(\tau)}{\dd\tau}X_k^r(t)\Bigr)\dd t\\
   &+C_\chi(\tau)\dd Y(t)+\sum_{i=1}^nB_{\chi,i}(\tau)\dd v_i(t)+\sum_{k=1}^{n_r}E_{\chi,k}(\tau)\dd X_k^r(t)\Bigr\},\\
  \dd\langle\chi(t),M(t)\rangle ={}& M(t)\sum_{i=1}^nv_i(t)\Bigl\{c_i(\tau)a_i(\tau)
    +\bigl[c_i(\tau)B_{\psi,i}(\tau)+a_i(\tau)B_{\chi,i}(\tau)\bigr]\rho_i\sigma_{v,i}\\
   &\hphantom{M(t)\sum_{i=1}^nv_i(t)\Bigl\{}+\sigma_{v,i}^2B_{\psi,i}(\tau)B_{\chi,i}(\tau)\Bigr\}\dd t\\
   &+M(t)\Bigl\{E_\chi(\tau)^\top H_0E_\psi(\tau)+\sum_{k=1}^{n_r}X_k^r(t)\,E_\chi(\tau)^\top H_kE_\psi(\tau)\Bigr\}\dd t .
\end{align*}
Proceeding as before, we can insert the dynamics of $X^r(t)$, $Y(t)$ and $v(t)$, from which, by separation of variables, we get the ODEs in \eqref{eq:12c} and \eqref{eq:12d}. Here the terms proportional to $X_k^r(t)$ show that $E_\chi$ solves
the linear homogeneous system $\dd E_{\chi,k}(\tau)/\dd\tau=(\gamma_r^\top E_\chi(\tau))_k+E_\chi(\tau)^\top H_kE_\psi(\tau)$
with $E_{\chi,k}(0;\omega)=0$, so that $E_\chi(\tau;\omega)=0$, while $C_\chi$ has the solution
$C_\chi(\tau;\omega)=1$, such that $c_i(\tau)=\varphi_{Y,i}(\tau)$, and $a_i$ is as derived above. Evaluating
$\mathcal H(\omega)=\chi(\omega)e^{\psi(\omega)}$ at $\omega=0$ gives $\mathcal H(0)=P(t,T)\bar Y(t,T)$, the second part of \eqref{eq:11dd}.
 
The price of the put option is then given by $KG(K)-H(K)$, from which we can derive the formula \eqref{eq:11d} by
inserting \eqref{eq:35} and \eqref{eq:37} and rearranging. 
\end{proof}

\subsection{Proof of Proposition 3}
\label{sub:proof3}
\begin{proof}
By Proposition~\ref{prop:optionprice}, $A_\psi(\tau;\omega)+B_\psi(\tau;\omega)\cdot v(t)=\log\mathbb{E}^{\mathbb{Q}}_t\{e^{\ii\omega(Y(T)-Y(t))}\}$,
so its coefficient of $(\ii\omega)^m/m!$ is the $m$th cumulant of $Y(T)-Y(t)$. Since $A_\psi(\tau;0)=0$ and
$B_\psi(\tau;0)=0$, write
\begin{equation*}
A_\psi(\tau;\omega)=\sum_{m=1}^{3}\frac{(\ii\omega)^m}{m!}A_m(\tau)+o(\omega^3),\qquad
B_\psi(\tau;\omega)=\sum_{m=1}^{3}\frac{(\ii\omega)^m}{m!}B_m(\tau)+o(\omega^3).
\end{equation*}
In powers of $\ii\omega$,
\begin{equation*}
R_{\psi,i}=\frac{\sigma_{v,i}^2}{2}B_{\psi,i}^2+\ii\omega\,\rho_i\sigma_{v,i}\varphi_{Y,i}(\tau)B_{\psi,i}
+\frac{(\ii\omega)^2}{2}\varphi_{Y,i}^2(\tau),
\end{equation*}
and matching the coefficients of $(\ii\omega)^m/m!$ in \eqref{eq:tfA_ode}--\eqref{eq:tfB_ode} gives
\eqref{eq:momode}, with $R_m$ the coefficient of $(\ii\omega)^m/m!$ in $R_\psi[\tau,B_\psi(\tau;\omega);\omega]$. For
$m=1$ this coefficient is zero, so $A_1=B_1=0$, $Y(T)-Y(t)$ has zero mean, and its second and third cumulants
are $\varkappa_2$ and $\varkappa_3$. With $B_1=0$, $B_{\psi,i}^2$ starts at $(\ii\omega)^4$, and the last two
terms of $R_{\psi,i}$ give \eqref{eq:forcing}.
\end{proof}
 

%% file: appendix_results.tex

\section{Additional results}\label{app}
This appendix collects the results that the main text refers to. As in the main text, the skew
forcing enters only the estimation, and every price, error and hedge is computed without it.

\subsection{Estimates not in the main text}\label{app:est}
\begin{figure}[!tbp]
\centering
\includegraphics[width=\textwidth]{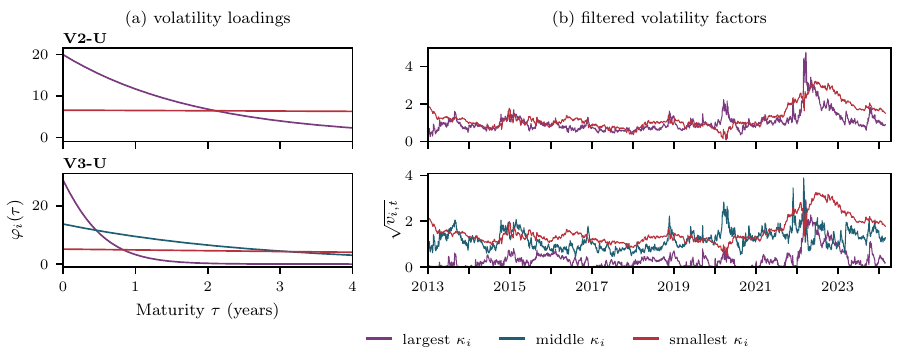}
\caption[Volatility loadings and filtered volatility factors of V2-U and V3-U]{\textbf{Volatility loadings and filtered volatility factors of V2-U and V3-U}\\ As \autoref{fig:res_factors}, for V2-U and V3-U. (a)~Volatility loadings $\varphi_i(\tau)$. (b)~Square roots of the filtered volatility factors.}
\label{fig:app_factors_chi0}
\end{figure}

\autoref{tab:app_est0} reports the estimates of the two models without the skew forcing, V2-U
and V3-U, and \autoref{fig:app_factors_chi0} their volatility loadings and filtered volatility
factors. Without the forcing, the log-likelihood is lower by 63{,}278 with two factors and by
39{,}555 with three. The volatility factors must then reproduce the skew as well, through
the leverage correlations described in Section~\ref{sec:res_est}.

\begin{table}[!tbp]
\centering\footnotesize
\setlength{\aboverulesep}{0pt}\setlength{\belowrulesep}{0pt}%
\renewcommand{\arraystretch}{1.05}%
\begin{tabular}{@{}l cc| ccc@{}}
\toprule
 & \multicolumn{2}{c|}{V2-U} & \multicolumn{3}{c}{V3-U}\\
\cmidrule(lr){2-3}\cmidrule(l){4-6}
factor $i$ & 1 & 2 & 1 & 2 & 3\\
\midrule
$\kappa_i$ & 0.594 & 0.011 & 3.026 & 0.434 & 0.058\\
 & {\scriptsize (0.0070)} & {\scriptsize ( --- )} & {\scriptsize (0.3057)} & {\scriptsize (0.0064)} & {\scriptsize (0.0052)}\\[2pt]
$\beta_{0,i}$ & 19.956 & 6.574 & 28.826 & 13.751 & 5.140\\
 & {\scriptsize (0.2657)} & {\scriptsize (0.1625)} & {\scriptsize (3.1777)} & {\scriptsize (0.2594)} & {\scriptsize (0.1906)}\\[2pt]
$\beta_{1,i}$ & 1.202 & --- & 38.464 & 0.881 & ---\\
 & {\scriptsize (0.0387)} &  & {\scriptsize (20.6118)} & {\scriptsize (0.1117)} & \\[2pt]
\midrule
$\sigma_{v,i}$ & 3.092 & 0.889 & 3.777 & 2.650 & 0.649\\
 & {\scriptsize (0.1123)} & {\scriptsize (0.0212)} & {\scriptsize (0.7435)} & {\scriptsize (0.1495)} & {\scriptsize (0.0164)}\\[2pt]
$\rho_i$ & $-$0.137 & 0.340 & 0.894 & $-$0.703 & $-$0.236\\
 & {\scriptsize (0.0103)} & {\scriptsize (0.0407)} & {\scriptsize (0.0057)} & {\scriptsize (0.0164)} & {\scriptsize (0.0309)}\\[2pt]
$\Gamma_{1,i}$ & $-$5.821 & 3.065 & $-$0.358 & 0.000 & 0.265\\
 & {\scriptsize (0.2715)} & {\scriptsize (0.2042)} & {\scriptsize (0.1415)} & {\scriptsize ( --- )} & {\scriptsize ( --- )}\\[2pt]
$\Gamma_{2,i}$ & --- & $-$0.396 & --- & $-$7.034 & 5.206\\
 &  & {\scriptsize (0.0189)} &  & {\scriptsize (0.3375)} & {\scriptsize ( --- )}\\[2pt]
$\Gamma_{3,i}$ & --- & --- & --- & --- & $-$0.536\\
 &  &  &  &  & {\scriptsize (0.0249)}\\[2pt]
$\Gamma^{\mathbb P}_{i,i}$ & $-$8.216 & $-$0.400 & $-$1.812 & $-$5.903 & $-$0.400\\
 & {\scriptsize (1.3241)} & {\scriptsize ( --- )} & {\scriptsize (1.1693)} & {\scriptsize (0.8138)} & {\scriptsize ( --- )}\\[2pt]
$\lambda_{x,i}$ & 0.775 & $-$0.245 & $-$0.495 & $-$0.559 & $-$0.286\\
 & {\scriptsize (0.2355)} & {\scriptsize (0.1869)} & {\scriptsize (0.3895)} & {\scriptsize (0.2549)} & {\scriptsize (0.1464)}\\[2pt]
\midrule
$F_0$ & \multicolumn{2}{c|}{20.356} & \multicolumn{3}{c}{65.029}\\
 & \multicolumn{2}{c|}{{\scriptsize (2.6823)}} & \multicolumn{3}{c}{{\scriptsize (1.4098)}}\\[2pt]
$\log\sigma_F$ & \multicolumn{2}{c|}{$-$0.797} & \multicolumn{3}{c}{$-$1.702}\\
 & \multicolumn{2}{c|}{{\scriptsize (0.0123)}} & \multicolumn{3}{c}{{\scriptsize (0.0205)}}\\[2pt]
$\sigma_V$ & \multicolumn{2}{c|}{0.0164} & \multicolumn{3}{c}{0.0170}\\
 & \multicolumn{2}{c|}{{\scriptsize (0.0004)}} & \multicolumn{3}{c}{{\scriptsize (0.0004)}}\\[2pt]
$a_V$ & \multicolumn{2}{c|}{3.683} & \multicolumn{3}{c}{2.054}\\
 & \multicolumn{2}{c|}{{\scriptsize (0.1168)}} & \multicolumn{3}{c}{{\scriptsize (0.0663)}}\\[2pt]
$\chi_V$ & \multicolumn{2}{c|}{---} & \multicolumn{3}{c}{---}\\
 & \multicolumn{2}{c|}{} & \multicolumn{3}{c}{}\\[2pt]
$\log\mathcal L$ & \multicolumn{2}{c|}{354{,}943.11} & \multicolumn{3}{c}{441{,}871.71}\\
\bottomrule
\end{tabular}
\caption[Parameter estimates of V2-U and V3-U]{\textbf{Parameter estimates of V2-U and V3-U}\\ As \autoref{tab:res_est}, for V2-U and V3-U, estimated without the skew forcing, $\chi_V=0$.}
\label{tab:app_est0}
\end{table}

\subsection{Variance shares by maturity}\label{app:varshare}
\autoref{tab:app_varshare} shows how the factors divide the volatility of the curve by contract.
In V3, the three factors divide the curve by maturity. The transitory factor contributes mainly at
the front, with 35 percent at M1 and 4 percent at Q1, the middle factor carries the largest share
from M1 to Y1, and the share of the level factor rises with maturity to 80 percent at Y3. In V2,
the middle factor carries most of the variance out to Q3, and the level factor from Y1 out.

In VC3, the transitory factor contributes little at any maturity, and the middle factor carries
most of the variance of the front contracts and more than the level factor out to Y1. Including
the calendar spread options in the estimation thus moves the volatility of the front contracts
from the transitory to the middle factor, the reassignment described in Section~\ref{sec:res_est}.

\begin{table}[!tbp]
\centering\footnotesize
\setlength{\aboverulesep}{0pt}\setlength{\belowrulesep}{0pt}%
\renewcommand{\arraystretch}{1.05}%
\begin{tabular}{@{}lr|rr|rrr|rrr@{}}
\toprule
 & & \multicolumn{2}{c|}{V2} & \multicolumn{3}{c|}{V3} & \multicolumn{3}{c}{VC3}\\
Contract & $\tau$ & $v_1$ & $v_2$ & $v_1$ & $v_2$ & $v_3$ & $v_1$ & $v_2$ & $v_3$\\
\midrule
M1 & 0.10 & 84 & 16 & 35 & 47 & 17 & 9 & 72 & 19\\
M2 & 0.18 & 83 & 17 & 28 & 53 & 19 & 6 & 73 & 21\\
M3 & 0.26 & 82 & 18 & 22 & 57 & 21 & 4 & 74 & 22\\
M4 & 0.35 & 81 & 19 & 17 & 60 & 23 & 2 & 74 & 23\\
M5 & 0.43 & 79 & 21 & 13 & 63 & 24 & 1 & 74 & 25\\
M6 & 0.51 & 78 & 22 & 10 & 65 & 26 & 1 & 73 & 26\\
M7 & 0.59 & 76 & 24 & 7 & 66 & 27 & 1 & 72 & 27\\
Q1 & 0.76 & 73 & 27 & 4 & 67 & 29 & 0 & 71 & 29\\
Q2 & 1.01 & 68 & 32 & 2 & 66 & 32 & 0 & 67 & 33\\
Q3 & 1.26 & 62 & 38 & 1 & 64 & 35 & 0 & 64 & 36\\
Y1 & 1.89 & 47 & 53 & 0 & 55 & 45 & 0 & 55 & 45\\
Y2 & 2.89 & 24 & 76 & 0 & 37 & 63 & 0 & 39 & 61\\
Y3 & 3.89 & 9 & 91 & 0 & 20 & 80 & 0 & 25 & 75\\
\bottomrule
\end{tabular}
\caption[Variance shares by maturity]{\textbf{Variance shares by maturity}\\ Share of each volatility factor in the instantaneous futures variance $\sum_i\varphi_i(\tau)^2v_i$ of V2, V3 and VC3, in percent, by contract, summed over the sample at the previous date's filtered state. $\tau$ is the contract's average maturity in years.}
\label{tab:app_varshare}
\end{table}

\subsection{Pricing diagnostics}\label{app:diag}
\autoref{tab:app_futerr} reports the median and mean signed futures errors by contract. The
median errors are far smaller than the RMSEs of \autoref{tab:res_futrmse}, 0.33 percent at M1 for
V3 against an RMSE of 1.09, which indicates that a minority of large errors drives the RMSE. The
medians of V3 are below those of V2 at every contract, so the gain from the third factor is not
confined to the extreme days. The mean signed errors of V2, V3 and V3-U are significant at most
contracts and change sign along the curve, so each leaves a systematic error in the shape of the
curve, of up to 0.78 percent for V2 and 0.19 percent for V3 and V3-U. Those of VC3 are significant
only at Q1 and Q2.

\begin{table}[!tbp]
\centering\footnotesize
\setlength{\aboverulesep}{0pt}\setlength{\belowrulesep}{0pt}%
\renewcommand{\arraystretch}{1.05}%
\begin{tabular}{@{}lrrrrrrrr@{}}
\toprule
 & \multicolumn{4}{c}{median absolute error} & \multicolumn{4}{c}{mean signed error}\\
\cmidrule(lr){2-5}\cmidrule(l){6-9}
Contract & V2 & V3 & VC3 & V3-U & V2 & V3 & VC3 & V3-U\\
\midrule
M1 & 0.94 & 0.33 & 0.23 & 0.29 & $0.78^{***}$ & $0.19^{**}$ & $0.09$ & $0.19^{*}$\\
M2 & 0.51 & 0.11 & 0.09 & 0.13 & $0.24^{**}$ & $-0.02$ & $-0.02$ & $-0.03^{*}$\\
M3 & 0.27 & 0.15 & 0.16 & 0.15 & $-0.05$ & $-0.09^{**}$ & $-0.04$ & $-0.10^{***}$\\
M4 & 0.24 & 0.19 & 0.16 & 0.17 & $-0.21^{***}$ & $-0.11^{***}$ & $-0.05$ & $-0.11^{***}$\\
M5 & 0.34 & 0.20 & 0.14 & 0.17 & $-0.28^{***}$ & $-0.09^{***}$ & $-0.04$ & $-0.08^{**}$\\
M6 & 0.41 & 0.19 & 0.12 & 0.15 & $-0.29^{***}$ & $-0.05$ & $-0.01$ & $-0.04$\\
M7 & 0.46 & 0.15 & 0.12 & 0.14 & $-0.27^{***}$ & $-0.00$ & $0.01$ & $0.01$\\
Q1 & 0.51 & 0.11 & 0.11 & 0.13 & $-0.17^{*}$ & $0.07^{***}$ & $0.04^{**}$ & $0.08^{***}$\\
Q2 & 0.54 & 0.17 & 0.16 & 0.18 & $0.01$ & $0.12^{***}$ & $0.06^{**}$ & $0.13^{***}$\\
Q3 & 0.59 & 0.27 & 0.22 & 0.25 & $0.16^{*}$ & $0.11^{***}$ & $0.05$ & $0.12^{***}$\\
Y1 & 0.57 & 0.27 & 0.24 & 0.24 & $0.23^{***}$ & $-0.11^{**}$ & $-0.08$ & $-0.08^{*}$\\
Y2 & 0.38 & 0.19 & 0.14 & 0.18 & $0.21^{***}$ & $-0.12^{***}$ & $-0.02$ & $-0.09^{***}$\\
Y3 & 0.96 & 0.32 & 0.33 & 0.30 & $-0.16$ & $0.12^{**}$ & $0.04$ & $0.05$\\
\midrule
All & 0.46 & 0.19 & 0.16 & 0.18 & & & & \\
\bottomrule
\end{tabular}
\caption[Median and signed futures pricing errors]{\textbf{Median and signed futures pricing errors}\\ Relative futures pricing errors, in percent, by contract: the median absolute error and the mean signed error, model less data, of V2, V3, VC3 and V3-U, over the dates on which each contract is observed. The row ``All'' aggregates every contract. $^{*}$, $^{**}$ and $^{***}$ denote significance of the mean signed error at the 10, 5 and 1 percent levels, on the Newey--West $t$-statistic (50 lags).}
\label{tab:app_futerr}
\end{table}

\begin{table}[!tbp]
\centering\footnotesize\setlength{\tabcolsep}{4pt}
\setlength{\aboverulesep}{0pt}\setlength{\belowrulesep}{0pt}%
\renewcommand{\arraystretch}{1.05}%
\begin{tabular}{@{}lrrrrrrrrrr|rrr|r@{}}
\toprule
Model & M1 & M2 & M3 & M4 & M5 & M6 & M7 & Q1 & Q2 & Q3 & OTM puts & ATM & OTM calls & All\\
\midrule
V2 & 0.68 & 0.45 & 0.46 & 0.48 & 0.49 & 0.48 & 0.48 & 0.45 & 0.45 & 0.52 & 0.69 & 0.28 & 0.64 & 0.49\\
V3 & 0.53 & 0.47 & 0.45 & 0.42 & 0.41 & 0.41 & 0.43 & 0.42 & 0.39 & 0.39 & 0.67 & 0.20 & 0.62 & 0.43\\
VC3 & 0.71 & 0.47 & 0.47 & 0.49 & 0.50 & 0.49 & 0.48 & 0.45 & 0.45 & 0.52 & 0.71 & 0.29 & 0.65 & 0.50\\
V3-U & 0.67 & 0.33 & 0.30 & 0.33 & 0.35 & 0.34 & 0.34 & 0.35 & 0.39 & 0.45 & 0.55 & 0.25 & 0.41 & 0.37\\
\bottomrule
\end{tabular}
\caption[Median vanilla pricing errors]{\textbf{Median vanilla pricing errors}\\ Median absolute errors of the implied Bachelier volatilities, in \$/bbl a year, of V2, V3, VC3 and V3-U. By contract over all delta buckets, and by moneyness over all contracts: out-of-the-money puts, delta buckets $-4$ to $-2$, at the money, $-1$ to $1$, and out-of-the-money calls, $2$ to $4$. The column ``All'' aggregates every option.}
\label{tab:app_vanerr}
\end{table}

\autoref{tab:app_vanerr} reports the median absolute errors of the vanilla options. Here too the
medians are well below the means, 0.43 against 0.79 for V3, and those of V3 are below those of V2
at every contract except M2.

\autoref{tab:app_signed_van} reports the mean signed errors by delta bucket at M1, M4 and Q1, the
contracts of \autoref{fig:res_smile}, and at M2. At M1 and M2, the forced models price both far
wings too low, V3 by the least, so the front smiles in the data are more curved than the models
price. At M4 and Q1, they price the far call wing too low, V3 by 1.02 at M4, and at Q1 they also
price the far put wing too high. This indicates that at longer maturities the smile in the data is
tilted further toward the calls than the factors and their $\rho_i$ price on average. V3-U,
estimated without the forcing, removes most of the error in the far call wing at M4 and Q1, but at
M1 its errors are larger than those of V3 in every bucket except the far call wing.

\begin{table}[!tbp]
\centering\footnotesize\setlength{\tabcolsep}{3pt}
\setlength{\aboverulesep}{0pt}\setlength{\belowrulesep}{0pt}%
\renewcommand{\arraystretch}{1.05}%
\begin{tabular}{@{}llrrrrrrrrr@{}}
\toprule
Contract & Model & $-4$ & $-3$ & $-2$ & $-1$ & $0$ & $1$ & $2$ & $3$ & $4$\\
\midrule
M1 & V2 & $-1.31^{***}$ & $-0.61^{***}$ & $-0.28^{*}$ & $-0.09$ & $0.06$ & $0.11$ & $0.07$ & $-0.21$ & $-1.26^{***}$\\
 & V3 & $-0.75^{***}$ & $-0.44^{**}$ & $-0.24^{*}$ & $-0.07$ & $0.08^{*}$ & $0.16^{**}$ & $0.20$ & $0.07$ & $-0.56$\\
 & VC3 & $-1.49^{***}$ & $-0.72^{***}$ & $-0.34^{**}$ & $-0.11$ & $0.07$ & $0.14$ & $0.12$ & $-0.15$ & $-1.19^{***}$\\
 & V3-U & $-1.63^{***}$ & $-0.77^{***}$ & $-0.36^{***}$ & $-0.10$ & $0.10$ & $0.20^{*}$ & $0.25^{*}$ & $0.12$ & $-0.53$\\
\midrule
M2 & V2 & $-0.59^{***}$ & $-0.26$ & $-0.11$ & $0.01$ & $0.11^{***}$ & $0.14^{*}$ & $0.09$ & $-0.19$ & $-1.15^{**}$\\
 & V3 & $-0.18$ & $-0.21$ & $-0.14$ & $-0.05$ & $0.07^{**}$ & $0.12$ & $0.13$ & $-0.01$ & $-0.58$\\
 & VC3 & $-0.75^{***}$ & $-0.36^{**}$ & $-0.16$ & $-0.01$ & $0.12^{***}$ & $0.17^{**}$ & $0.13$ & $-0.13$ & $-1.11^{**}$\\
 & V3-U & $-0.88^{***}$ & $-0.38^{***}$ & $-0.16^{***}$ & $-0.01$ & $0.13^{***}$ & $0.19^{***}$ & $0.22^{**}$ & $0.13$ & $-0.29$\\
\midrule
M4 & V2 & $0.00$ & $-0.08$ & $-0.09$ & $-0.03$ & $0.06$ & $0.06$ & $-0.02$ & $-0.33$ & $-1.41^{***}$\\
 & V3 & $0.27$ & $-0.05$ & $-0.12$ & $-0.07$ & $0.03^{*}$ & $0.04$ & $-0.01$ & $-0.25$ & $-1.02^{**}$\\
 & VC3 & $-0.08$ & $-0.15$ & $-0.13$ & $-0.05$ & $0.05$ & $0.07$ & $-0.00$ & $-0.32$ & $-1.40^{***}$\\
 & V3-U & $-0.21$ & $-0.17^{**}$ & $-0.14^{***}$ & $-0.08^{*}$ & $0.02$ & $0.04$ & $0.04$ & $-0.04$ & $-0.32^{*}$\\
\midrule
Q1 & V2 & $0.58^{***}$ & $-0.00$ & $-0.19^{**}$ & $-0.15^{***}$ & $-0.01$ & $0.06$ & $0.05$ & $-0.30$ & $-1.45^{***}$\\
 & V3 & $0.76^{***}$ & $0.07$ & $-0.14^{*}$ & $-0.13^{***}$ & $-0.01$ & $0.04$ & $0.01$ & $-0.34$ & $-1.39^{***}$\\
 & VC3 & $0.62^{***}$ & $0.02$ & $-0.18^{**}$ & $-0.15^{**}$ & $-0.01$ & $0.05$ & $0.04$ & $-0.31$ & $-1.48^{***}$\\
 & V3-U & $0.56^{***}$ & $0.00$ & $-0.20^{***}$ & $-0.20^{***}$ & $-0.09^{***}$ & $-0.03$ & $0.04$ & $-0.03$ & $-0.19$\\
\bottomrule
\end{tabular}
\caption[Signed vanilla pricing errors]{\textbf{Signed vanilla pricing errors}\\ Mean signed error of the implied Bachelier volatilities, model less data, in \$/bbl a year, by delta bucket, at M1, M2, M4 and Q1, for V2, V3, VC3 and V3-U, over the dates on which each cell is observed. $^{*}$, $^{**}$ and $^{***}$ denote significance at the 10, 5 and 1 percent levels, on the Newey--West $t$-statistic (50 lags).}
\label{tab:app_signed_van}
\end{table}

\begin{figure}[!tbp]
\centering
\includegraphics[width=\textwidth]{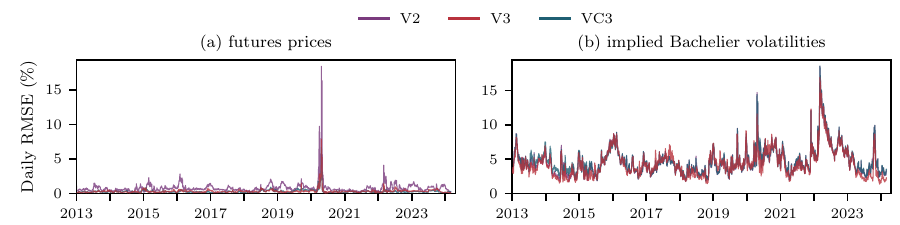}
\caption[Daily pricing errors]{\textbf{Daily pricing errors}\\ Daily RMSE of the relative pricing errors in percent, for V2, V3 and VC3. (a)~Futures prices. (b)~Implied Bachelier volatilities of the vanilla options.}
\label{fig:app_rmse}
\end{figure}

\begin{table}[!tbp]
\centering\footnotesize
\setlength{\aboverulesep}{0pt}\setlength{\belowrulesep}{0pt}%
\renewcommand{\arraystretch}{1.05}%
\begin{tabular}{@{}lr@{\,}lr@{\,}l|r@{\,}lr@{\,}l|r@{\,}lr@{\,}l@{}}
\toprule
 & \multicolumn{4}{c|}{V3 $-$ V2} & \multicolumn{4}{c|}{VC3 $-$ V3} & \multicolumn{4}{c}{V3-U $-$ V3}\\
 & \multicolumn{2}{c}{futures} & \multicolumn{2}{c|}{vanillas} & \multicolumn{2}{c}{futures} & \multicolumn{2}{c|}{vanillas} & \multicolumn{2}{c}{futures} & \multicolumn{2}{c}{vanillas}\\
\midrule
M1 & $-$0.89 & ($-$5.7) & $-$0.23 & ($-$4.7) & $-$0.12 & ($-$3.6) & 0.27 & (5.5) & 0.01 & (0.4)\textsuperscript{n.s.} & 0.07 & (1.2)\textsuperscript{n.s.}\\
M2 & $-$0.53 & ($-$9.6) & $-$0.06 & ($-$2.1)\textsuperscript{\dag} & $-$0.01 & ($-$1.4)\textsuperscript{n.s.} & 0.09 & (3.1) & 0.03 & (4.7) & $-$0.28 & ($-$3.4)\\
M3 & $-$0.13 & ($-$4.6) & $-$0.06 & ($-$2.9) & 0.01 & (0.9)\textsuperscript{n.s.} & 0.08 & (3.9) & 0.00 & (0.6)\textsuperscript{n.s.} & $-$0.38 & ($-$3.7)\\
M4 & $-$0.06 & ($-$4.0) & $-$0.08 & ($-$4.6) & $-$0.03 & ($-$1.9)\textsuperscript{n.s.} & 0.10 & (5.3) & $-$0.01 & ($-$1.7)\textsuperscript{n.s.} & $-$0.40 & ($-$3.5)\\
M5 & $-$0.18 & ($-$5.2) & $-$0.08 & ($-$5.5) & $-$0.06 & ($-$4.3) & 0.09 & (6.1) & $-$0.02 & ($-$2.3)\textsuperscript{\dag} & $-$0.38 & ($-$3.2)\\
M6 & $-$0.30 & ($-$6.5) & $-$0.06 & ($-$5.0) & $-$0.06 & ($-$5.2) & 0.07 & (5.5) & $-$0.02 & ($-$2.1)\textsuperscript{\dag} & $-$0.39 & ($-$3.4)\\
M7 & $-$0.41 & ($-$7.7) & $-$0.03 & ($-$3.8) & $-$0.04 & ($-$4.1) & 0.05 & (4.7) & $-$0.00 & ($-$0.1)\textsuperscript{n.s.} & $-$0.37 & ($-$3.6)\\
Q1 & $-$0.49 & ($-$10.0) & $-$0.00 & ($-$0.6)\textsuperscript{n.s.} & $-$0.00 & ($-$0.1)\textsuperscript{n.s.} & 0.02 & (2.8) & 0.03 & (4.4) & $-$0.35 & ($-$3.8)\\
Q2 & $-$0.45 & ($-$9.5) & $-$0.01 & ($-$1.2)\textsuperscript{n.s.} & $-$0.00 & ($-$0.0)\textsuperscript{n.s.} & 0.02 & (2.7) & 0.02 & (2.9) & $-$0.23 & ($-$3.3)\\
Q3 & $-$0.37 & ($-$7.1) & $-$0.07 & ($-$4.0) & $-$0.03 & ($-$1.3)\textsuperscript{n.s.} & 0.08 & (4.8) & $-$0.01 & ($-$0.7)\textsuperscript{n.s.} & $-$0.11 & ($-$2.0)\textsuperscript{\dag}\\
Y1 & $-$0.27 & ($-$6.3) &  &  & $-$0.04 & ($-$1.7)\textsuperscript{n.s.} &  &  & $-$0.04 & ($-$2.4)\textsuperscript{\dag} &  & \\
Y2 & $-$0.27 & ($-$5.8) &  &  & $-$0.06 & ($-$4.7) &  &  & $-$0.01 & ($-$0.7)\textsuperscript{n.s.} &  & \\
Y3 & $-$0.70 & ($-$10.1) &  &  & 0.00 & (0.1)\textsuperscript{n.s.} &  &  & $-$0.01 & ($-$0.6)\textsuperscript{n.s.} &  & \\
\midrule
All & $-$0.48 & ($-$9.4) & $-$0.08 & ($-$4.6) & $-$0.04 & ($-$3.1) & 0.10 & (5.6) & 0.00 & (0.1)\textsuperscript{n.s.} & $-$0.28 & ($-$3.2)\\
\midrule
OTM puts &  &  & $-$0.04 & ($-$1.6)\textsuperscript{n.s.} &  &  & 0.08 & (3.0) &  &  & $-$0.11 & ($-$2.3)\textsuperscript{\dag}\\
ATM &  &  & $-$0.11 & ($-$7.7) &  &  & 0.12 & (8.3) &  &  & 0.06 & (4.2)\\
OTM calls &  &  & $-$0.09 & ($-$3.5) &  &  & 0.10 & (4.0) &  &  & $-$0.56 & ($-$3.6)\\
\bottomrule
\end{tabular}
\caption[Diebold--Mariano tests]{\textbf{Diebold--Mariano tests}\\ Mean difference between the daily RMSEs of two models, negative where the first model has the smaller errors, for V3 against V2, VC3 against V3 and V3-U against V3, each date's RMSE taken over the cells both models price. By contract, over all contracts and, for the vanilla options, by moneyness as in \autoref{tab:res_vanmae}. Futures relative errors in percent, implied Bachelier volatilities of the vanilla options in \$/bbl a year. Diebold--Mariano $t$-statistics, with Newey--West standard errors (50 lags), in parentheses. Unless marked, every difference is significant at the 1 percent level; \textsuperscript{\dag} marks significance at the 5 percent level only and \textsuperscript{n.s.} no significance at the 5 percent level.}
\label{tab:app_dm}
\end{table}

\begin{table}[!tbp]
\centering\footnotesize
\setlength{\aboverulesep}{0pt}\setlength{\belowrulesep}{0pt}%
\renewcommand{\arraystretch}{1.05}%
\begin{tabular}{@{}l|rrrrr|rrrrr@{}}
\toprule
 & \multicolumn{5}{c|}{ATM vol} & \multicolumn{5}{c}{risk reversal}\\
Contract & data & V2 & V3 & VC3 & V3-U & data & V2 & V3 & VC3 & V3-U\\
\midrule
M1 & 0.13 & 0.14 & 0.16 & 0.14 & 0.11 & 0.45 & 0.20 & 0.30 & 0.22 & 0.41\\
M2 & 0.11 & 0.13 & 0.14 & 0.13 & 0.11 & 0.37 & 0.14 & 0.21 & 0.17 & 0.39\\
M3 & 0.16 & 0.16 & 0.16 & 0.15 & 0.15 & 0.30 & 0.05 & 0.08 & 0.09 & 0.35\\
M4 & 0.14 & 0.15 & 0.14 & 0.14 & 0.14 & 0.17 & $-$0.02 & 0.01 & 0.00 & 0.30\\
M5 & 0.15 & 0.16 & 0.15 & 0.15 & 0.15 & 0.12 & $-$0.04 & 0.01 & $-$0.03 & 0.29\\
M6 & 0.10 & 0.11 & 0.09 & 0.10 & 0.10 & $-$0.05 & $-$0.09 & $-$0.09 & $-$0.09 & 0.19\\
M7 & 0.23 & 0.22 & 0.21 & 0.22 & 0.22 & $-$0.01 & $-$0.08 & $-$0.06 & $-$0.08 & 0.22\\
Q1 & 0.16 & 0.21 & 0.17 & 0.21 & 0.20 & $-$0.04 & $-$0.09 & $-$0.06 & $-$0.09 & 0.24\\
Q2 & 0.06 & 0.10 & 0.07 & 0.09 & 0.06 & $-$0.16 & $-$0.11 & $-$0.13 & $-$0.12 & 0.07\\
Q3 & 0.27 & 0.35 & 0.34 & 0.36 & 0.36 & $-$0.31 & $-$0.17 & $-$0.17 & $-$0.19 & 0.14\\
\bottomrule
\end{tabular}
\caption[Implied volatility and the futures]{\textbf{Implied volatility and the futures}\\ Correlation of daily changes in the at-the-money implied Bachelier volatility and in the risk reversal, bucket $+4$ less bucket $-4$, with daily changes in the underlying future, by contract, in the data and for V2, V3, VC3 and V3-U. Days on which one of the three options changes underlying contract are left out.}
\label{tab:app_ivfut}
\end{table}

\autoref{fig:app_rmse} shows the daily RMSEs of V2, V3 and VC3 over time. In the futures, the
errors of V2 are larger than those of V3 and VC3 throughout the sample, most of all in spring 2020.
In the vanilla options the three models nearly coincide and rise together in the turbulent episodes
of 2020 and 2022, with median daily RMSEs of 4.2 percent for V2, 4.1 for V3 and 4.3 for VC3.

Since a few episodes dominate the aggregated errors, \autoref{tab:app_dm} tests the differences between
the models date by date, with the test of \cite{Diebold1995} on a Newey--West variance
\citep{Newey1987}.

\autoref{tab:app_ivfut} turns from the levels of the errors to their dynamics, with the correlation
of daily changes in the at-the-money implied volatility and in the risk reversal with daily changes
in the underlying future. At the money, V2, V3 and VC3 reproduce the sign and roughly the size of
the correlation in the data. The correlation of the risk reversal in the data falls with maturity,
from 0.45 at M1 to $-0.31$ at Q3. V2, V3 and VC3 follow this pattern with a smaller amplitude, 0.30
at M1 for V3. V3-U, two of whose $\rho_i$ are large, comes closer to the data at the front but keeps
a positive correlation at every contract, where the data turn negative from M6 out.

\subsection{The model-free moments}\label{app:cm}
For the volatility, we interpolate the data's $\mathcal M_2^2=\varkappa_2/\tau$,
computed as in Appendix~\ref{sub:data_construction}, linearly in
maturity onto constant maturities from 42 to 210 business days, while the
model's is evaluated at these maturities directly. The simulation in
\autoref{fig:res_cm2}(b) starts from the volatility factors that best fit the
term structure of $\varkappa_2$ on 2 January 2013, found by non-negative least
squares on its affine form in Proposition~\ref{prop:modelmom}, and simulates
them under $\mathbb P$ in daily steps.

For the skew, we fit the common shift $u_t$ of the leverage correlations of V3 to
the day's term structure of $\mathcal M_3$. Let $y_{t,j}$ be the observed
$\mathcal M_3$ of contract $j$ of M1--M7 on date $t$, computed in the same
way, and $c_{t,j,i}$ the model's $\mathcal M_3$ of that contract per unit of
$\rho_i$ at the filtered state. Since $\varkappa_3$ is linear in the $\rho_i$
and $\varkappa_2$ does not depend on them, the model's $\mathcal M_3$ is
$\sum_i c_{t,j,i}\rho_i$, and the shift is the least-squares fit
\begin{equation*}
  u_t=\argmin_{|u|\le2}\ \sum_j\Bigl(y_{t,j}-\sum_i c_{t,j,i}\tanh\bigl(\operatorname{atanh}\hat\rho_i+u\bigr)\Bigr)^2 .
\end{equation*}
The model's $\mathcal M_3$ is its exact value from
Proposition~\ref{prop:modelmom}, which the skew forcing does not enter.

At the estimate of V3, every $c_{t,j,i}$ is positive, so the model's
$\mathcal M_3$ rises in $u$, and we take the minimum over a fine grid of $u$ as
the fit.

The shift is at its bound on 5.2 percent of the dates and has a half-life of 83 business days. It
moves closely with the skew forcing of V3, with a correlation of 0.86. This indicates that the
shift, fitted to V3 after its estimation, follows the same movement of the skew that the forcing
carries in the estimation.

\subsection{The hedging construction}\label{app:hedge}
On each date, the portfolio holds every option used as a measurement that day, held to the next business day.
Each option, in the portfolio and among the hedge options, is delta-hedged in its own futures contract or
futures spread, which removes its exposure to $x$. This alone is the hedge with $k=0$. The hedge options
are the at-the-money options on M4, Q2 and M1 for the vanilla portfolio and on M2 and M4 for the calendar spread
portfolio, or the option nearest the money where there is none. For each option, their amounts solve
\begin{equation*}
	\min_\xi \; (r - V\xi)^\top W (r - V\xi) + \xi^\top N \xi,
	\qquad
	W = \Diag\!\left(\sigma_{v,i}^2\, v_{i,t}\right)/252,
	\qquad
	N = \frac{2 \cdot 0.01^2}{12}\, I,
\end{equation*}
with $r$ and $V$ the sensitivities to $v$ of the option and the hedge options, $W$ the model's variance of
the daily change in $v$, and $N$ that of rounding two settlements to the cent. All sensitivities are taken
at the observed underlying and the filtered volatility factors.

The hedged P\&L of an option is
\begin{equation*}
	\Pi = \Delta O - \delta\, \Delta U - \sum_{j=1}^k \xi_j \left( \Delta O^H_j - \delta^H_j\, \Delta U^H_j \right)
	- r_f \Delta t \left( O - \sum_{j=1}^k \xi_j O^H_j \right),
\end{equation*}
with $\delta$ the deltas, $U$ the underlyings, $H$ marking the hedge options, $r_f$ the zero-coupon rate
and $\Delta t$ the length of the holding period in years,
and that of the portfolio is the sum over its options. The hedging effectiveness is
$1-\operatorname{Var}(\Pi)/\operatorname{Var}(\Pi^u)$ \citep{ederington1979}, or
$1-(\mathrm{MAD}(\Pi)/\mathrm{MAD}(\Pi^u))^2$, for the hedged and unhedged P\&L of the portfolio per unit of its
total vega, and the mean absolute daily P\&L per option is the mean over the dates of each day's mean of
$|\Pi|$. The tests use Newey--West standard errors over 50 lags.

The ten-day hedges of the calendar spread options keep the daily positions and sum each option's
daily P\&L over overlapping ten-day windows, with Newey--West standard errors over 60 lags.

\autoref{tab:app_hedge_oi} weights each option by its open interest. The unhedged portfolio then
has a slightly smaller mean absolute daily P\&L per option, 24.12 cents against 25.38, and the
models rank as with equal weights at every number of hedge options.

\begin{table}[!tbp]
\centering\footnotesize
\setlength{\aboverulesep}{0pt}\setlength{\belowrulesep}{0pt}%
\renewcommand{\arraystretch}{1.05}%
\begin{tabular}{@{}l|c|c|ccc@{}}
\toprule
 & & \multicolumn{4}{c}{hedge options $k$}\\
\cmidrule(l){3-6}
model & unhedged & 0 & 1 & 2 & 3\\
\midrule
V2 & 24.12 & 7.79 & 4.20 & \textbf{3.57} & ---\\[2pt]
V3 & 24.12 & 7.77 & 4.87 & 3.63 & \textbf{3.06}\\[2pt]
VC3 & 24.12 & 7.80 & 4.19 & 3.58 & \textbf{3.26}\\[2pt]
V2-U & 24.12 & 7.80 & 4.19 & \textbf{3.56} & ---\\[2pt]
V3-U & 24.11 & 7.89 & 4.13 & 3.52 & \textbf{5.36}\\[2pt]
\bottomrule
\end{tabular}
\caption[Hedging the vanilla options, open-interest weights]{\textbf{Hedging the vanilla options, open-interest weights}\\ The hedges of \autoref{tab:res_hedge} over the same 2{,}809 dates: the mean absolute daily P\&L per option, in cents per barrel, with each option weighted by its open interest on the day the hedge is set. Bold where $k=n$. Each change from the column to the left is tested with the Newey--West $t$-statistic (50 lags). Every change is significant at the 1 percent level.}
\label{tab:app_hedge_oi}
\end{table}

\subsection{Out-of-sample pricing and hedging}\label{app:oos}
We split the sample into two halves of 1{,}405 trading days, from January 2013 to July 2018 and
from August 2018 to February 2024, and estimate V3 on each half alone, with its constraints
evaluated on that half. \autoref{tab:app_oos_est} reports the estimates, which keep the signs of
the full sample's leverage correlations in both halves. With the parameters held fixed, we then
price and hedge each half as in Section~\ref{sec:res_pricing}, with the estimate from that half, in
sample, with the estimate from the other half, out of sample, and with the full-sample estimate
for reference. \autoref{tab:app_oos_eval} reports the results.

Out of sample, the aggregated vanilla errors do not rise significantly in either half, although those
at the money rise in the earlier half and those of the out-of-the-money calls in the later half.
The futures errors rise in both halves, significantly only in the earlier, and the error in the
level of $\mathcal M_2$ rises significantly in both. The out-of-sample hedges are close to the in-sample
ones with every number of hedge options, and the third hedge option lowers the mean absolute P\&L
per option in both halves, in sample and out of sample.

\begin{table}[!tbp]
\centering\footnotesize
\setlength{\aboverulesep}{0pt}\setlength{\belowrulesep}{0pt}%
\renewcommand{\arraystretch}{1.1}%
\begin{tabular}{@{}l ccc| ccc| ccc@{}}
\toprule
 & \multicolumn{3}{c|}{full sample} & \multicolumn{3}{c|}{Jan 2013--Jul 2018} & \multicolumn{3}{c}{Aug 2018--Feb 2024}\\
\cmidrule(lr){2-4}\cmidrule(lr){5-7}\cmidrule(l){8-10}
factor $i$ & 1 & 2 & 3 & 1 & 2 & 3 & 1 & 2 & 3\\
\midrule
$\kappa_i$ & 2.237 & 0.702 & 0.060 & 3.344 & 0.965 & 0.039 & 2.346 & 0.874 & 0.135\\
 & {\scriptsize (0.0521)} & {\scriptsize (0.0137)} & {\scriptsize (0.0020)} & {\scriptsize (0.0981)} & {\scriptsize (0.0144)} & {\scriptsize (0.0017)} & {\scriptsize (0.1311)} & {\scriptsize (0.0368)} & {\scriptsize (0.0076)}\\[2pt]
$\beta_{0,i}$ & 16.476 & 11.814 & 5.946 & 11.368 & 6.467 & 4.672 & 21.396 & 13.021 & 7.431\\
 & {\scriptsize (0.3873)} & {\scriptsize (0.1365)} & {\scriptsize (0.1580)} & {\scriptsize (0.3001)} & {\scriptsize (0.3432)} & {\scriptsize (0.1390)} & {\scriptsize (0.9882)} & {\scriptsize (0.3793)} & {\scriptsize (0.3684)}\\[2pt]
$\beta_{1,i}$ & 2.578 & 7.685 & --- & 30.759 & 17.561 & --- & 1.728 & 14.193 & ---\\
 & {\scriptsize (0.3406)} & {\scriptsize (0.4215)} &  & {\scriptsize (2.4809)} & {\scriptsize (0.5449)} &  & {\scriptsize (0.1096)} & {\scriptsize (1.1576)} & \\[2pt]
\midrule
$\sigma_{v,i}$ & 14.794 & 1.027 & 0.901 & 9.919 & 1.394 & 0.958 & 13.605 & 0.914 & 0.975\\
 & {\scriptsize (0.5464)} & {\scriptsize (0.0484)} & {\scriptsize (0.0339)} & {\scriptsize (0.4837)} & {\scriptsize (0.0740)} & {\scriptsize (0.0327)} & {\scriptsize (0.5567)} & {\scriptsize (0.0324)} & {\scriptsize (0.0368)}\\[2pt]
$\rho_i$ & $-$0.071 & $-$0.274 & 0.268 & $-$0.139 & $-$0.289 & 0.244 & $-$0.021 & $-$0.242 & 0.245\\
 & {\scriptsize (0.0123)} & {\scriptsize (0.0260)} & {\scriptsize (0.0224)} & {\scriptsize (0.0127)} & {\scriptsize (0.0467)} & {\scriptsize (0.0449)} & {\scriptsize (0.0193)} & {\scriptsize (0.0343)} & {\scriptsize (0.0306)}\\[2pt]
$\Gamma_{1,i}$ & $-$5.544 & 3.800 & 0.351 & $-$5.317 & 3.721 & 0.260 & $-$4.703 & 3.167 & 0.355\\
 & {\scriptsize (0.4121)} & {\scriptsize ( --- )} & {\scriptsize ( --- )} & {\scriptsize (0.3178)} & {\scriptsize ( --- )} & {\scriptsize ( --- )} & {\scriptsize (0.3854)} & {\scriptsize ( --- )} & {\scriptsize ( --- )}\\[2pt]
$\Gamma_{2,i}$ & --- & $-$2.585 & 1.931 & --- & $-$4.320 & 3.228 & --- & $-$2.741 & 2.048\\
 &  & {\scriptsize (0.0721)} & {\scriptsize ( --- )} &  & {\scriptsize (0.1585)} & {\scriptsize ( --- )} &  & {\scriptsize (0.1762)} & {\scriptsize ( --- )}\\[2pt]
$\Gamma_{3,i}$ & --- & --- & $-$0.554 & --- & --- & $-$0.527 & --- & --- & $-$0.476\\
 &  &  & {\scriptsize (0.0168)} &  &  & {\scriptsize (0.0325)} &  &  & {\scriptsize (0.0360)}\\[2pt]
$\Gamma^{\mathbb P}_{i,i}$ & $-$6.601 & $-$2.411 & $-$0.400 & $-$9.894 & $-$3.710 & $-$0.400 & $-$5.063 & $-$2.023 & $-$0.400\\
 & {\scriptsize (4.8574)} & {\scriptsize (0.1798)} & {\scriptsize ( --- )} & {\scriptsize (1.6417)} & {\scriptsize (0.7878)} & {\scriptsize (0.0014)} & {\scriptsize (3.7730)} & {\scriptsize (0.4278)} & {\scriptsize ( --- )}\\[2pt]
$\lambda_{x,i}$ & $-$0.449 & $-$0.188 & $-$0.311 & $-$0.558 & $-$0.035 & $-$0.417 & $-$1.078 & $-$0.126 & $-$0.213\\
 & {\scriptsize (0.2206)} & {\scriptsize (0.0634)} & {\scriptsize (0.0549)} & {\scriptsize (0.2641)} & {\scriptsize (0.1897)} & {\scriptsize (0.3270)} & {\scriptsize (0.3312)} & {\scriptsize (0.2604)} & {\scriptsize (0.3629)}\\[2pt]
\midrule
$F_0$ & \multicolumn{3}{c|}{68.575} & \multicolumn{3}{c|}{80.541} & \multicolumn{3}{c}{52.373}\\
 & \multicolumn{3}{c|}{{\scriptsize (0.6721)}} & \multicolumn{3}{c|}{{\scriptsize (1.1342)}} & \multicolumn{3}{c}{{\scriptsize (0.2845)}}\\[2pt]
$\log\sigma_F$ & \multicolumn{3}{c|}{$-$1.669} & \multicolumn{3}{c|}{$-$1.680} & \multicolumn{3}{c}{$-$1.835}\\
 & \multicolumn{3}{c|}{{\scriptsize (0.0156)}} & \multicolumn{3}{c|}{{\scriptsize (0.0200)}} & \multicolumn{3}{c}{{\scriptsize (0.0294)}}\\[2pt]
$\sigma_V$ & \multicolumn{3}{c|}{0.0091} & \multicolumn{3}{c|}{0.0077} & \multicolumn{3}{c}{0.0098}\\
 & \multicolumn{3}{c|}{{\scriptsize (0.0002)}} & \multicolumn{3}{c|}{{\scriptsize (0.0001)}} & \multicolumn{3}{c}{{\scriptsize (0.0003)}}\\[2pt]
$a_V$ & \multicolumn{3}{c|}{4.614} & \multicolumn{3}{c|}{4.072} & \multicolumn{3}{c}{4.618}\\
 & \multicolumn{3}{c|}{{\scriptsize (0.0945)}} & \multicolumn{3}{c|}{{\scriptsize (0.0852)}} & \multicolumn{3}{c}{{\scriptsize (0.1383)}}\\[2pt]
$\chi_V$ & \multicolumn{3}{c|}{0.123} & \multicolumn{3}{c|}{0.177} & \multicolumn{3}{c}{0.114}\\
 & \multicolumn{3}{c|}{{\scriptsize (0.0015)}} & \multicolumn{3}{c|}{{\scriptsize (0.0022)}} & \multicolumn{3}{c}{{\scriptsize (0.0018)}}\\[2pt]
$\log\mathcal L$ & \multicolumn{3}{c|}{481{,}426.46} & \multicolumn{3}{c|}{265{,}804.38} & \multicolumn{3}{c}{235{,}885.97}\\
\bottomrule
\end{tabular}
\caption[Out-of-sample estimates of V3]{\textbf{Out-of-sample estimates of V3}\\ As \autoref{tab:res_est}, for V3 estimated on the full sample and on each half of the sample alone, each half's estimation started from the full-sample estimate.}
\label{tab:app_oos_est}
\end{table}

\begin{table}[!tbp]
\centering\footnotesize
\setlength{\aboverulesep}{0pt}\setlength{\belowrulesep}{0pt}%
\renewcommand{\arraystretch}{1.05}%
\begin{tabular}{@{}l|cccc|cccc@{}}
\toprule
 & \multicolumn{4}{c|}{Jan 2013--Jul 2018} & \multicolumn{4}{c}{Aug 2018--Feb 2024}\\
\cmidrule(lr){2-5}\cmidrule(l){6-9}
 & IS & OOS & full & $t$ & IS & OOS & full & $t$\\
\midrule
\multicolumn{9}{@{}l}{\emph{A. Pricing errors}}\\
Futures & 0.29 & 0.49 & 0.30 & 5.47 & 0.52 & 0.80 & 0.57 & 1.86\\
Vanilla, all & 0.53 & 0.56 & 0.54 & 1.36 & 1.02 & 1.07 & 1.04 & 1.56\\
Vanilla, OTM puts & 0.72 & 0.75 & 0.74 & 1.56 & 1.11 & 1.10 & 1.11 & $-$1.06\\
Vanilla, ATM & 0.21 & 0.24 & 0.23 & 5.13 & 0.37 & 0.36 & 0.36 & $-$1.87\\
Vanilla, OTM calls & 0.67 & 0.70 & 0.67 & 0.58 & 1.59 & 1.77 & 1.67 & 2.12\\
$\mathcal M_2$ level & 0.38 & 0.65 & 0.40 & 3.29 & 1.51 & 1.79 & 1.63 & 2.25\\
$\mathcal M_2$ slope & 0.34 & 0.48 & 0.43 & 3.84 & 0.81 & 0.81 & 0.79 & 0.01\\
\midrule
\multicolumn{9}{@{}l}{\emph{B. Hedging effectiveness}}\\
$k=0$ & 0.099 & 0.100 & 0.100 & $-$0.21 & 0.093 & 0.086 & 0.090 & 1.28\\
 & ($-$2.42) & ($-$2.62) & ($-$2.52) &  & ($-$1.41) & ($-$1.29) & ($-$1.35) & \\[2pt]
$k=1$ & 0.917 & 0.934 & 0.934 & $-$2.60 & 0.943 & 0.922 & 0.940 & 3.15\\
 & ($-$5.35) & ($-$5.23) & ($-$5.26) &  & ($-$3.56) & ($-$3.56) & ($-$3.56) & \\[2pt]
$k=2$ & 0.964 & 0.961 & 0.962 & 1.86 & 0.957 & 0.953 & 0.956 & 1.02\\
 & ($-$3.78) & ($-$3.93) & ($-$3.86) &  & ($-$2.36) & ($-$3.47) & ($-$2.55) & \\[2pt]
$k=3$ & 0.975 & 0.973 & 0.974 & 3.93 & 0.966 & 0.963 & 0.965 & 1.14\\
 & ($-$4.21) & ($-$3.98) & ($-$4.17) &  & ($-$2.18) & ($-$2.59) & ($-$2.28) & \\[2pt]
\midrule
\multicolumn{9}{@{}l}{\emph{C. Mean absolute daily P\&L per option, cents}}\\
unhedged & 21.33 & 21.33 & 21.33 &  & 29.43 & 29.43 & 29.43 & \\
$k=0$ & 6.26 & 6.26 & 6.26 & $-$0.44 & 9.39 & 9.43 & 9.40 & 3.19\\
 & ($-$19.90) & ($-$19.90) & ($-$19.90) &  & ($-$13.19) & ($-$13.19) & ($-$13.19) & \\[2pt]
$k=1$ & 3.55 & 3.75 & 3.73 & 5.49 & 5.36 & 5.19 & 5.35 & $-$4.03\\
 & ($-$12.44) & ($-$12.71) & ($-$12.87) &  & ($-$9.19) & ($-$9.15) & ($-$9.26) & \\[2pt]
$k=2$ & 2.56 & 2.66 & 2.65 & 7.44 & 3.97 & 3.94 & 3.97 & $-$1.09\\
 & ($-$6.71) & ($-$6.35) & ($-$6.24) &  & ($-$8.80) & ($-$9.17) & ($-$8.45) & \\[2pt]
$k=3$ & 2.30 & 2.30 & 2.27 & $-$0.50 & 3.60 & 3.69 & 3.59 & 2.78\\
 & ($-$12.75) & ($-$12.06) & ($-$12.35) &  & ($-$9.10) & ($-$9.68) & ($-$10.17) & \\[2pt]
\bottomrule
\end{tabular}
\caption[Out-of-sample pricing and hedging of V3]{\textbf{Out-of-sample pricing and hedging of V3}\\ V3 evaluated on each half of the sample, with the estimate from that half (IS), from the other half (OOS) and from the full sample (full). The column $t$ is the Newey--West $t$-statistic (50 lags) of the per-date difference in the loss, OOS less IS. (A)~Pricing errors over the 1{,}405 dates of each half: for the futures, the RMSE of the relative pricing errors, in percent, for the vanilla options, the MAE of the implied Bachelier volatilities, in \$/bbl a year, aggregated and by moneyness as in \autoref{tab:res_vanmae}, and for $\mathcal M_2$, the RMSE of its level and slope, in \$/bbl a year, as in \autoref{fig:res_cm2}. The loss in $t$ is the squared error for an RMSE and the absolute error for an MAE. (B)~Hedging effectiveness per unit of portfolio vega and (C)~mean absolute daily P\&L per option, in cents per barrel, of the hedges of \autoref{tab:res_hedge} with $k$ hedge options, over 1{,}404 dates in each half. In parentheses, the Newey--West $t$-statistic of the change from the row above, in the variance of the P\&L per unit of portfolio vega (B) or in the absolute P\&L per option (C), and the column $t$ tests the same loss. A negative $t$ is a reduction.}
\label{tab:app_oos_eval}
\end{table}

\subsection{Calendar spread options}\label{app:cso}
\autoref{tab:app_cso_hedge_periods} reports the ten-day hedges of Section~\ref{sec:res_cso} by
subperiod, and over the full sample without the 1 percent of dates on which the unhedged
portfolio's daily P\&L deviates most from its mean. With one or two hedge options, the
median-based effectiveness of VC3 is the higher in every subperiod. Without the extreme dates,
the hedges of VC3 with two hedge options also have a significantly lower variance and mean
absolute P\&L per option than those of V3, with $t$-statistics of $-3.46$ and $-3.14$.

\autoref{tab:app_cso_vanilla} compares V3 and VC3 on the futures and vanilla options. Filtered on
these alone, the parameters of VC3 give a log-likelihood of 446{,}720, against 481{,}426 for V3,
which is what including the calendar spread options in the estimation costs on the futures and
vanilla options. At its own filtered states, VC3 has larger vanilla pricing errors in every
moneyness group, and its model-implied $\mathcal M_2$ follows the data less closely.

\begin{table}[!tbp]
\centering\footnotesize\setlength{\tabcolsep}{3pt}
\setlength{\aboverulesep}{0pt}\setlength{\belowrulesep}{0pt}%
\renewcommand{\arraystretch}{1.05}%
\resizebox{\textwidth}{!}{%
\begin{tabular}{@{}l|ccc|ccc|ccc|ccc|ccc@{}}
\toprule
 & \multicolumn{3}{c|}{2013--2024} & \multicolumn{3}{c|}{2013--2019} & \multicolumn{3}{c|}{2020--2021} & \multicolumn{3}{c|}{2022--2024} & \multicolumn{3}{c}{top 1\% of dates removed}\\
 & \multicolumn{3}{c|}{2{,}796 windows} & \multicolumn{3}{c|}{1{,}754 windows} & \multicolumn{3}{c|}{496 windows} & \multicolumn{3}{c|}{528 windows} & \multicolumn{3}{c}{2{,}657 windows}\\
\cmidrule(lr){2-4}\cmidrule(lr){5-7}\cmidrule(lr){8-10}\cmidrule(lr){11-13}\cmidrule(l){14-16}
hedge & V3 & VC3 & $t$ & V3 & VC3 & $t$ & V3 & VC3 & $t$ & V3 & VC3 & $t$ & V3 & VC3 & $t$\\
\midrule
\multicolumn{16}{@{}l}{\emph{A. Hedging effectiveness}}\\
$k=0$, spreads & 0.38\textsuperscript{\dag} & 0.48\textsuperscript{n.s.} & $-$1.06 & 0.20\textsuperscript{n.s.} & 0.14\textsuperscript{n.s.} & 2.14 & 0.55\textsuperscript{n.s.} & 0.58\textsuperscript{n.s.} & $-$1.18 & 0.27\textsuperscript{n.s.} & 0.48\textsuperscript{n.s.} & $-$1.07 & 0.28\textsuperscript{\dag} & 0.20\textsuperscript{n.s.} & 3.47\\
 & (0.34) & (0.25) &  & (0.41) & (0.35) &  & (0.39) & (0.39) &  & (0.03) & ($-$0.28) &  & (0.35) & (0.26) & \\[2pt]
$k=1$, M2 & 0.64\textsuperscript{\dag} & 0.69\textsuperscript{\dag} & $-$0.92 & 0.80\textsuperscript{\dag} & 0.78\textsuperscript{\dag} & 0.43 & 0.86\textsuperscript{n.s.} & 0.84\textsuperscript{n.s.} & 1.11 & 0.42\textsuperscript{n.s.} & 0.54\textsuperscript{n.s.} & $-$1.28 & 0.71 & 0.71 & 0.06\\
 & (0.57) & (0.63) &  & (0.56) & (0.60) &  & (0.65) & (0.70) &  & (0.51) & (0.62) &  & (0.52) & (0.60) & \\[2pt]
$k=2$, M2 and M4 & 0.85\textsuperscript{n.s.} & 0.86\textsuperscript{n.s.} & $-$0.61 & 0.88 & 0.91 & $-$1.64 & 0.87\textsuperscript{n.s.} & 0.85\textsuperscript{n.s.} & 0.96 & 0.82\textsuperscript{n.s.} & 0.84\textsuperscript{n.s.} & $-$2.13 & 0.79 & 0.84 & $-$3.46\\
 & (0.72) & (0.79) &  & (0.75) & (0.80) &  & (0.72) & (0.82) &  & (0.62) & (0.72) &  & (0.70) & (0.77) & \\[2pt]
\midrule
\multicolumn{16}{@{}l}{\emph{B. Mean absolute 10-day P\&L per option, cents}}\\
unhedged & 5.61 & 5.61 &  & 4.45 & 4.45 &  & 7.12 & 7.12 &  & 8.10 & 8.10 &  & 4.78 & 4.78 & \\[2pt]
$k=0$, spreads & 3.42 & 3.20 & $-$2.44 & 2.63 & 2.46 & $-$2.86 & 4.50 & 4.41 & $-$1.18 & 5.03 & 4.56 & $-$1.17 & 2.76 & 2.65 & $-$2.62\\[2pt]
$k=1$, M2 & 2.90 & 2.79 & $-$1.56 & 2.19 & 2.00 & $-$1.99 & 4.27\textsuperscript{n.s.} & 4.44\textsuperscript{n.s.} & 0.99 & 3.99 & 3.86 & $-$1.13 & 2.39 & 2.25 & $-$2.10\\[2pt]
$k=2$, M2 and M4 & 2.65 & 2.50 & $-$2.24 & 2.05 & 1.81 & $-$2.86 & 3.88\textsuperscript{n.s.} & 4.02\textsuperscript{n.s.} & 0.89 & 3.47\textsuperscript{n.s.} & 3.39\textsuperscript{\dag} & $-$1.22 & 2.25 & 2.07 & $-$3.14\\[2pt]
\bottomrule
\end{tabular}}
\caption[Hedging the calendar spread options over ten days]{\textbf{Hedging the calendar spread options over ten days}\\ The hedges of \autoref{tab:res_cso_hedge} with their daily P\&L summed over overlapping ten-business-day windows: over the full sample, within each subperiod, and over the full sample without the 1 percent of dates on which the unhedged portfolio's daily P\&L deviates most from its mean, dropping every window that contains one of them. (A)~Hedging effectiveness per unit of portfolio vega, with the version based on the median absolute deviation in parentheses. (B)~Mean absolute ten-day P\&L per option, in cents per barrel. Each change from the row above is tested with the Newey--West $t$-statistic (60 lags). Unless marked, every change is significant at the 1 percent level; \textsuperscript{\dag} marks significance at the 5 percent level only and \textsuperscript{n.s.} no significance at the 5 percent level. The column $t$ is the Newey--West $t$-statistic of the paired difference between the models, VC3 less V3, in the variance of the P\&L per unit of portfolio vega (A) or in the absolute P\&L per option (B).}
\label{tab:app_cso_hedge_periods}
\end{table}

\subsection{Arithmetic and geometric models}\label{app:geom}
\autoref{tab:app_geom_est} reports the estimates of the two-factor arithmetic model and its geometric
counterpart of Section~\ref{sec:res_geom}, whose estimation is described in
Appendix~\ref{sub:geometric}. Since the comparison measures the vanilla options differently and leaves out a few of them, its
arithmetic model is a separate estimate from V2-U.

\begin{table}[p]
\centering\footnotesize
\setlength{\aboverulesep}{0pt}\setlength{\belowrulesep}{0pt}%
\renewcommand{\arraystretch}{1.05}%
\begin{tabular}{@{}lcc@{}}
\toprule
 & V3 & VC3\\
\midrule
$\log\mathcal L$, futures and vanilla options & 481{,}426.46 & 446{,}720.37\\
$\mathcal M_2$ level, RMSE / $R^2$ & 1.19 / 0.981 & 1.27 / 0.978\\
$\mathcal M_2$ slope, RMSE / $R^2$ & 0.63 / 0.974 & 0.84 / 0.954\\
share of dates in the 99 percent band, level / slope & 0.91 / 0.99 & 0.89 / 0.98\\
vanilla MAE, all / bucket 0 & 0.79 / 0.18 & 0.86 / 0.31\\
vanilla MAE, bucket $-4$ / $4$ & 1.27 / 1.78 & 1.34 / 1.91\\
\bottomrule
\end{tabular}
\caption[V3 and VC3 on the futures and vanilla options]{\textbf{V3 and VC3 on the futures and vanilla options}\\ The first row is the log-likelihood of the futures and vanilla options alone, for VC3 at its estimate with the calendar spread options left out of the filter. The other rows are at each model's own filtered states, for VC3 with the calendar spread options in the filter, as in the main text. For $\mathcal M_2$, the rows give the RMSE, in \$/bbl a year, and the $R^2$ of the model-implied level and slope, as in \autoref{fig:res_cm2}(a), and the share of dates on which the observed level and slope lie in the 99 percent band of the paths simulated under $\mathbb P$, as in \autoref{fig:res_cm2}(b). The last two rows give the MAE of the implied Bachelier volatilities, in \$/bbl a year, aggregated and in delta bucket 0, and in the outermost delta buckets.}
\label{tab:app_cso_vanilla}
\end{table}

\begin{table}[p]
\centering\footnotesize
\setlength{\aboverulesep}{0pt}\setlength{\belowrulesep}{0pt}%
\renewcommand{\arraystretch}{1.0}%
\begin{tabular}{@{}l cc| cc@{}}
\toprule
 & \multicolumn{2}{c|}{Arithmetic} & \multicolumn{2}{c}{Geometric}\\
\cmidrule(lr){2-3}\cmidrule(l){4-5}
factor $i$ & 1 & 2 & 1 & 2\\
\midrule
$\kappa_i$ & 0.707 & 0.011 & 0.708 & 0.080\\
 & {\scriptsize (0.0358)} & {\scriptsize ( --- )} & {\scriptsize (0.0647)} & {\scriptsize (0.0095)}\\[2pt]
$\beta_{0,i}$ & 14.451 & 9.340 & 0.358 & 0.240\\
 & {\scriptsize (0.7687)} & {\scriptsize (0.2304)} & {\scriptsize (0.0130)} & {\scriptsize (0.0268)}\\[2pt]
$\beta_{1,i}$ & 2.699 & --- & 0.019 & ---\\
 & {\scriptsize (0.6012)} &  & {\scriptsize (0.0013)} & \\[2pt]
\midrule
$\sigma_{v,i}$ & 17.995 & 0.808 & 17.580 & 0.798\\
 & {\scriptsize (3.2387)} & {\scriptsize (0.0349)} & {\scriptsize (1.3834)} & {\scriptsize (0.0384)}\\[2pt]
$\rho_i$ & $-$0.123 & 0.226 & $-$0.674 & $-$0.408\\
 & {\scriptsize (0.0344)} & {\scriptsize (0.0741)} & {\scriptsize (0.0695)} & {\scriptsize (0.0493)}\\[2pt]
$\Gamma_{1,i}$ & $-$20.396 & 14.023 & $-$30.026 & 22.520\\
 & {\scriptsize (4.7813)} & {\scriptsize ( --- )} & {\scriptsize (5.2766)} & {\scriptsize ( --- )}\\[2pt]
$\Gamma_{2,i}$ & --- & $-$0.584 & --- & $-$0.319\\
 &  & {\scriptsize (0.0371)} &  & {\scriptsize (0.0306)}\\[2pt]
$\Gamma^{\mathbb P}_{i,i}$ & $-$23.643 & $-$0.400 & $-$11.172 & $-$0.400\\
 & {\scriptsize (4.8017)} & {\scriptsize ( --- )} & {\scriptsize (4.7809)} & {\scriptsize ( --- )}\\[2pt]
$\lambda_{x,i}$ & 0.027 & $-$0.145 & $-$0.151 & $-$0.010\\
 & {\scriptsize (0.2062)} & {\scriptsize (0.1415)} & {\scriptsize (0.2433)} & {\scriptsize (0.0648)}\\[2pt]
\midrule
$F_0$ & \multicolumn{2}{c|}{23.247} & \multicolumn{2}{c}{52.840}\\
 & \multicolumn{2}{c|}{{\scriptsize (2.8001)}} & \multicolumn{2}{c}{{\scriptsize (1.3952)}}\\[2pt]
$\log\sigma_F$ & \multicolumn{2}{c|}{$-$0.795} & \multicolumn{2}{c}{$-$4.365}\\
 & \multicolumn{2}{c|}{{\scriptsize (0.0131)}} & \multicolumn{2}{c}{{\scriptsize (0.0678)}}\\[2pt]
$\sigma_V$ & \multicolumn{2}{c|}{0.0084} & \multicolumn{2}{c}{0.0159}\\
 & \multicolumn{2}{c|}{{\scriptsize (0.0007)}} & \multicolumn{2}{c}{{\scriptsize (0.0050)}}\\[2pt]
$a_V$ & \multicolumn{2}{c|}{2.749} & \multicolumn{2}{c}{3.722}\\
 & \multicolumn{2}{c|}{{\scriptsize (0.3015)}} & \multicolumn{2}{c}{{\scriptsize (1.6683)}}\\[2pt]
$\log\mathcal L$ & \multicolumn{2}{c|}{552{,}125.60} & \multicolumn{2}{c}{495{,}264.74}\\
\bottomrule
\end{tabular}
\caption[Parameter estimates of the arithmetic and geometric models]{\textbf{Parameter estimates of the arithmetic and geometric models}\\ As \autoref{tab:res_est}, for the two-factor arithmetic model and its geometric counterpart, both estimated without the skew forcing on the comparison panel of \autoref{tab:res_geom}. The loadings and the futures noise of the geometric model are in log units.}
\label{tab:app_geom_est}
\end{table}

\FloatBarrier

%% file: bib.bib
@article{SeokBrorsenNiyibizi2018,
  author  = {Seok, Juheon and Brorsen, B. Wade and Niyibizi, Bart},
  title   = {Modeling Calendar Spread Options},
  journal = {Agricultural Finance Review},
  year    = {2018},
  volume  = {78},
  number  = {5},
  pages   = {551--570},
  doi     = {10.1108/AFR-09-2017-0088},
}

@inproceedings{SchmitzWangKimn2013,
  author    = {Schmitz, Adam and Wang, Zhiguang and Kimn, Jung-Han},
  title     = {Pricing and Hedging Calendar Spread Options on Agricultural Grain Commodities},
  booktitle = {Proceedings of the NCCC-134 Conference on Applied Commodity Price Analysis, Forecasting, and Market Risk Management},
  address   = {St.\ Louis, MO},
  year      = {2013},
  doi       = {10.22004/ag.econ.285794},
}

@article{BenthDiNunnoKhedherSchmeck2015,
  author  = {Benth, Fred Espen and Di Nunno, Giulia and Khedher, Asma and Schmeck, Maren Diane},
  title   = {Pricing of Spread Options on a Bivariate Jump Market and Stability to Model Risk},
  journal = {Applied Mathematical Finance},
  year    = {2015},
  volume  = {22},
  number  = {1},
  pages   = {28--62},
}

@article{BenthPiccirilliVargiolu2019,
  author  = {Benth, Fred Espen and Piccirilli, Marco and Vargiolu, Tiziano},
  title   = {Mean-Reverting Additive Energy Forward Curves in a {H}eath--{J}arrow--{M}orton Framework},
  journal = {Mathematics and Financial Economics},
  year    = {2019},
  volume  = {13},
  number  = {4},
  pages   = {543--577},
}

@article{LatiniPiccirilliVargiolu2019,
  author  = {Latini, Luca and Piccirilli, Marco and Vargiolu, Tiziano},
  title   = {Mean-Reverting No-Arbitrage Additive Models for Forward Curves in Energy Markets},
  journal = {Energy Economics},
  year    = {2019},
  volume  = {79},
  pages   = {157--170}
}

@article{PiccirilliSchmeckVargiolu2021,
  author  = {Piccirilli, Marco and Schmeck, Maren Diane and Vargiolu, Tiziano},
  title   = {Capturing the Power Options Smile by an Additive Two-Factor Model for Overlapping Futures Prices},
  journal = {Energy Economics},
  year    = {2021},
  volume  = {95},
  pages   = {105006},

}

@phdthesis{Piccirilli2018,
  author = {Piccirilli, Marco},
  title  = {Additive Models for Energy Markets},
  school = {Universit{\`a} degli Studi di Padova, Dipartimento di Matematica},
  year   = {2018}
}

@article{LuciaSchwartz2002,
  author  = {Lucia, Julio J. and Schwartz, Eduardo S.},
  title   = {Electricity Prices and Power Derivatives: Evidence from the {N}ordic {P}ower {E}xchange},
  journal = {Review of Derivatives Research},
  year    = {2002},
  volume  = {5},
  number  = {1},
  pages   = {5--50}
}

@article{Bachelier1900,
  author  = {Bachelier, Louis},
  title   = {Th{\'e}orie de la sp{\'e}culation},
  journal = {Annales scientifiques de l'{\'E}cole Normale Sup{\'e}rieure},
  year    = {1900},
  volume  = {17},
  pages   = {21--86},
  series  = {3}
}

@article{ChoiKwakTeeWang2022,
  author  = {Choi, Jaehyuk and Kwak, Minsuk and Tee, Chyng Wen and Wang, Yumeng},
  title   = {A {B}lack--{S}choles User's Guide to the {B}achelier Model},
  journal = {Journal of Futures Markets},
  year    = {2022},
  volume  = {42},
  number  = {5},
  pages   = {959--980},
  doi     = {10.1002/fut.22315},
}

@misc{CME2020Bachelier,
  author       = {{CME Group}},
  title        = {Switch to {B}achelier Options Pricing Model Effective {A}pril 22, 2020},
  howpublished = {Clearing Advisory 20-171},
  year         = {2020},
  month        = apr,
  url          = {https://www.cmegroup.com/content/cmegroup/en/notices/clearing/2020/04/Chadv20-171.html}
}

@techreport{Gabillon1991,
  author      = {Gabillon, Jacques},
  title       = {The Term Structures of Oil Futures Prices},
  institution = {Oxford Institute for Energy Studies},
  type        = {Working Paper},
  year        = {1991},
  number      = {WPM 17}
}

@article{BorovkovaGeman2006,
  author  = {Borovkova, Svetlana and Geman, H{\'e}lyette},
  title   = {Seasonal and Stochastic Effects in Commodity Forward Curves},
  journal = {Review of Derivatives Research},
  year    = {2006},
  volume  = {9},
  number  = {2},
  pages   = {167--186}
}

@article{ZikesBarunik2016,
  author  = {{\v{Z}}ike{\v{s}}, Filip and Barun{\'\i}k, Jozef},
  title   = {Semi-Parametric Conditional Quantile Models for Financial Returns and Realized Volatility},
  journal = {Journal of Financial Econometrics},
  year    = {2016},
  volume  = {14},
  number  = {1},
  pages   = {185--226},
}

@article{ballestra2026multifactor,
  author  = {Ballestra, Luca Vincenzo and Tezza, Christian},
  title   = {A multi-factor model for improved commodity pricing: calibration and an application to the oil market},
  journal = {Quantitative Finance},
  year    = {2026},
  volume  = {26},
  number  = {3},
  pages   = {449--466},
  doi     = {10.1080/14697688.2026.2619531}
}

@article{hughen2010maximal,
  author  = {Hughen, W. Keener},
  title   = {A maximal affine stochastic volatility model of oil prices},
  journal = {Journal of Futures Markets},
  year    = {2010},
  volume  = {30},
  number  = {2},
  pages   = {101--133}
}

@article{schone2017fourfactor,
  author  = {Sch{\"o}ne, Max F. and Spinler, Stefan},
  title   = {A four-factor stochastic volatility model of commodity prices},
  journal = {Review of Derivatives Research},
  year    = {2017},
  volume  = {20},
  number  = {2},
  pages   = {135--165},
  doi     = {10.1007/s11147-016-9126-y}
}

@article{yan2002valuation,
  author  = {Yan, Xuemin Sterling},
  title   = {Valuation of commodity derivatives in a new multi-factor model},
  journal = {Review of Derivatives Research},
  year    = {2002},
  volume  = {5},
  number  = {3},
  pages   = {251--271},
  doi     = {10.1023/A:1020871616158}
}

@article{gilpelaez1951,
  title = {Note on the inversion theorem},
  volume = {38},
  ISSN = {1464-3510},
  url = {http://dx.doi.org/10.1093/biomet/38.3-4.481},
  DOI = {10.1093/biomet/38.3-4.481},
  number = {3–4},
  journal = {Biometrika},
  publisher = {Oxford University Press (OUP)},
  author = {Gil-Pelaez,  J.},
  year = {1951},
  pages = {481–482}
}

@article{Tavin2020,
journal={Economics Bulletin},
author={Guillaume Coqueret and Bertrand Tavin},
title={A note on implied correlation for bivariate contracts},
year={2020},
pages={1388-1396},
volume={40},
number={2},
url={https://ideas.repec.org/a/ebl/ecbull/eb-20-00109.html},
}

@article{FRAU2025108809,
title = {Energy commodities and calendar spread options},
journal = {Energy Economics},
volume = {151},
pages = {108809},
year = {2025},
issn = {0140-9883},
doi = {https://doi.org/10.1016/j.eneco.2025.108809},
url = {https://www.sciencedirect.com/science/article/pii/S014098832500636X},
author = {Carme Frau and Gianluca Fusai and Ioannis Kyriakou}
}

@article{Heath1992,
  title = {Bond Pricing and the Term Structure of Interest Rates: A New Methodology for Contingent Claims Valuation},
  volume = {60},
  ISSN = {0012-9682},
  url = {http://dx.doi.org/10.2307/2951677},
  DOI = {10.2307/2951677},
  number = {1},
  journal = {Econometrica},
  publisher = {JSTOR},
  author = {Heath,  David and Jarrow,  Robert and Morton,  Andrew},
  year = {1992},
  month = jan,
  pages = {77-105}
}

@article{Cheng2018,
  title = {Pricing of long-dated commodity derivatives: Do stochastic interest rates matter?},
  volume = {95},
  ISSN = {0378-4266},
  url = {http://dx.doi.org/10.1016/j.jbankfin.2017.05.012},
  DOI = {10.1016/j.jbankfin.2017.05.012},
  journal = {Journal of Banking \& Finance},
  publisher = {Elsevier BV},
  author = {Cheng,  Benjamin and Nikitopoulos,  Christina Sklibosios and Schl\"{o}gl,  Erik},
  year = {2018},
  month = oct,
  pages = {148–166}
}

@BOOK{Duffie2001,
  title     = "Dynamic asset pricing theory",
  author    = "Duffie, Darrell",
  publisher = "Princeton University Press",
  series    = "Princeton Series in Finance",
  edition   = "Third",
  month     =  oct,
  year      =  2001,
  address   = "Princeton, NJ",
  language  = "en"
}

@article{Duffee2002,
  title = {Term Premia and Interest Rate Forecasts in Affine Models},
  volume = {57},
  ISSN = {1540-6261},
  DOI = {10.1111/1540-6261.00426},
  number = {1},
  journal = {The Journal of Finance},
  publisher = {Wiley},
  author = {Duffee,  Gregory R.},
  year = {2002},
  month = feb,
  pages = {405–443}
}

@article{Gibson1990,
 ISSN = {00221082, 15406261},
 URL = {http://www.jstor.org/stable/2328801},
 author = {Rajna Gibson and Eduardo S. Schwartz},
 journal = {The Journal of Finance},
 number = {3},
 pages = {959--976},
 publisher = {[American Finance Association, Wiley]},
 title = {Stochastic Convenience Yield and the Pricing of Oil Contingent Claims},
 urldate = {2024-09-17},
 volume = {45},
 year = {1990}
}

@article{gsw2007,
  title = {The {U}.{S}. {T}reasury yield curve: 1961 to the present},
  volume = {54},
  ISSN = {0304-3932},
  url = {http://dx.doi.org/10.1016/j.jmoneco.2007.06.029},
  DOI = {10.1016/j.jmoneco.2007.06.029},
  number = {8},
  journal = {Journal of Monetary Economics},
  publisher = {Elsevier BV},
  author = {G\"{u}rkaynak,  Refet S. and Sack,  Brian and Wright,  Jonathan H.},
  year = {2007},
  month = nov,
  pages = {2291–2304}
}

@BOOK{Bouchouev2023,
  title     = "{Virtual Barrels: Quantitative Trading in the Oil Market}",
  author    = "Bouchouev, Ilia",
  publisher = "Springer International Publishing",
  year      =  2023,
  address   = "Cham, Switzerland",
  language  = "en"
}

@article{trolle2009,
    author = {Trolle, Anders B. and Schwartz, Eduardo S.},
    title = "{Unspanned stochastic volatility and the pricing of commodity derivatives}",
    journal = {The Review of Financial Studies},
    volume = {22},
    number = {11},
    pages = {4423-4461},
    year = {2009},
    month = {05},
    issn = {0893-9454},
    doi = {10.1093/rfs/hhp036},
    url = {https://doi.org/10.1093/rfs/hhp036},
}

@BOOK{Sarkka2013,
  title     = "Bayesian
               filtering and smoothing",
  author    = "S{\"a}rkk{\"a}, Simo",
  publisher = "Cambridge University Press",
  month     =  sep,
  year      =  2013,
  address   = "Cambridge, England"
}

@BOOK{Harvey1991,
  title     = "Forecasting, structural time series models and the Kalman filter",
  author    = "Harvey, Andrew C",
  publisher = "Cambridge University Press",
  year      =  1989,
  address   = "Cambridge, England"
}

@article{Bjork2006,
  title = {ON FINITE DIMENSIONAL REALIZATIONS FOR THE TERM STRUCTURE OF FUTURES PRICES},
  volume = {9},
  ISSN = {1793-6322},
  url = {http://dx.doi.org/10.1142/S0219024906003639},
  DOI = {10.1142/s0219024906003639},
  number = {3},
  journal = {International Journal of Theoretical and Applied Finance},
  publisher = {World Scientific Pub Co Pte Lt},
  author = {Tomas Björk and Magnus Blix and Camilla Landén},
  year = {2006},
  month = may,
  pages = {281–314}
}

@article{Andersen2010,
  title = {Markov models for commodity futures: Theory and practice},
  volume = {10},
  ISSN = {1469-7696},
  url = {http://dx.doi.org/10.1080/14697680903493599},
  DOI = {10.1080/14697680903493599},
  number = {8},
  journal = {Quantitative Finance},
  publisher = {Informa UK Limited},
  author = {Andersen,  Leif},
  year = {2010},
  month = oct,
  pages = {831–854}
}

@article{chiarella2013,
title = {Humps in the volatility structure of the crude oil futures market: New evidence},
journal = {Energy Economics},
volume = {40},
pages = {989-1000},
year = {2013},
issn = {0140-9883},
doi = {https://doi.org/10.1016/j.eneco.2013.05.019},
url = {https://www.sciencedirect.com/science/article/pii/S0140988313001072},
author = {Carl Chiarella and Boda Kang and Christina Sklibosios Nikitopoulos and Thuy-Duong Tô}
}

@article{Dai2000,
  title = {Specification Analysis of Affine Term Structure Models},
  volume = {55},
  ISSN = {1540-6261},
  url = {http://dx.doi.org/10.1111/0022-1082.00278},
  DOI = {10.1111/0022-1082.00278},
  number = {5},
  journal = {The Journal of Finance},
  publisher = {Wiley},
  author = {Dai,  Qiang and Singleton,  Kenneth J.},
  year = {2000},
  month = oct,
  pages = {1943–1978}
}

@article{Schneider2018,
  title = {From the {S}amuelson volatility effect to a {S}amuelson correlation effect: {A}n analysis of crude oil calendar spread options},
  volume = {95},
  ISSN = {0378-4266},
  url = {http://dx.doi.org/10.1016/j.jbankfin.2016.12.001},
  DOI = {10.1016/j.jbankfin.2016.12.001},
  journal = {Journal of Banking \& Finance},
  publisher = {Elsevier BV},
  author = {Schneider,  Lorenz and Tavin,  Bertrand},
  year = {2018},
  month = oct,
  pages = {185–202}
}

@article{Schneider2015,
  title = {Seasonal volatility in agricultural markets: {M}odelling and empirical investigations},
  volume = {334},
  ISSN = {1572-9338},
  url = {http://dx.doi.org/10.1007/s10479-021-04241-7},
  DOI = {10.1007/s10479-021-04241-7},
  number = {1–3},
  journal = {Annals of Operations Research},
  publisher = {Springer Science and Business Media LLC},
  author = {Schneider,  L. and Tavin,  B.},
  year = {2024},
  month = sep,
  pages = {7–58}
}

@article{Duffie2000,
  title = {Transform Analysis and Asset Pricing for Affine Jump-diffusions},
  volume = {68},
  ISSN = {1468-0262},
  url = {http://dx.doi.org/10.1111/1468-0262.00164},
  DOI = {10.1111/1468-0262.00164},
  number = {6},
  journal = {Econometrica},
  publisher = {The Econometric Society},
  author = {Duffie,  Darrell and Pan,  Jun and Singleton,  Kenneth},
  year = {2000},
  month = nov,
  pages = {1343–1376}
}

@techreport{CollinDufresne2003,
  title = {Generalizing the Affine Framework to {HJM} and Random Field Models},
  DOI = {10.2139/ssrn.410421},
  institution = {SSRN},
  type = {Working Paper},
  author = {Collin-Dufresne, Pierre and Goldstein, Robert S.},
  year = {2003}
}

@article{Black1976,
  title = {The pricing of commodity contracts},
  volume = {3},
  ISSN = {0304-405X},
  url = {http://dx.doi.org/10.1016/0304-405X(76)90024-6},
  DOI = {10.1016/0304-405x(76)90024-6},
  number = {1–2},
  journal = {Journal of Financial Economics},
  publisher = {Elsevier BV},
  author = {Black,  Fischer},
  year = {1976},
  month = jan,
  pages = {167–179}
}

@inproceedings{Wan2004,
  title = {The unscented {K}alman filter for nonlinear estimation},
  url = {http://dx.doi.org/10.1109/ASSPCC.2000.882463},
  DOI = {10.1109/asspcc.2000.882463},
  booktitle = {Proceedings of the IEEE 2000 Adaptive Systems for Signal Processing, Communications, and Control Symposium},
  publisher = {IEEE},
  year = {2000},
  author = {Wan, E. A. and {van der Merwe}, R.},
  pages = {153--158}
}

@article{Julier2004,
  title = {Unscented Filtering and Nonlinear Estimation},
  volume = {92},
  ISSN = {0018-9219},
  url = {http://dx.doi.org/10.1109/JPROC.2003.823141},
  DOI = {10.1109/jproc.2003.823141},
  number = {3},
  journal = {Proceedings of the IEEE},
  publisher = {Institute of Electrical and Electronics Engineers (IEEE)},
  author = {Julier, S. J. and Uhlmann, J. K.},
  year = {2004},
  month = mar,
  pages = {401–422}
}

@article{filipovic2005time,
  author  = {Filipovi{\'c}, Damir},
  title   = {Time-Inhomogeneous Affine Processes},
  journal = {Stochastic Processes and their Applications},
  volume  = {115},
  number  = {4},
  pages   = {639--659},
  year    = {2005},
  doi     = {10.1016/j.spa.2004.11.006}
}

@article{duffie2003affine,
  author  = {Duffie, Darrell and Filipovi{\'c}, Damir and Schachermayer, Walter},
  title   = {Affine Processes and Applications in Finance},
  journal = {The Annals of Applied Probability},
  volume  = {13},
  number  = {3},
  pages   = {984--1053},
  year    = {2003},
  doi     = {10.1214/aoap/1060202833}
}

@book{filipovic2009term,
  title={Term-structure models: A graduate course},
  author={Filipovi{\'c}, Damir},
  year={2009},
  publisher={Springer}
}

@book{benth2008stochastic,
  title={Stochastic modelling of electricity and related markets},
  author={Benth, Fred Espen and {\v{S}}altyt{\.e} Benth, J{\=u}rat{\.e} and Koekebakker, Steen},
  series={Advanced Series on Statistical Science and Applied Probability},
  volume={11},
  year={2008},
  publisher={World Scientific}
}

@book{benth2023stochastic,
  title={Stochastic models for prices dynamics in energy and commodity markets},
  author={Benth, Fred Espen and Kr{\"u}hner, Paul},
  year={2023},
  publisher={Springer}
}

@techreport{clewlow1999multi,
  title={A multi-factor model for energy derivatives},
  author={Clewlow, Les and Strickland, Chris},
  year={1999},
institution={Quantitative Finance Research Centre, University of Technology Sydney},
  type={Research Paper},
  number={28}
}

@article{carmonaludkovski2004,
  title={Spot convenience yield models for the energy markets},
  author={René Carmona and Michael Ludkovski},
  journal={Contemporary Mathematics},
  volume={351},
  pages={65--79},
  year={2004},
  publisher={The American Mathematical Society}
}

@article{MIRANTES2013,
title = {The stochastic seasonal behavior of energy commodity convenience yields},
journal = {Energy Economics},
volume = {40},
pages = {155-166},
year = {2013},
issn = {0140-9883},
doi = {https://doi.org/10.1016/j.eneco.2013.06.011},
url = {https://www.sciencedirect.com/science/article/pii/S0140988313001242},
author = {Garc{\'\i}a Mirantes, Andr{\'e}s and Poblaci{\'o}n, Javier and Serna, Gregorio}
}

@article{carr2001,
    author = {Carr, Peter and Madan, Dilip},
    title = {Optimal positioning in derivative securities},
    journal = {Quantitative Finance},
    volume = {1},
    number = {1},
    pages = {19--37},
    year = {2001}
}

@article{Hestenes1969,
  author  = {Hestenes, Magnus R.},
  title   = {Multiplier and gradient methods},
  journal = {Journal of Optimization Theory and Applications},
  volume  = {4},
  number  = {5},
  pages   = {303--320},
  year    = {1969}
}

@incollection{Powell1969,
  author    = {Powell, M. J. D.},
  title     = {A method for nonlinear constraints in minimization problems},
  booktitle = {Optimization},
  editor    = {Fletcher, R.},
  publisher = {Academic Press},
  address   = {London},
  pages     = {283--298},
  year      = {1969}
}

@article{Rockafellar1973,
  author  = {Rockafellar, R. Tyrrell},
  title   = {A dual approach to solving nonlinear programming problems by unconstrained optimization},
  journal = {Mathematical Programming},
  volume  = {5},
  number  = {1},
  pages   = {354--373},
  year    = {1973}
}

@article{Nelder1965,
  author  = {Nelder, J. A. and Mead, R.},
  title   = {A simplex method for function minimization},
  journal = {The Computer Journal},
  volume  = {7},
  number  = {4},
  pages   = {308--313},
  year    = {1965}
}

@article{Takahasi1974,
  author  = {Takahasi, Hidetosi and Mori, Masatake},
  title   = {Double exponential formulas for numerical integration},
  journal = {Publications of the Research Institute for Mathematical Sciences},
  volume  = {9},
  number  = {3},
  pages   = {721--741},
  year    = {1974}
}

@article{ederington1979,
  author  = {Ederington, Louis H.},
  title   = {The hedging performance of the new futures markets},
  journal = {The Journal of Finance},
  volume  = {34},
  number  = {1},
  pages   = {157--170},
  year    = {1979}
}

@article{Tsitouras2011,
  author  = {Tsitouras, Ch.},
  title   = {{R}unge--{K}utta pairs of order 5(4) satisfying only the first column simplifying assumption},
  journal = {Computers \& Mathematics with Applications},
  volume  = {62},
  number  = {2},
  pages   = {770--775},
  year    = {2011},
  doi     = {10.1016/j.camwa.2011.06.002}
}

@article{AndersenLake2019,
  author  = {Andersen, Leif B. G. and Lake, Mark},
  title   = {{H}eston\ldots{}Again?! {R}obust High-Precision Option Pricing by {F}ourier Transforms: Contour Deformations and Double-Exponential Quadrature},
  journal = {Wilmott},
  volume  = {2019},
  number  = {102},
  pages   = {22--41},
  year    = {2019},
  doi     = {10.1002/wilm.10774}
}

@article{Mueller2007,
  author  = {M{\"u}ller, Ulrich K.},
  title   = {A theory of robust long-run variance estimation},
  journal = {Journal of Econometrics},
  volume  = {141},
  number  = {2},
  pages   = {1331--1352},
  year    = {2007}
}

@article{Lazarus2018,
  author  = {Lazarus, Eben and Lewis, Daniel J. and Stock, James H. and Watson, Mark W.},
  title   = {{HAR} Inference: Recommendations for Practice},
  journal = {Journal of Business \& Economic Statistics},
  volume  = {36},
  number  = {4},
  pages   = {541--559},
  year    = {2018}
}

@article{Diebold1995,
  author  = {Diebold, Francis X. and Mariano, Roberto S.},
  title   = {Comparing Predictive Accuracy},
  journal = {Journal of Business \& Economic Statistics},
  volume  = {13},
  number  = {3},
  pages   = {253--263},
  year    = {1995}
}

@article{Newey1987,
  author  = {Newey, Whitney K. and West, Kenneth D.},
  title   = {A Simple, Positive Semi-Definite, Heteroskedasticity and Autocorrelation Consistent Covariance Matrix},
  journal = {Econometrica},
  volume  = {55},
  number  = {3},
  pages   = {703--708},
  year    = {1987}
}
